\documentclass[useAMS,usenatbib,usegraphicx]{mn2e}

\usepackage[british]{babel}
\usepackage{times}
\usepackage{latexsym,amssymb}
\usepackage[fleqn]{amsmath}
\usepackage{graphicx}
\usepackage{natbib}
\usepackage{journalnames}
\usepackage{color}
\usepackage[draft]{hyperref}
\usepackage{comment}

\usepackage[T1]{fontenc}
\usepackage{aecompl}

\usepackage{caption}
\usepackage{comment}
\usepackage{lipsum}
\usepackage{array}
\usepackage{dcolumn}
\usepackage{rotating}

\title[COHRS cloud catalogs]{The integrated properties of the molecular clouds \\from the JCMT CO(3-2) High Resolution Survey}

\author[D. Colombo et al.]{%
D. Colombo$^1$\thanks{E-mail: dcolombo@mpifr-bonn.mpg.de}, 
E. Rosolowsky$^2$,
A. Duarte-Cabral$^3$, A. Ginsburg$^4$,
\newauthor{J.~Glenn$^5$,  E.~Zetterlund$^5$, A. K. Hernandez$^6$, 
J. Dempsey$^7$, M. J. Currie$^{7,8}$}
\\
$^1$Max Planck Institute for Radioastronomy, Auf dem H\"ugel 69, D-53121, Bonn, Germany\\
$^2$Department of Physics, University of Alberta, 4-181 CCIS, Edmonton, AB T6G 2E1, Canada\\
$^3$School of Physics and Astronomy, Cardiff University, The Parade, Cardiff, CF24 3AA, UK\\
$^4$National Radio Astronomy Observatory, 1003 Lopezville Rd, Socorro, NM 87801, USA\\
$^5$CASA, Department of Astrophysical and Planetary Sciences, University of Colorado 389-UCB, Boulder, CO 80309, USA\\
$^6$Department of Astronomy, University of Wisconsin, 475 North Charter Street, Madison, WI 53706, USA\\
$^7$East Asian Observatory, 660 North A'ohoku Place, Hilo, Hawaii 96720, USA\\ 
$^8$RAL Space, Rutherford Appleton Laboratory, Harwell Oxford, Didcot, Oxfordshire OX11 0QX, UK\\
}

\pagerange{\pageref{firstpage}--\pageref{lastpage}} \pubyear{2018}

\begin{document}

\date{Draft \today \hfill\fbox{\textbf{\emph{VERSION 1}}}}

\label{firstpage}

\maketitle

\begin{abstract}
We define the molecular cloud properties of the Milky Way first quadrant using data from the JCMT CO(3-2) High Resolution Survey. We apply the Spectral Clustering for Interstellar Molecular Emission Segmentation (SCIMES) algorithm to extract objects from the full-resolution dataset, creating the first catalog of molecular clouds with a large dynamic range in spatial scale. We identify $>85\,000$ clouds with two clear sub-samples: $\sim35\,500$ well-resolved objects and $\sim540$ clouds with well-defined distance estimations. Only 35\% of the cataloged clouds (as well as the total flux encompassed by them) appear enclosed within the Milky Way spiral arms. The scaling relationships between clouds with known distances are comparable to the characteristics of the clouds identified in previous surveys.
However, these relations between integrated properties, especially from the full catalog, show a large intrinsic scatter ($\sim0.5$\,dex), comparable to other cloud catalogs of the Milky Way and nearby galaxies. The mass distribution of molecular clouds follows a truncated-power law relationship over three orders of magnitude in mass with a form $dN/dM \propto M^{-1.7}$ with a clearly defined truncation at an upper mass of $M_0 \sim 3 \times 10^6~M_\odot$, consistent with theoretical models of cloud formation controlled by stellar feedback and shear. Similarly, the cloud population shows a power-law distribution of size with $dN/dR \propto R^{-2.8}$ with a truncation at $R_0 = 70$\,pc.
\end{abstract}


\section{Introduction}
\label{S:intro}

Molecular clouds are the raw material for all star formation (SF) in the local Universe.  These clouds are the initial seeds for the SF process, and their internal conditions dictate the relevant physics for SF and therefore the evolution of galaxies. The star-forming molecular interstellar medium (ISM) is cold ($T=10$-$30$~K), relatively dense ($n_\mathrm{H}\gtrsim 10^{2}\mbox{ cm}^{-3}$), dominated by supersonic turbulence (Mach number $\mathcal{M}>10$; Reynolds numbers $>10^8$), and moderate magnetic fields (Alfv\'en speed comparable to flow speed $v_A\lesssim v$).  
Despite a rich interplay of physical processes, star formation yields a surprisingly uniform initial mass function for stars.
Given the wealth of physical processes at work, the star formation field has relied upon statistical characterization of the molecular gas to understand the range of conditions necessary for star formation and the processes regulating gas evolution.  In particular, molecular gas
has been observed to organize into discrete molecular clouds (MCs) bounded by the photodissociation-regulated transition from atomic to molecular gas and the chemical creation of molecular gas tracers (most commonly CO).  This chemical boundary provides a means by which molecular clouds can be inventoried and characterized as a population of discrete objects.  

Historically, the molecular cloud ``paradigm'' has provided a way by which the complexity of the molecular ISM could be simplified into the form mimicking that of the population studies that successfully drove the understanding of stellar populations and their evolution. In particular, wide area mapping of the Galactic plane in CO emission presented an opportunity to describe the star forming molecular clouds across the Galactic disk.  Given these survey data, several groups provided statistical descriptions of the Milky Way molecular ISM \citep[e.g.,][]{scoville75, solomon79a,sanders1986, solomon87,scoville87}. Such cataloging approaches encountered complications when applied to observational data sets where nominally discrete clouds appeared blended. These seminal studies adopted several approaches for describing both the blending of emission and the resolution of the distance ambiguity that affects kinematic distance determinations in the inner Milky Way. While some of these approaches relied on by-eye assignment of molecular emission into clouds, later work used contour-based methods.  These contours considered the survey volume as a position-position-velocity (PPV) data cube of brightness temperature $T(l, b, v)$, where $l$, $b$, and $v$ indicate, respectively, the Galactic longitude, latitude, and line-of-sight velocity.  Clouds and their substructure were identified as discrete features of emission above fixed brightness temperature thresholds \citep{solomon87, scoville87}.

With these cloud definitions, the resulting analysis of molecular emission established the canonical scalings between the discrete features in the molecular ISM.  In general, cloud properties were determined by measuring the (emission-weighted) size of the features in the survey space and resolving the distances to these clouds.  These works showed that Milky Way molecular clouds followed a size-linewidth relationship suggestive of supersonic turbulence: $\sigma_\mathrm{v} \propto R^{\beta}; \beta \in[0.4,0.7]$, virial parameters $\alpha \sim 2$, and a top-heavy mass distribution with $dN/dM \propto M^{\gamma}; \gamma\in [-1.8,-1.5]$.

Interferometers showed that extragalactic studies of molecular clouds, when analyzed using similar techniques, followed similar relationships between their bulk properties \citep{bolatto2008,fukui10}, but high density systems \citep{oka2001, wilson03, rosolowsky_blitz05} showed significant departures from the scaling relationships seen in the Milky Way.  In particular, those clouds showed larger turbulent linewidths on a fixed physical scale.  Nonetheless, the cloud populations still showed $\alpha \sim 2$. A careful re-examination of the molecular cloud properties in the Milky Way by \citet{heyer2009} also found variation of the turbulent linewidths within the Galaxy.  This work proposed another fundamental relationship between the molecular gas surface density $\Sigma_{\rm mol}$ and the linewidth on a fixed physical scale: $\sigma_v / R^{1/2} \propto \Sigma_{\rm mol}$ \citep[see also][]{field2011}.

At the heart of these analyses is the definition of a discrete molecular cloud that provides a suitable basis for cataloging.  Since the approaches forwarded in the early works, the quality of survey data has improved dramatically.  These improved data reveal that objects are {\it blended} in both the inner Milky Way and in the relatively low physical resolution studies of nearby galaxies with interferometers \citep{bolatto08}.  The edge of the CO emission in PPV space, even at a specified threshold, no longer serves as a good boundary to define an object.  Several strategies have been proposed to deal with object identification in blended emission, and the primary approaches used historically fall into two main categories: functional fitting \citep[e.g., {\sc gaussclumps}][]{stutzki_guesten90} and watershed algorithms (e.g., {\sc clumpfind} by \citealt{williams94}, {\sc SExtractor} by \citealt{sextractor} and {\sc cprops} by \citealt{rl06}).  The shortcomings of these approaches emerge in their application to the emission from the molecular ISM: because molecular gas is permeated by turbulence, the emission in this medium has structures on a wide range of scales.  The emission structure is further filtered by chemical (e.g., CO destruction or depletion), opacity, and excitation effects. Furthermore, the shapes of large scale molecular ISM do not have specific functional forms, though objects smaller than the thermal scale for turbulence appear to be well represented by physically motivated models such as Bonnor-Ebert spheres \citep[i.e., cores;][]{difrancesco07}

In contrast to explicitly modeling the structure of the ISM emission, the watershed approach has the advantage of being model-free.  However, the major problem with this approach is that the blind application of watershed algorithms to emission with structure on a range of scales finds objects with scales comparable to the resolution element (\citealt{pineda09}, \citealt{leroy2016}). This shortcoming has been avoided with several strategies that rely on prior information about the expected scales to be recovered in the emission. \citet{heyer2009} simply use the cloud definitions established using lower resolution data of previous studies.  \citet{rathborne2009} smooth their data to a  resolution of $\sim 10$ pc scales before applying {\sc clumpfind}. Well-resolved studies of extragalactic clouds \citep{bolatto08} use the notion of a ``physical prior'' where the watershed algorithm is seeded on 10 pc scales comparable to the expected sizes of clouds.  These approaches facilitate comparing data sets of disparate qualities, and allow population-based approaches to studying the molecular ISM.  However, these decomposition approaches necessarily ignore the full dynamic range of information in the observational data set.

With several new surveys of the Galactic plane in emission from the molecular ISM, the gap between the quality of the available data and the tools used to define molecular clouds has grown particularly large.  Recent studies have revisited the definition of cloud identification in the Milky Way. The combined survey of CO(1-0) emission over the Galaxy by \citet{dame01} (hereafter the Dame survey) provides a uniform reference for the molecular ISM at 0.125$^{\circ}$.  \citet{rice2016} use a {\it dendrogram} approach to create a cloud catalog of this emission.  The dendrogram representation transforms PPV data cubes into a tree-like graph that is defined by the connectivity of their emission contours \citep{rosolowsky08}. Identification of cloud structures is done by breaking the graph into individual trees based on specified criteria.  In this case, the authors fix the amount of substructure that can present in a tree and tune this parameter to match the structures seen in the Dame survey.  They use the \citet{reid2016} distance determination code, which is based on trigonometric parallaxes to map PPV space to the three dimensional structure of the Galaxy.  This mapping assumes that all emission is concentrated in the spiral structure of the Galaxy and that each location in PPV space can be assigned to a unique distance.

\citet{miville_deschenes2017} also use the Dame survey but adopt a complementary approach that first decomposes individual spectra into a family of Gaussian line components.  Clouds are identified by clustering these components together into groups using assignment guided by a watershed analysis of the original PPV data set. The clustering approach defines clusters based on the brightest emission components and associates other components with these peaks if their coordinates are within the scatter of the coordinates for the peaks.   Distances are assigned by assuming a zero-intrinsic-scatter size-linewidth relationship that accounts for variations in cloud surface density following \citet{heyer2009}.  Clouds are assigned to a kinematic distance that gives a size most consistent with the (distance-independent) linewidth.

Both of these recent works present new approaches for identifying objects on scales larger than the resolution element, avoiding the main problems of watershed-based decomposition algorithms.  Since the resolution of the underlying Dame survey data is only 0.125$^{\circ}$, these studies necessarily find objects significantly larger than this scale with little information on smaller scales.  Other datasets have much better angular resolution, though they do not cover the entire Galactic plane uniformly.  These include the Galactic Ring Survey (GRS, with $45''$ resolution in $^{12}$CO(1-0); \citealt{jackson2006}); the Three-mm Ultimate Mopra Milky Way Survey (ThRUMMS, with $66''$ resolution in $^{12}$CO (1-0) and other species; \citealt{barnes2015}); the JCMT $^{12}$CO(3-2) High Resolution Survey (COHRS, with $16.6''$ resolution in $^{12}$CO(3-2); \citealt{dempsey2013}); the $^{13}$CO/C$^{18}$O(3-2) Heterodyne Inner Milky Way Plane Survey (CHIMPS, with $15''$ resolution in $^{13}$CO(3-2) and C$^{18}$O(3-2); \citealt{rigby2016}); the Structure, excitation, and dynamics of the inner Galactic interstellar medium (SEDIGISM, with $30''$ resolution in $^{13}$CO(2-1) and C$^{18}$O(2-1); \citealt{schuller2017}) survey; and the FOREST Unbiased Galactic plane Imaging survey with the Nobeyama 45-m telescope (FUGIN, with $20''$ resolution in $^{12}$CO(1-0), $^{13}$CO(1-0) and C$^{18}$O(1-0); \citealt{umemoto2017}). The high spatial resolution and sensitivity of these datasets make it now possible to obtain vastly improved information about the molecular cloud population not only by being able to detect smaller and lower-mass objects than those seen in previous studies, but also by the ability to detect variations in the substructure of the clouds. It becomes, therefore, essential to be able to extract molecular clouds by retaining the maximum amount of information on the hierarchical structure of the gas. This is one of the key advantages of the cloud extraction algorithm we employ here.

In this study, we present a new catalog of molecular clouds that emphasizes the spatial dynamic range within recovered objects.  We use the COHRS survey of \citet{dempsey2013} as the underlying data set because of its excellent spatial resolution.  We then decompose these data using the Spectral Clustering in Molecular Emission Surveys algorithm (SCIMES, \citealt{colombo15}), which uses graph-based image processing to decompose a dendrogram representation of the emission into individual structures.  Finally, we combine this decomposition with the distance catalog generated in \citet{ellsworth_bowers2013}.

We present this approach in the following sections.  In Section \ref{S:cohrs}, we briefly describe the COHRS surveys.  We summarize the {\sc SCIMES} decomposition approach in Section \ref{S:dendro_scimes} and its particular application to the COHRS data in Section \ref{S:scimes_cohrs}.
Sections \ref{S:r31} and \ref{S:cohrs_distance} detail the distance determination procedure, and the determination of the $^{12}$CO(3-2)-to-H$_2$ conversion factor, respectively.  We describe the cloud properties in Section \ref{S:props}. Section~\ref{S:catalog} shows the content of the catalog, the analysis of the cloud property distributions with the comparison with \citet{roman_duval2010} catalog (Section~\ref{SS:properties}), and the fit to the mass and size cumulative spectra (Section~\ref{SS:cum_dists}). We study those properties in relation to the cloud location within the Milky Way in Section~\ref{S:mw_distribution} and we describe the correlations among the properties in Section \ref{S:scaling_relations}. In Section~\ref{S:surveys} we contextualize the results from our catalog considering other molecular cloud surveys of the Milky Way and nearby galaxies. Our findings and perspective for the future research with SCIMES are summarized in Section~\ref{S:summary}. 

\section{The JCMT $^{12}$CO(3-2) High Resolution Survey}
\label{S:cohrs}

The JCMT $^{12}$CO(3-2) High Resolution Survey (COHRS) is a large-scale CO survey that observed the inner Galactic plane in $^{12}$CO\,($J = 3\rightarrow 2$) emission using the Heterodyne Array Receiver Programme B-band (HARP-B) instrument on the James Clerk Maxwell Telescope (JCMT). The current data release mapped a strip of the Milky Way $|b| \leq 0^{\circ}.5$ between $10^{\circ}.25<l<17^{\circ}.5$ and $50^{\circ}.25<l<55^{\circ}.25$, and $|b| \leq 0^{\circ}.25$ between $17^{\circ}.5<l<50^{\circ}.25$.  The survey covers a velocity range of $-30$\,km\,s$^{-1}<v_\mathrm{LSR}<155$\,km\,s$^{-1}$. The data have a spectral resolution of 1 km\,s$^{-1}$ and an angular resolution of $\theta_{\mathrm{FWHM}}=16.6$ arcsec, achieving a mean noise level of $\sigma_{\mathrm{RMS}}\sim 1$ K. The $J = 3$-2 transition of the $^{12}$CO molecule traces the warm molecular medium (10-50\,K) around the active star formation regions. For full details about COHRS, refer to \cite{dempsey2013}.

\section{Spectral Clustering for Interstellar Molecular Emission Segmentation}
\label{S:dendro_scimes}
To decompose clouds in the COHRS data we use the publicly available Spectral Clustering for Interstellar Molecular Emission Segmentation (SCIMES) algorithm\footnote{\url{https://github.com/Astroua/SCIMES}}. The method has been explained fully in \cite{colombo15}, and here we give only a brief description and note some changes relative to the original work in Appendix~\ref{A:new_scimes}. In general, SCIMES finds relevant objects within a dendrogram of emission using spectral clustering. A {\em dendrogram} is a tree representation of image data that encodes the hierarchical structure emission (e.g. \citealt{rosolowsky08}). The dendrogram is composed of two types of structures: {\em branches}, which are structures which split into multiple substructures, and {\em leaves}, which are structures that have no substructure.  Leaves are associated with the local maxima in the emission. We also consider the {\em trunk}, which is the super-structure that has no parent structure, and contains all branches and leaves.

SCIMES uses graph theory to analyze dendrograms. A {\em graph} is a collection of objects ({\em nodes}) that possess defined relationships ({\em edges}). In this case, the edges connect a given branch to the sub-structures of a branch and the structures containing that branch.  Under this interpretation, each edge in the graph corresponds to an isosurface (i.e., contour) in the PPV data.  We specifically use weighted graphs, where each graph edge carries a numerical value called the {\em affinity} where larger values of the affinity represent more similarity between two parts of the graph.  In SCIMES, we consider two different affinities corresponding to the properties of the isosurface based on the PPV {\em volume} and {\em luminosity}. The volume is defined as $V=\sigma_v\pi R_\mathrm{eff}^2$, where $\sigma_v$ is the velocity dispersion and $R_\mathrm{eff}$ is the effective radius of the isosurfaces (see Section~\ref{S:props} for further details). The luminosity is calculated as $L_{\rm CO}=F_{\rm CO}d^2$, where $F_{\rm CO}$ is the integrated emission within the isosurface (the {\em flux}) and $d$ is the distance to the structure. If the distance to the structure is unknown the flux is considered. The affinity between two parts of the graph is defined as the inverse of the volume or the luminosity for the bigger or more luminous object.  SCIMES generates an {\em affinity matrix}, where the element $A_{ij}$ is the affinity between leaf $i$ and leaf $j$, which correspond to the graph nodes.

The final part of SCIMES is to use the affinity matrix to divide the graph into separate components using {\em spectral clustering}, corresponding to segmenting the emission into individual clouds.  The SCIMES algorithm considers the eigenvalues of the affinity matrix and finds the $k$ most significant vectors that represent clusters in the spectral decomposition of the affinity matrix.  Selecting these $k$ vectors divides the graph into $k$ regions, corresponding to structures within the dendrogram, which in turn correspond to connected regions of emission in PPV space.  We generically call these objects {\em molecular gas clusters} since the literature can describe them as clouds, clumps or cores depending on the scale of the emission.

While the SCIMES method is abstract and complex, it features the major advantage of being able to utilize data with wide spatial dynamic range (i.e., many resolution elements across a cloud).  SCIMES has been developed to mimic the action of by-eye decomposition, but it is automated and requires no manual tuning. It relies on natural transitions in the emission structure to define objects and is robust across scales.  In particular, SCIMES is a multi-scale decomposition approach that explicitly takes the hierarchical nature of the ISM into account.

\begin{figure*}
\centering
\includegraphics[width=1\textwidth]{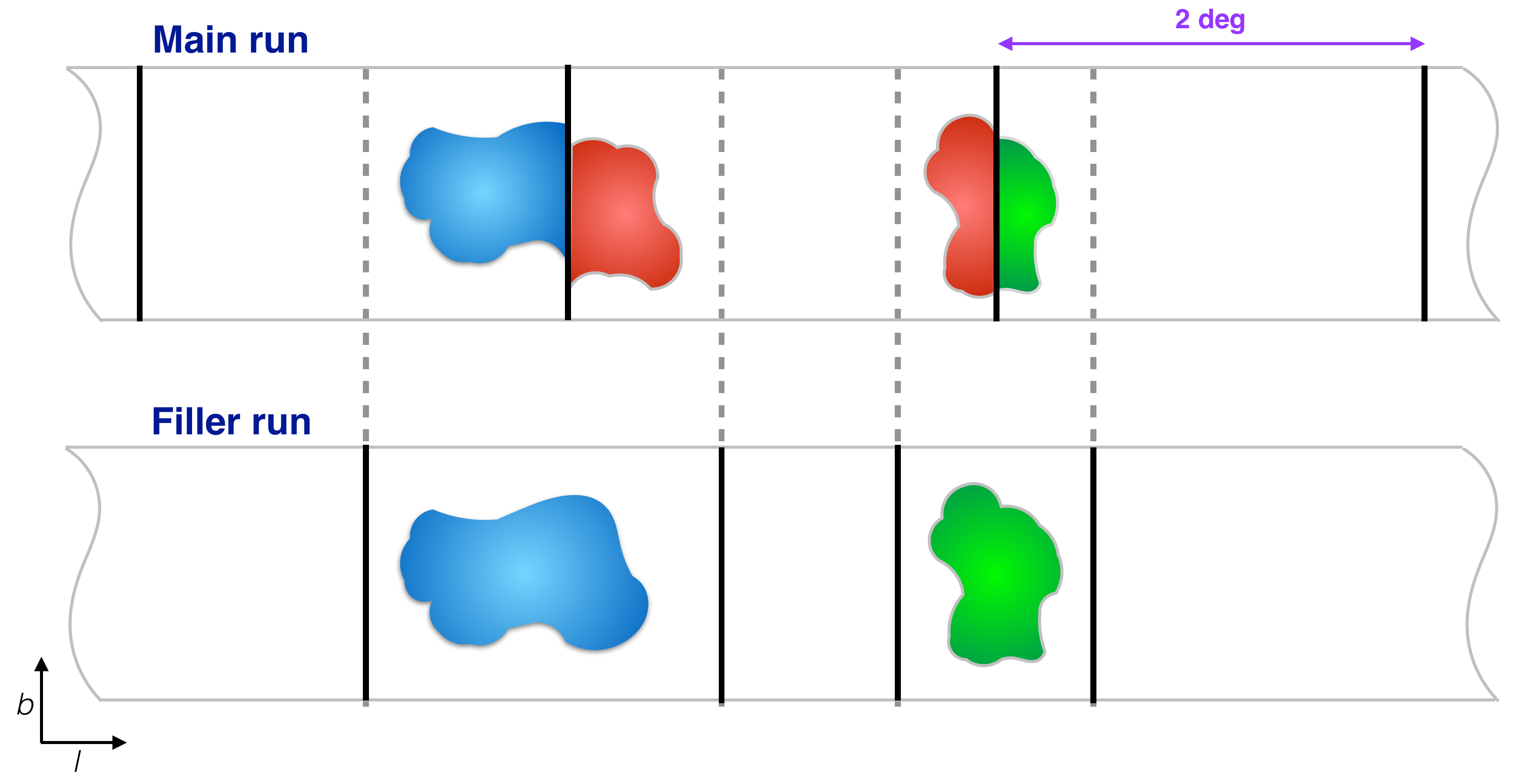}
\caption{Scheme of the full COHRS data (gray, full lines) division into sub-cubes that define the ``main'' (top) and ``filler'' (bottom) runs. Sub-cubes of the ``main'' run span all approximately 2$^{\circ}$ in longitude. In the figure, sub-cube edges are indicated with black vertical lines. Clouds on the longitudinal edges are removed from the ``main'' run cubes since they do not have closed iso-contours. In the figure, clouds in the same sub-cube are indicated with the same color. To recover the clouds on the longitudinal edges of the ``main'' run cubes we build customized, ``filler'' run, sub-cubes which span the full extend of the largest clouds on two consecutive edges of two adjacent ``main'' run sub-cubes, allowing a padding of 20 pixels on both sides. In some cases, the clouds recovered in the ``filler'' run do not always perfectly correspond to the two chunks of edge clouds of the ``main'' run.}
\label{F:cohrs_runs}
\end{figure*}

\section{Application of SCIMES to COHRS data}
\label{S:scimes_cohrs}

The full first data release of the COHRS data is publicly available\footnote{\url{http://dx.doi.org/10.11570/13.0002}}. The data are provided in tiles of $0.5^{\circ}$ in longitude. Before proceeding with the cloud identification we built a single data cube of the full survey using the {\sc spectral-cube}\footnote{\url{https://doi.org/10.5281/zenodo.1213217}} and {\sc montage} software.
Given the large data set, we first construct a signal mask of the data using the  technique discussed in \citet{rl06}.  The mask is built using a two-step process, first including all PPV pixels that have emission above $10\sigma_{\mathrm{RMS}}$ where $\sigma_{\mathrm{RMS}}$ is the local noise level. This mask is then expanded to all connected pixels that have emission above $3\sigma_{\mathrm{RMS}}$ in two consecutive channels. In this way small clumps with SNR$<10$ are incorporated within a larger structure avoiding noisy regions. The noise ($\sigma_\mathrm{RMS}$) is estimated by calculating the standard deviation along each line-of-sight from the first and the last 10 line-free channels of the data cube.  Using this mask, we can then extract sub-cubes from the full data cube that contain connected regions for decomposition with SCIMES. Those cubes span 1200 pixels in longitude, i.e. $\sim2^{\circ}$ given the pixel size of the COHRS data of 6 arcsec. In practice, we perform this extraction in two stages, pulling out subcubes, processing those cubes with SCIMES and identifying objects on the longitude edges.  Longitude edge objects are then rejected from the first catalog pass (``main'' SCIMES run) since their contours are by definition not closed (see Fig.~\ref{F:cohrs_runs}).  We then extract sub-cubes around the rejected edge objects and carry out a SCIMES decomposition (``filler'' SCIMES run).  We then include all objects that do not overlap between the two decompositions and retain the larger of any  overlapping objects that occur in both passes. These two passages are performed to overcome the difficulty to generate a single dendrogram from the full COHRS dataset which is computationally expensive. The same is true for the affinity matrix analysis performed by SCIMES where each additional required cluster is equivalent to an additional dimension in the clustering space. Several objects are also found along the survey actual (upper and lower) latitudinal edges. Those clouds are instead retained within the catalog and marked as ``edge'' clouds (see below).

For each sub-cube, we generate a dendrogram of the emission using SCIMES parameters that: (1) require each branch of the dendrogram to be defined by an intensity change of $>3\sigma_{\mathrm{RMS}}$ (\verb"min_delta"$=3\langle\sigma_\mathrm{RMS}\rangle$), (2) contain all of the emission in the mask (\verb"min_value"=0\,K), and (3) contain at least three resolution elements worth of pixels (\verb"min_npix"$=3\Omega_{\mathrm{bm}}$, where $\Omega_{\mathrm{bm}}$ is the solid angle of the beam expressed in pixels).
We do not know the distance of the all dendrogram structures {\em a priori}, so the \emph{volume} and \emph{luminosity} affinity matrices are generated using the PPV volumes and integrated intensity values instead of spatial volumes and intrinsic luminosities. The scaling parameter (see \citealt{colombo15}) is searched above $3\sigma_\mathrm{RMS}$ (see Appendix~\ref{A:new_scimes} for further details). SCIMES searches for the dendrogram branches that can be considered as single independent ``molecular gas clusters''. 
As discussed in \cite{colombo15}, leaves that do not form isolated clusters, are grouped all together in sparse clusters without any neighbors in PPV space between constituent objects. In the original implementation of SCIMES, a sparse cluster was pruned by those leaves and only the largest branch (i.e. the branch that contains the largest number of leaves) was retained as representative of the sparse cluster. The structures pruned from the sparse cluster can consist of isolated leaves or small branches. Elements that cannot be assigned to any cluster given the clustering criterion are called ``noise'' in clustering theory (e.g. \citealt{ester1996}). Here, however, we retain those branches and leaves since they are significant emission and well-resolved objects (given the dendrogram construction parameters) even if not assigned to another cluster.

\begin{figure}
\centering
\includegraphics[width=0.4\textwidth]{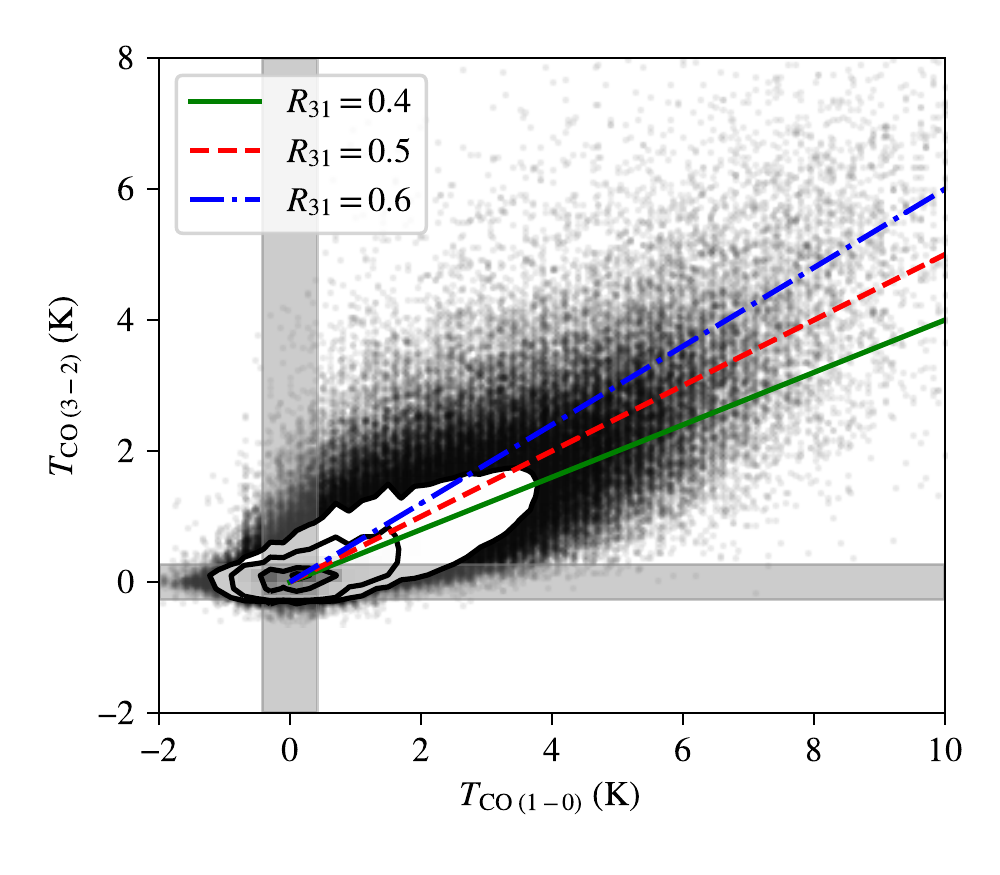}
\caption{Pixel-by-pixel brightness comparison between $^{12}$CO(1-0) from the Massachusetts Stony Brook Survey (\citealt{sanders1986}) and the $^{12}$CO(3-2) from COHRS. The grey shaded regions indicate the $\pm 1\sigma$ levels of the noise. We obtain an average $R_{31}=^{12}$CO(3-2)/$^{12}$CO(1-0) = 0.6 for the bright emission (blue dashed dot line), while a value $R_{31}=0.4$ is appropriate for faint emission (green full line). For the whole sample we choose the average value of $R_{31}=0.5$ (red dashed line).}
\label{F:r31}
\end{figure}

\section{Determination of $^{12}$CO(3-2)-to-$^{12}$CO(1-0) flux ratio}
\label{S:r31}
To calculate the masses and column densities of molecular clouds, we use a CO-to-H$_2$ conversion factor $X_\mathrm{CO}$, allowing us to scale the integrated intensities of the CO emission, $W_{\mathrm{CO}}$, directly to H$_2$ column densities, $N_\mathrm{H_2}$.  In general, $X_\mathrm{CO}$ is a parameter that depends on environmental conditions \citep[e.g.,][]{bolatto2013, barnes2018}. Even so, for the purpose of this paper, we assume a constant $X_\mathrm{CO}$ for  direct comparison to literature results, with $X_{\rm ^{12}CO(1-0)} \sim 2 \pm 1 \times 10^{20}$\,cm$^{-2}$\,K$^{-1}$\,km$^{-1}$\,s \citep[e.g.][]{dame01,bolatto2013,duarte-cabral2015}. We express masses in terms of the molecular gas luminosity using the scaled conversion factor $M_\mathrm{lum} = \alpha_\mathrm{CO} L_\mathrm{CO}$,  with $\alpha_{\rm ^{12}CO(1-0)} \sim 4.35$\,M$_{\odot}$\,pc$^{-2}$\,K$^{-1}$\,km$^{-1}$\,s, which assumes a mean molecular weight of $2.8m_{\mathrm{H}}$ per hydrogen molecule \citep[e.g.][]{bolatto2013}. These conversion factors are calibrated using the {${}^{12}\mathrm{CO}$(1-0)} transition, and therefore we must assume a line ratio {$R_{31}\equiv {}^{12}$CO(3-2)/$^{12}$CO(1-0)} to scale our calculated properties directly to physical properties.

To calculate $R_{31}$, we convolve the COHRS data to $45''$ and reproject it to the coordinate grid of the $^{12}$CO(1-0) data from the University of Massachusetts Stony Brook Survey \citep{sanders1986}. We show the pixel-by-pixel plot of the brightness in the two data cubes in Figure~\ref{F:r31}.  We use these relationships to estimate the typical line ratio across the survey.  There is not a single, well-defined line ratio across the Galaxy reflecting the variation in excitation conditions \citep[e.g.][]{barnes2015, penaloza2018}. On average, we observe $R_{31}\equiv{}^{12}$CO(3-2)/$^{12}$CO(1-0) = 0.4 for the faint emission, but $R_{31}=0.6$ describes the line ratio of the brightest emission.
As a global summary across the survey region, we adopt $R_{31}=0.5$ and use this line ratio to establish the CO-to-H$_2$ conversion based on the calibration work done for $\alpha_{\rm ^{12}CO(1-0)}$.  This line ratio value is consistent to the value measured in nearby galaxies by \cite{warren2010}. In this way we scale from the standard values for the CO(1-0) line \citep{bolatto2013} to get: 

\begin{equation}\label{E:XCO32}
X_{^{12}\mathrm{CO(3-2)}}= \frac{X_{^{12}\mathrm{CO(1-0)}}}{R_{31}} = 4\times10^{20} \frac{\rm cm^{-2}}{\rm K\,km\,s^{-1}},
\end{equation}
\noindent and:
\begin{equation}\label{E:aCO32}
\alpha_{^{12}\mathrm{CO(3-2)}}= \frac{\alpha_{^{12}\mathrm{CO(1-0)}}}{R_{31}} = 8.7 \frac{\rm M_{\odot}}{\rm K\,km\,s^{-1}\,pc^{2}}.
\end{equation}
The data in Figure \ref{F:r31} show significant scatter in the low end and this simple conversion of CO luminosity to mass becomes even less certain for these low luminosity objects. Given this, the line ratio alone suggests an uncertainty of at least 40\% in referencing to $\alpha_{\rm ^{12}CO(1-0)}$.

\section{Distance to the molecular clouds}
\label{S:cohrs_distance}

We use the distances defined by Zetterlund et al.~\citep[2018.; see also][]{ellsworth_bowers2013, ellsworth_bowers2015} to establish the physical properties to the decomposed molecular gas clusters.  These distance estimates are based on an analysis of the Bolocam Galactic Plane Survey (BGPS) version 2 \citep{ginsburg2013} and a suite of multiwaveband data.

The BGPS is a 1.1~mm dust continuum survey that largely overlaps with COHRS, covering the Galactic plane region between $-10^{\circ}\leq l\leq90^{\circ}$ and $|b| \leq 0^{\circ}.5$ with a spatial resolution of 33 arcseconds, making it particularly useful for our needs. The distances are obtained using a Bayesian approach that provides a posterior probability density function of distances to an object through a suite of techniques including kinematic distances from matching to molecular gas emission, maser distances, and a model of the infrared emission and absorption of different features in the plane. In most cases, the method allows the definition of a single distance through the maximum-likelihood distance or the probability-weighted mean distance.  Each object then has a distance assigned to it from one of these two methods according to its ability to resolve the kinematic distance ambiguity (see \citealt{ellsworth_bowers2013} for further details).  This method generates both estimates of the distances and known uncertainties. We focus on the 2202 objects with well constrained distances in the BGPS distance sample, i.e., those with distance uncertainties of $\pm 0.5~\mathrm{kpc}$ or better.

For BGPS objects with a single, well-constrained distance, this measurement corresponds to a single position (pixel) within the PPV data as their spectroscopic $v_{\rm LSR}$ are also defined (see \citealt{shirley2013}). That pixel can be associated with the dendrogram structure that contains the known-distance pixel, which then inherits that distance measurement. Around 140 decomposed molecular gas clusters ($\sim0.2\%$ of the total number of clouds) have substructures that have different distances from each other.  For these objects, we assign the cluster to the near distance corresponding to the brightest spot within the object, assuming that that sub-structure carries the largest amount of cloud mass.

We call this distance attribution method {\em exact}.  In contrast, a molecular gas cluster may not contain any pixels with a distance measurement.  These objects may be the substructures of larger connected emission features with distance assignments.  In this case, we assign the cluster the closest distance in PPV space contained within the larger structure and describe this assignment method as a {\em broadcast}.  For isolated objects without an assigned distance or a parent structure with a distance, we assign the object a distance to be the same as the {\em closest} PPV pixel with a known distance. In this way we are able to break the kinematic distance degeneracy at least for the objects with better defined distance (see Appendix~\ref{A:kda} for further details).
We also report a parameter, called {\em broadcast inaccuracy}, which indicates (in pixels) how far is a distance pixel to the outer edge of the cloud. By definition {\em exact} distance clouds have broadcast inaccuracy equal to zero.

\section{Molecular cloud integrated properties}
\label{S:props}
For most of the cloud decomposition in the literature, the size of clouds is governed by the resolution element of the survey. Therefore, the internal structure of these objects remains elusive and cataloguing focuses on  the ``integrated'' properties obtained by summing across the cloud pixels. While SCIMES is able to separate these two structures, we focus this work on the integrated properties for comparison with the existing literature and we will present studies based on the well-resolved nature of COHRS clouds in future work.

\subsection{Coordinates}
\label{SS:coordinates}

For each cloud in our catalog we provide four sets of coordinates in different projections: pixel, Galactic, heliocentric, and Galactocentric.

The pixel coordinates ($x_\mathrm{cen}$, $y_\mathrm{cen}$, $v_\mathrm{cen}$) are the mean positions of the clouds related to the sub-cube they have been segmented from. Galactic coordinates are listed as longitude ($l$), latitude ($b$), and velocity with respect to the local standard of rest ($v_\mathrm{LSR}$). Heliocentric coordinates are defined with the $x$-axis on the line that connects the Sun and the Galactic centre:
\begin{equation}\label{E:suncoord}
\begin{array}{lcl} 
x_\odot & = & d\cos l \cos b, \\ 
y_\odot & = & d\sin l \cos b, \\ 
z_\odot & = & d\sin b, 
\end{array} 
\end{equation}
where $d$ is the cloud assigned distance. 

Following \cite{ellsworth_bowers2013} and \cite{rice2016} we define the Galactocentric coordinates as: 
\begin{equation}\label{E:galcoord}
\begin{array}{lcl} 
x_\mathrm{Gal} & = & R_0\cos\theta - d(\cos l \cos b \cos\theta + 
\sin b \sin\theta), \\ 
y_\mathrm{Gal} & = & - d\sin l \cos b, \\ 
z_\mathrm{Gal} & = & R_0\sin\theta - d(\cos l \cos b \cos\theta - 
\sin b \cos\theta), 
\end{array} 
\end{equation}
where $\theta=\arcsin(z_0/R_0)$, and $z_0=25$\,pc is the height of the Sun above the midplane of the Milky Way \citep{goodman2014}, and $R_0=8.51$\,kpc is the radius of the Solar circle, i.e. the distance of the Sun to the Galactic Center (\citealt{ellsworth_bowers2013}). 

\subsection{Pixel-based properties}
\label{SS:props_pixel}

Many properties of the isosurfaces are already defined by the dendrogram implementation we used. Flux and Volume are used to generate the affinity matrix necessary for the cluster analysis. The properties considered by the dendrogram are ``pixel-based'' if distances are not provided a priori. The statistics offered by the {\sc astrodendro} software have been defined by \cite{rl06}.

The basic properties of the dendrogram structures are calculated through the moment technique. This technique has been shown to perform better than the area method to recover the actual effective radius of the clouds (see \citealt{rl06} for details). The centroid of the clouds is given by the intensity-weighted mean of the pixel coordinates along the two spatial and the velocity axes of the datacube. The principal axes of the emission map using intensity-weighted second moments over position defines the major ($\sigma_\mathrm{maj}$) and minor  axis ($\sigma_\mathrm{min}$) sizes of the cloud as well as the position angle ($\phi$). In a similar way, we calculate the velocity dispersion ($\sigma_v$) from the intensity-weighted second moment along the velocity axis. Through the major and minor semi-axis measurements, we also define an area measurement for clouds based on the elliptical area:
\begin{equation}\label{E:a_ellispe}
A_\mathrm{ellipse} =(8\pi\ln 2) \sigma_\mathrm{maj}\sigma_\mathrm{min}.
\end{equation}
We also measure the area of each cloud from the number of pixels that the cloud occupies in the spatial dimensions ($A_\mathrm{exact}$). Finally, the flux ($F_\mathrm{CO}$) is given by the sum of all the pixel values within the isosurface that contains the cloud.

We enrich the dendrogram-generated catalog with several other properties related to the hierarchical structure itself (see Appendix~\ref{A:new_scimes}). Each final cloud has: the number of pixels and leaves within the structure ($N_\mathrm{pixel}$ and $N_\mathrm{leaves}$, respectively); the identifier for the parental structure that fully contains the structure (\emph{parent}); the identifier for the structure at the bottom of the hierarchy that fully contains the structure under analysis (\emph{ancestor}); and classification flag (\emph{type}) that indicates which kind of structure we are dealing with (in dendrogram terminology): ``L'' (leaf) a structure without children; ``B'' (branch) a structure with children and parent; and ``T'' (trunk) structure with children and without parent (bottom of the hierarchy).

\subsection{Physical properties}
\label{SS:props_phys}

By assigning a distance to each molecular gas cluster as per Section~\ref{S:cohrs_distance}, we can convert the pixel-based properties into physical properties of the molecular structures.  The semi-major and semi-minor axes are 
converted into parsecs via the usual small angle formula and the world-coordinate-system information from the data file. The effective radius of the cloud ($R_\mathrm{eff}$) is generated from quadrature sum of the semi-major and semi-minor axes:
\begin{equation}\label{E:reff}
R_\mathrm{eff} = \eta\sqrt{\sigma_\mathrm{maj}^2 + \sigma_\mathrm{min}^2}.
\end{equation}
Here, $\eta=1.91$ is assumed from \cite{rl06} (and \citealt{solomon87}) to relate the quadrature sum of the two semi-axes and the radius of a spherical cloud. The final velocity dispersion of the cloud is obtained by multiplying the channel width, $\Delta v$, by the velocity dispersion measured in pixels.\\

The CO luminosity of the cloud is obtained by:
\begin{equation}
\label{E:luminosity}
L_\mathrm{CO} = \sum_i T^{i}_\mathrm{CO}\, \Delta v \Omega_\mathrm{pix} d^2,
\end{equation}
where $d$ is the distance to the cloud (in parsecs), $\Omega_\mathrm{pix}$ is the solid angle subtended by a pixel, $\Delta v$ is the channel width (in km/s), and $T^i_{\rm CO}$ is the brightness temperature of each voxel $i$ within the cloud (in K).

The effective radius, velocity dispersion, and CO luminosity are the basis for all other properties presented in the catalog. The mass is derived from the CO luminosity (or luminosity mass) by assuming a $^{12}$CO(1-0)-to-H$_2$ conversion factor $\alpha_\mathrm{CO}$ and $R_{31}$ (see Section~\ref{S:r31}). 

We also make a dynamical measurement of the cloud mass assuming the clouds are virialized, spherical objects and ignoring external pressure and magnetic fields \citep{rl06} and an internal density profile that scales like $\rho(r)\propto r^{-1}$. The virial mass is then:
\begin{equation}\label{E:mvir}
M_\mathrm{vir} = 1040\sigma_v^2 R_\mathrm{eff}.
\end{equation}

The ratio between virial and luminous mass gives the virial parameter, $\alpha$, which is often used to characterize the deviation of a cloud from virial equilibrium. Variations on the estimated $\alpha$ can be due to a true unbalance between physical effects such as gravity, pressure, and magnetic fields, as well as due to time evolution, or simply due to observational biases, such as the possible variations of the assumed $\alpha_\mathrm{CO}$ \citep{bertoldi92,houlahanscalo92}:
\begin{equation}\label{E:alpha_vir}
\alpha = 1.12\frac{\mathrm{M}_\mathrm{vir}}{\mathrm{M}_\mathrm{lum}}.
\end{equation}
In general, $\alpha > 2$ would indicate that the cloud is stabilized against collapse, while finding $\alpha << 2$ might suggest a significant magnetic support \citep[e.g.][]{kauffmann2013}. Generally $\alpha \approx 1$ means that the cloud is virialized.

Through the measurements of $M_\mathrm{lum}$ and $R_\mathrm{eff}$ we measure the mean molecular mass surface density and volume density by assuming clouds have an uniform density and a spherical shape with radius $R_\mathrm{eff}$:

\begin{eqnarray}
\Sigma_\mathrm{mol} &=& \frac{M_\mathrm{lum}}{\pi R_\mathrm{eff}^2}, \label{E:surf_dens} \\
\rho_\mathrm{mol} & =& \frac{3 M_\mathrm{lum}}{4\pi R_\mathrm{eff}^3}.\label{E:vol_dens}
\end{eqnarray}

\subsection{Extrapolation and deconvolution}
\label{SS:props_exdc}

The dendrogram implementation we choose assumes a bijection paradigm to calculate the properties of the structures (see \citealt{rosolowsky08}), i.e., there is a direct connection between pixel intensities in PPV space and the corresponding emission in real space. In this approach, a clump of emission is associated with a physical structure above a certain column density threshold. \cite{rl06} showed that cloud properties are strongly dependent on the brightness level at which they are identified above the surrounding emission. Following \cite{rl06}, we consider an alternative measure of cloud properties that attempts to correct for the biases introduced by finite sensitivity and resolution.

The first step in this procedure is {\em extrapolation} (indicated as ``ex'' in the catalog), which infers the moments of the cloud ($\sigma_\mathrm{maj}$, $\sigma_\mathrm{min}$, $\sigma_v$, and $F_\mathrm{CO}$) that would be measured with infinite sensitivity.  The extrapolation works by considering the scaling of cloud properties as a function of brightness threshold and fitting a linear relation between the measured moments ($\sigma_\mathrm{maj}$, $\sigma_\mathrm{min}$, and $\sigma_v$) and the brightness.  This relationship is extrapolated to the 0\,K contour. A similar procedure is used to derive the extrapolated flux, except a quadratic extrapolation is used. We indicate the extrapolated properties as $\sigma_\mathrm{maj}$(0\,K), $\sigma_\mathrm{min}$(0\,K), $\sigma_v$(0\,K), and $F_\mathrm{CO}$\,(0\,K).

The second step consists in the deconvolution of the survey beam and channel width from the extrapolated moments. The deconvolution is performed by subtracting the beam width in quadrature from the measured radius:
\begin{equation}\label{E:smaj_smin_dc}
\sigma_\mathrm{maj, dc}^2(0\,\mathrm{K}) = \sigma_\mathrm{maj, }^2(0\,\mathrm{K}) - \left(\frac{\theta_\mathrm{FWHM}}{\sqrt{8\ln(2)}}\right)^2  
\end{equation}
with a similar expression for the minor-axis.  The channel width is also deconvolved from the linewidth:
\begin{equation}\label{E:sv_dc}
\sigma_\mathrm{v; dc}^2(0\,\mathrm{K}) = \sigma_\mathrm{v}^2(0\,\mathrm{K}) - \frac{\Delta v^2}{2\pi} 
\end{equation}
\noindent where the subscript ``dc'' indicates deconvolved properties, $\theta_\mathrm{FWHM}$ is the survey beam, and $\Delta v$ is the channel width. All the other properties are then recalculated using these extrapolated, deconvolved measurements. We use these extrapolated, deconvolved properties as the basis for our analysis.  In Appendix \ref{A:props_exdec}, we explore how the extrapolation and deconvolution affects the inferred cloud properties. Generally, the deconvolution affects mostly low values of $R_{\rm eff}$ and $\sigma_{\rm v}$. Extrapolation, instead, shifts some of the velocity dispersion values by up to $\sim0.5$\,dex and $L_{\rm CO}$ by to $\sim1$\,dex, while it does not change measured effective radii significantly.

We choose to apply the extrapolation corrections so that all cloud properties are referenced to a common intensity threshold, which facilitates comparing the cloud properties to each other.  Without a common reference threshold, each cloud would be subject to different biases in the measured properties \citep{rl06}.  However, this application then engenders a specific scientific interpretation of the results, namely we are estimating cloud properties, defining such objects as bounded by a 0~K intensity isosurface. Since these emission structures are part of larger, hierarchical ISM, enforcing this interpretation can obscure the true complexity of the ISM. Indeed, as we note in Section~\ref{S:catalog}, where the emission is heavily blended, SCIMES will segment structures high above the noise level of the data, and this extrapolation may effectively over-correct for the amount of emission of each cloud and its extent. Nevertheless, this cloud-segmentation approach is selected so we can create a catalogued set of objects, which can then be compared to other work executing similar analyses.  Our catalog provides both corrected and uncorrected properties so that other work could use the same SCIMES decomposition without these corrections, provided such an interpretation suits the question being investigated.

\subsection{Uncertainties on cloud properties}
\label{SS:props_err}

The uncertainties on the physical properties in our cloud catalog are dominated by two sources of errors: the distance ($d$) and the CO-to-H$_2$ conversion ($\alpha_{\rm CO}$). Therefore, we use error propagation by taking into account the uncertainty on the distances provided by \cite{zetterlund2018} and by assuming 40\% error on our calculated $\alpha_{\rm CO}$ (see Section~\ref{S:r31}).  Using the CO-to-H$_2$ conversion factor method also introduces systematic uncertainties at the factor of $\sim 2$ level \citep{bolatto2013}. 
For the ``closest'' distance objects we use the near-far distance ambiguity as distance uncertainty given by:
\begin{equation}
\delta d  = \left|R_0 \cos l \right|,
\end{equation}
where $R_0 = 8.51 \times 10^3$\,pc (\citealt{ellsworth_bowers2013}) and $l$ is the Galactic longitude of the cloud centroid.

For purely pixel quantities ($\sigma_\mathrm{maj}$,  $\sigma_\mathrm{min}$,  $\sigma_\mathrm{v}$, and $F_\mathrm{CO}$) we use the bootstrap approach described in \cite{rl06}. This method generates several synthetic clouds by considering a cloud as a set of $N$ volumetric pixels with coordinates $x_i$, $y_i$, $v_i$, and $T_i$; i.e., two spatial coordinates, one velocity coordinate, and the brightness value, respectively. At each iteration, $N$ sets of the cloud data are sampled randomly from the observed values allowing for repeated draws. The sets of bootstrapped $\sigma_{\rm maj}$,  $\sigma_{\rm min}$,  $\sigma_{\rm v}$, and $F_{\rm CO}$ are measured at each iteration. The uncertainty is given by the standard deviation of the bootstrapped quantities. We also rescaled each uncertainty by an oversampling rate, given by the square root of the pixels in the beam. The oversampling rate accounts for not all pixels in each cloud being independent.  These bootstrap uncertainties are summed in quadrature with the  uncertainties induced by the distance and conversion factors.  While the distance and conversion factors are both typically 40\%, the uncertainties in the sizes ($\delta \sigma_\mathrm{maj}$, $\delta \sigma_\mathrm{min}$) are typically 15\% and the flux uncertainty ($\delta F_\mathrm{CO}$) is typically 6\%.

\begin{figure*}
\centering
\includegraphics[width=1\textwidth]{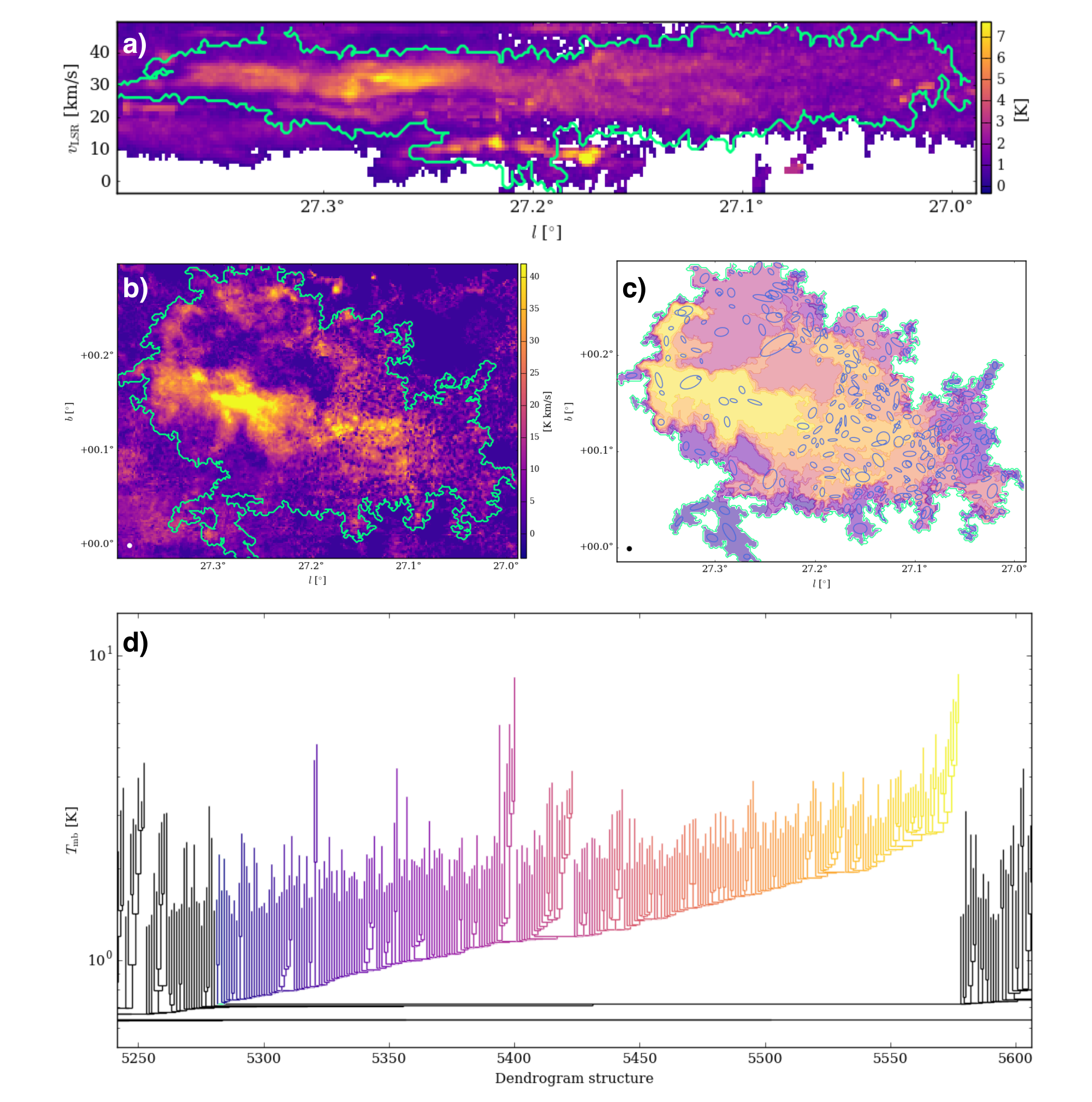}
\caption{Peak temperature longitude-velocity map (panel $a$) and integrated intensity map (panel $b$) of the cloud ID 3076 from the cohrs\_19\_22800\_24000 sub-cube. Data are masked as explained in Section~\ref{S:scimes_cohrs}. Green contours indicate the full extend of the cloud. Cloud ID 3076 hierarchical structure rebuilt through the dendrogram (panel $c$). In the figure every second dendrogram level sub-structure is shown. Blue ellipses indicate the leaves within the cloud. The ellipses parametrize the intensity-weighted major and minor axis of the leaves and their orientation with respect to the longitudinal direction. The section of the dendrogram corresponding to Cloud ID 3076 is highlighted in color in panel $d$, where each dendrogram structure is color-encoded as in panel $c$. The circle in the bottom left corner of panels $b$ and $c$ show the beam of the COHRS data. Dendrogram section corresponding to cloud ID 3076, where branches are color-encoded as the sub-structures in panel $c$, is shown. SCIMES identifies clouds considering and preserving the multi scale and hierarchical nature of the ISM.}
\label{F:cohrs_decomp}
\end{figure*}

\section{Molecular cloud catalog}
\label{S:catalog}

Figure \ref{F:cohrs_decomp} shows an example of a cloud segmented by SCIMES. The figure illustrates how SCIMES naturally works on a multi-scale data. In Appendix~\ref{A:full_survey_decomp} the longitude-latitude and longitude-velocity masks for the clouds in the full survey are given. Those figures show that the SCIMES approach identifies a variety of cloud morphologies, complexity, and sizes. The segmented structures are mostly coherent, however multiple velocity components are sometimes merged in the same object. This behavior is especially associated with clouds on the  border of the data cubes that do not have closed contours. While the objects on the right and left edges of the sub-cubes are removed by construction, the clouds on the lower and upper edges are retained in the catalog since the span in latitude is an intrinsic limitation of the survey rather than the algorithm. The same is true for the clouds on the left edge of the first sub-cube and on the right edge of the last sub-cube which constitute the outermost sections of the survey.

\begin{table*}
\begin{tabular}{cclc}
\hline
\hline
Quantity & Unit & Description & Catalog entry \\
\hline
(1) & (2) & (3) & (4) \\
\hline
ID & & Identification number & \texttt{cloud\_id} \\
DS & & Structure number within the dendrogram & \texttt{dendro\_structure} \\
File name & & Sub-cube assignment file name & \texttt{orig\_file} \\
\hline
$x_\mathrm{cen}$ & pixel & Centroid position along the sub-cube x-axis & \texttt{xcen\_pix} \\
$y_\mathrm{cen}$ & pixel& Centroid position along the sub-cube y-axis & \texttt{ycen\_pix} \\
$v_\mathrm{cen}$ & pixel & Centroid position along the sub-cube velocity & \texttt{vcen\_pix} \\
$l$ & degree & Mean Galactic longitude & \texttt{glon\_deg} \\
$b$ & degree & Mean Galactic latitude & \texttt{glat\_deg} \\
$v_\mathrm{LSR}$ & km\,s$^{-1}$ & Mean velocity w.r.t. the local standard of rest & \texttt{vlsr\_kms} \\
$x_\mathrm{\odot}$ & pc & Heliocentric coordinate X & \texttt{xsun\_pc} \\
$y_\mathrm{\odot}$ & pc & Heliocentric coordinate Y & \texttt{ysun\_pc} \\
$z_\mathrm{\odot}$ & pc & Heliocentric coordinate Z & \texttt{zsun\_pc} \\
$x_\mathrm{Gal}$ & pc & Galactocentric coordinate X & \texttt{xgal\_pc} \\
$y_\mathrm{Gal}$ & pc & Galactocentric coordinate Y & \texttt{ygal\_pc} \\
$z_\mathrm{Gal}$ & pc & Galactocentric coordinate Z & \texttt{zgal\_pc} \\
\hline
$\sigma_\mathrm{maj}$ & pixel/arcsec & Major semi-axis size & \texttt{major\_sigma} \\
$\sigma_\mathrm{min}$ & pixel/arcsec & Minor semi-axis size & \texttt{minor\_sigma} \\
$\phi$ & deg & Position angle w.r.t the cube x-axis & \texttt{pa\_deg} \\
$F_{\rm CO}$ & K & Integrated flux & \texttt{flux\_K} \\
$A_{\rm exact}$ & pixel/arcsec$^2$ & Area defined as projected total number of pixel & \texttt{area\_exact} \\
$A_{\rm ellipse}$ & pixel/arcsec$^2$ & Area of the ellipse from $\sigma_{\rm maj,min}$ & \texttt{area\_ellipse} \\
\hline
$T_{\rm peak}$ & K & Peak $T_{\rm mb}$ within the cloud& \texttt{t\_peak\_K} \\
$T_{\rm mean}$ & K & Mean $T_{\rm mb}$ within the cloud& \texttt{t\_mean\_K} \\
$SNR_{\rm peak}$ & & Peak signal-to-noise within the cloud& \texttt{peak\_snr} \\
$SNR_{\rm mean}$ & & Mean signal-to-noise within the cloud& \texttt{mean\_snr} \\
\hline
$d$ & pc & Object distance & \texttt{distance\_pc} \\
\emph{Broadcast type} &  & Distance quality (0 = exact, 1 = broadcasted, 2 = closest)& \texttt{broadcast\_type} \\
\emph{Broad. inaccuracy} & pixel & Broadcast inaccuracy & \texttt{broadcast\_inaccuracy\_pix} \\
\hline
$R_\mathrm{eff}$ & pc & Effective radius & \texttt{radius\_pc} \\ 
$\sigma_\mathrm{v}$ & km\,s$^{-1}$ & Velocity dispersion & \texttt{sigv\_kms} \\ 
$L_\mathrm{CO}$ & K\,km\,s$^{-1}$\,pc$^{2}$ & CO luminosity & \texttt{lco\_kkms\_pc2} \\
$M_\mathrm{lum}$ & M$_\odot$ & Mass from the CO luminosity & \texttt{mlum\_msun} \\
$M_\mathrm{vir}$ & M$_\odot$ & Mass from the virial theorem & \texttt{mvir\_msun} \\
$\sigma_0^2$ & (km~s$^{-1}$)$^{2}$~pc$^{-1}$ & Scaling parameter & \texttt{scalpar\_kms2\_pc} \\
$I_\mathrm{CO}$ & K\,km\,s$^{-1}$ & Integrated CO luminosity & \texttt{surf\_bright\_k\_kms} \\
$N_\mathrm{H_2}$ & cm$^{-2}$ & H$_2$ column density & \texttt{col\_dens\_cm2} \\
$\Sigma_\mathrm{mol}$ & M$_\odot$\,pc$^{-2}$ & Surface density & \texttt{surf\_dens\_msun\_pc2} \\
$\rho_\mathrm{mol}$ & M$_\odot$\,pc$^{-3}$ & Volumetric density & \texttt{dens\_msun\_pc3} \\
\emph{Volume} & pc$^2$\,km\,s$^{-1}$ & Volume & \texttt{volume\_pc2\_kms} \\
$\alpha$ &  & Virial parameter & \texttt{alpha} \\
\hline
$N_\mathrm{pix}$ &  & Number of pixel within the cloud & \texttt{n\_pixel} \\
$N_\mathrm{leaves}$ &  & Number of leaves within the cloud & \texttt{n\_leaves} \\
\emph{Edge} &  & The cloud is on the cube lower or upper border & \texttt{edge} \\
\emph{Parent} &  & Cloud parental structure ID & \texttt{parent} \\
\emph{Ancestor} &  & Cloud parental structure ID at the bottom of the hierarchy& \texttt{ancestor} \\
\emph{Struct. type} &  & Structure type ($T=$\,trunk, $B=$\,branch, $L=$\,leaf) & \texttt{structure\_type} \\
\hline
\emph{Spiral arm} & & Spiral arm associated to the cloud: ($Sa=$Sagittarius, & \texttt{assoc\_sparm} \\
& & $Sc=$Scutum, $Lo=$Local, $Pe=$Perseus, $No=$Norma) & \\
\emph{Dist. to arm} & pc & Distance to the associated spiral arm & \texttt{dist\_to\_sparm} \\
\hline
\end{tabular}
\caption{Contents of the COHRS cloud catalog. In the following analysis we consider also the clouds on the edge of the data cubes. Removing them does not significantly alter the results. The catalog includes uncorrected (without suffix, e.g. \texttt{radius\_pc}), extrapolated (``ex'' suffix, e.g. \texttt{radius\_ex\_pc}), deconvolved (``dc'' suffix, e.g. \texttt{radius\_dc\_pc}), extrapolated and deconvolved (``ex\_dc'' suffix, e.g. \texttt{radius\_ex\_dc\_pc}) properties. For properties that depended only on the flux (e.g., $L_{\rm CO}$, $M_{\rm lum}$, $\Sigma_{\rm mol}$) only uncorrected and extrapolated properties are defined. In the catalog, uncertainties on the properties are specified with the prefix ``\texttt{err}'', e.g. \texttt{err\_radius\_pc}. The electronic version of the catalog is available online.}
\label{T:catalog_entry}
\end{table*}

The catalog we produced contains the data listed in Table~\ref{T:catalog_entry}. The whole catalog is made of 85020 objects: 73140 (86\%) are leaves and 11880 (12\%) are branches. Dendrogram leaves dominate the catalog. These leaves are generally small, isolated structures with sizes comparable to the imposed minimum size limit for the inclusion in the dendrogram (i.e. only a few resolution elements) that cannot be uniquely associated with any other cluster in the catalog, and are therefore retained as independent entities.  While these features are not consistent with the definition of {\em molecular gas cluster} proposed in \cite{colombo15} (since they do not have substructures within them), they can correspond to clumps collected in the BGPS sample. 

This cloud segmentation contains 36\% of the total flux of the survey. This percentage is slightly higher than the flux attributed to GMCs of the full Milky Way catalog designed by \cite{rice2016} (25\%). We attribute the higher fraction to CO(3-2) tracing higher densities than CO(1-0) and is more likely to be associated with compact objects.  This measurement represents just those pixels that are identified with cataloged objects.  The extrapolated flux is 94\% of the total flux in the survey.  Note that the extrapolated flux is not bounded to be less than 100\% and our recovery of a fraction near 100\% does not mean we are characterizing all the flux in the COHRS data. In heavily blended, bright emission where the SCIMES decomposition segments structures high above the noise level of the data, this extrapolation may effectively over-correct for the amount of emission and its extent. Most of the flux that is missed by algorithm is the low-brightness emission near cataloged objects.

In our catalog, 406 objects are ``exact'' distance clouds, 41\,896 are ``broadcasted'', while 42\,718 have a ``closest'' distance association. Given this, across the analysis we will distinguish between the {\em full sample} (all clouds), and a {\em fiducial sample}, consisting of those objects that have a \emph{broadcast inaccuracy} below 5, i.e. the distance pixel is fewer than 5 pixels away from the cloud surface. The latter criterion should compensate for possible mismatches between COHRS and BGPS astrometries. The fiducial sample consists of 597 clouds 
for which we have an accurate measurement of their distances. Clouds in the full sample have a median peak SNR$\sim10$, while for the objects in the fiducial sample the median peak SNR$\sim50$.

For the dendrogram construction, we required that a local maximum had to be separated from other local maxima in space by least 3$\theta_{\rm FWHM}$ to be considered independent. Nevertheless, the effective radius of the clouds can be formally smaller than this limit since it is intensity-weighted. The same is true for the velocity dispersion, which is derived following the same philosophy. For the analyses of the paper we consider only objects with extrapolated $\sigma_{\rm maj}, \sigma_\mathrm{min}>\theta_{\rm FWHM}/\sqrt{8\ln(2)}$ and $\sigma_{\rm v}>\Delta v/\sqrt{2\pi}$, where for COHRS $\theta_{\rm FWHM} = 17''$ and $\Delta v = 1$\,km~s$^{-1}$. s
This restricts the full and the fiducial samples to 35446 and 542 well resolved entries, respectively. For approximately 75\% of the excluded objects the extrapolation failed to derive proper semi-major, semi-minor and/or velocity dispersion; since those structures have generally low signal-to-noise (typically peak $SNR<4$). 

\begin{figure*}
\centering
\includegraphics[width=0.85\textwidth]{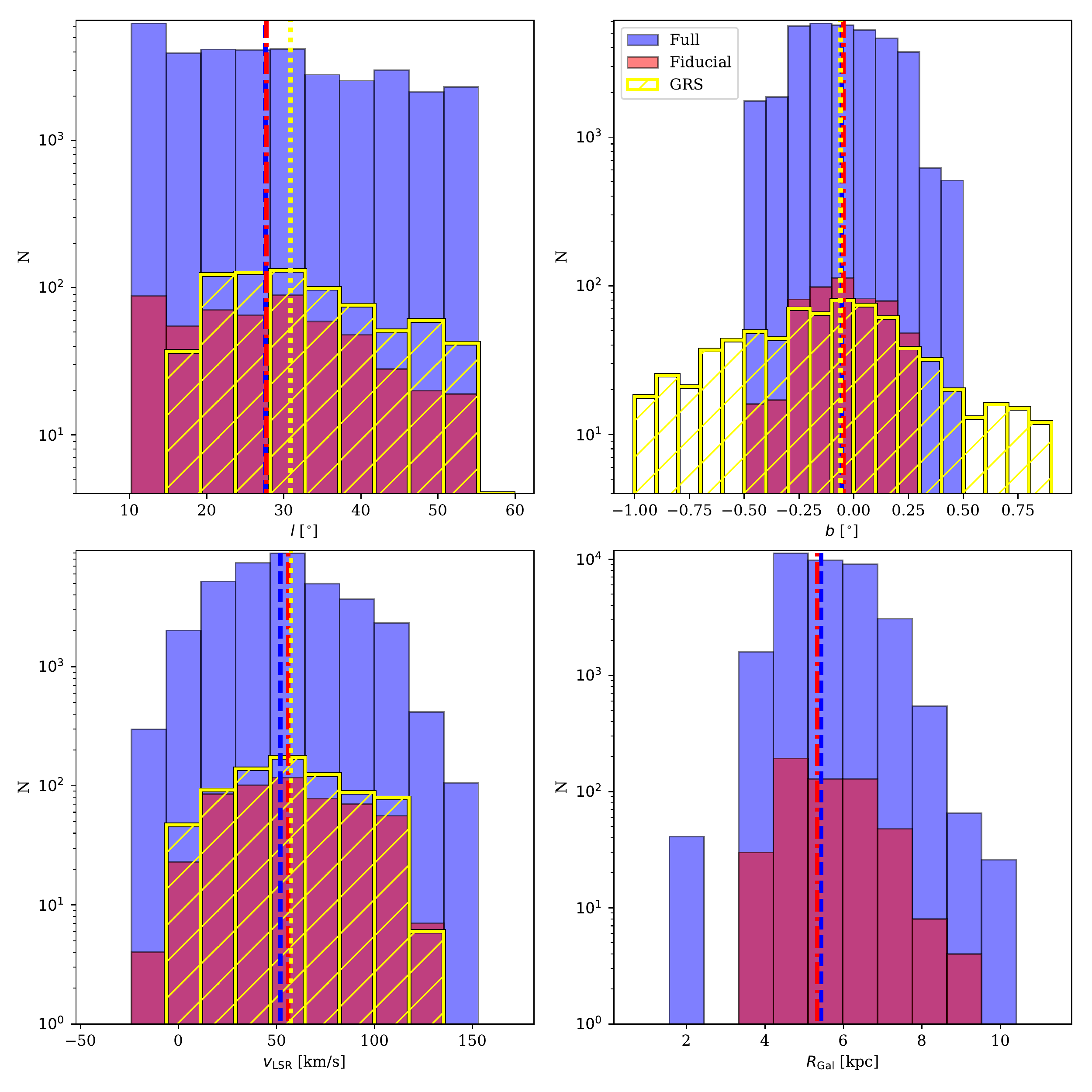}
\caption{Distribution of the COHRS clouds with respect to Galactic longitude ($l$), Galactic latitude ($b$), systemic velocity ($v_\mathrm{LSR}$), and Galacto-centric distance ($R_\mathrm{Gal}$) for the fiducial sample (red), the full sample (blue), and GRS catalog from Roman-Duval et al. 2010 (yellow). The vertical red dashed-dotted lines indicate the median of the fiducial sample, the blue dashed lines the median of the full sample, and yellow dotted lines the median of GRS property distributions. The COHRS fiducial catalog is comparable in scope to the GRS catalog and the differences between these distributions can be attributed to differences in the survey coverages.}
\label{F:cohrs_hist_coor}
\end{figure*}

\subsection{Ensemble properties}
\label{SS:properties}

\begin{figure*}
\centering
\includegraphics[width=0.8\textwidth]{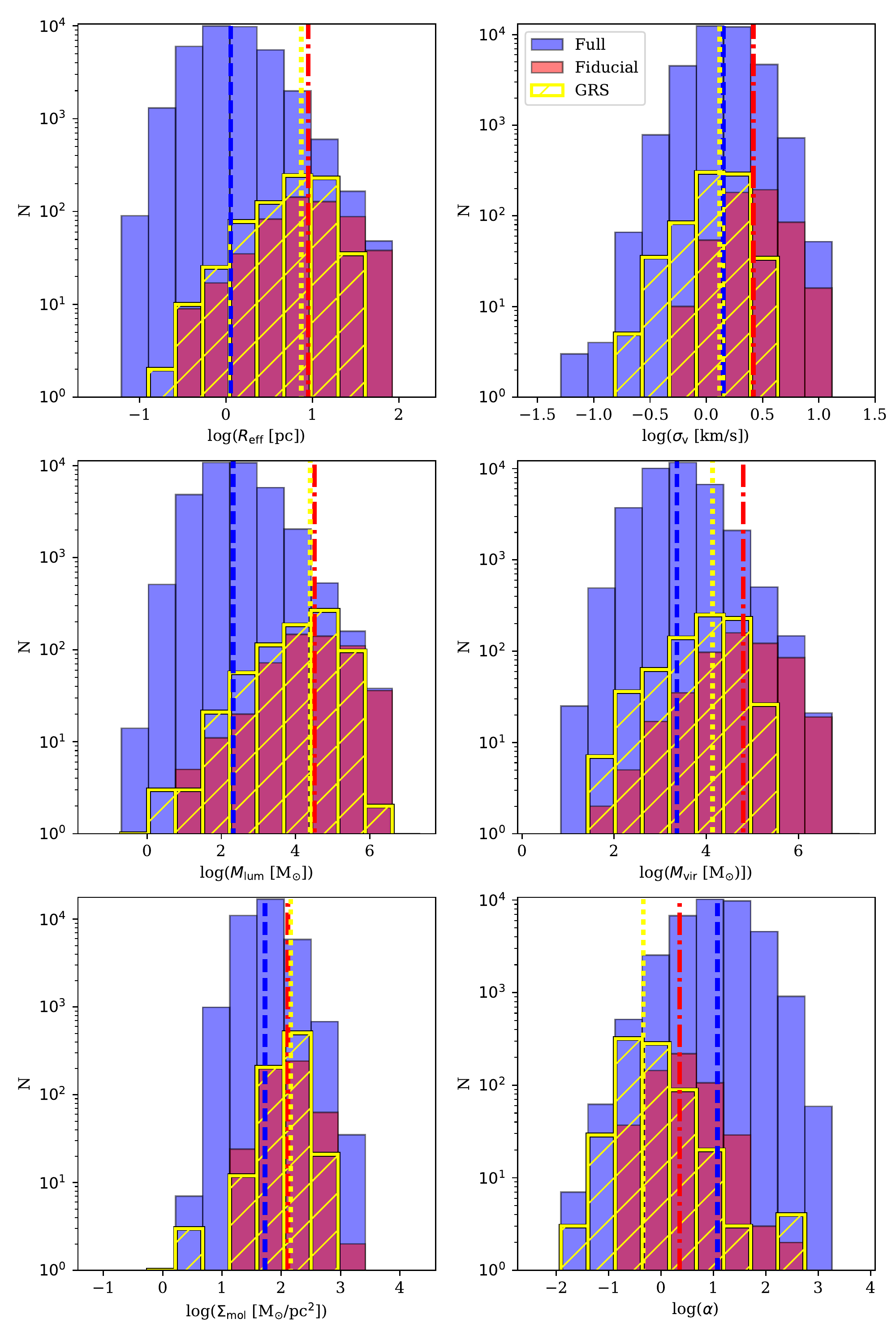}
\caption{COHRS object ensemble properties (from left to right, top to bottom): effective radius ($R_\mathrm{eff}$), velocity dispersion ($\sigma_v$), mass from CO luminosity ($M_\mathrm{lum}$), virial mass ($M_\mathrm{vir}$), molecular gas mass surface density ($\Sigma_{\rm mol}$), virial parameter ($\alpha$). Symbol conventions follow Fig.~\ref{F:cohrs_hist_coor}. The COHRS fiducial sample is comparable to the GRS catalog but the coarse velocity resolution (1 km~s$^{-1}$) of the COHRS data strongly affects our results.}
\label{F:cohrs_hist_props}
\end{figure*}

Here we compare the properties of the objects identified in the COHRS data to the cloud catalog presented in \cite{rathborne2009} and \cite{roman_duval2010}. That catalog was obtained using {\sc Clumpfind} \citep{williams94} on the Galactic Ring Survey (GRS, \citealt{jackson2006}) data. The GRS observed ${}^{13}\mathrm{CO}$(1-0) emission over a large part of the first Galactic quadrant: $18^{\circ}<l<55^{\circ}.7$ and $|b|\leq1$, comparable to but larger than the COHRS survey.  The spatial resolution the GRS is $46''$ and the channel width is 0.21 km~s$^{-1}$. 

In Figure \ref{F:cohrs_hist_coor}, we compare spatial/velocity distribution the fiducial and full sample of our catalog to the GRS catalog. The distributions of GRS clouds and our fiducial sample appear very similar. Nevertheless, we identify three orders of magnitude more objects in the full catalog of the COHRS data. This large discrepancy is because \cite{rathborne2009} smooth the GRS data with a Gaussian kernel of $6'$ and 0.6 km~s$^{-1}$ meaning that one spatial resolution element in the GRS catalog contains 440 resolution elements of the COHRS data. The smoothing increases the signal-to-noise ratio of the GRS data but was primarily done to enable the identification of large, GMC-scale objects using the {\sc Clumpfind} algorithm.  Since {\sc Clumpfind} typically recovers objects a few resolution elements across \citep{pineda09}, it is necessary to suppress the small-scale local maxima with the $6'$ smoothing beam. SCIMES, instead, finds naturally clusters of emission across a wide range of scales and provides large complexes comparable to the GRS catalog without the need of data smoothing.

Clouds in the full sample are uniformly distributed by number along all Galactic longitudes surveyed by COHRS. At large longitudes ($l>40^{\circ}$), the number of the fiducial sample clouds drops by 30\% due to the fact that BGPS distances are less available there. The GRS catalog follows a similar trend, but the decrease of sources at increasing longitudes is less prominent. The median latitude of the three samples peak at latitudes slightly lower than $b=0^{\circ}$ because of the offset of the Sun above the Galactic plane \citep{goodman2014}. Because the latitude range of the GRS data is wider than that of COHRS over a wide longitude range, the latitude distribution of extracted sources is larger in the GRS catalog. In contrast, our fiducial sample has more clouds than the GRS catalog in longitudes between $|b| \leq 0.3^{\circ}$. The cloud samples we are comparing span similar velocity ranges. The GRS catalog and our fiducial sub-sample do not contain exactly the same clouds; however, the distribution of sources and number of clouds in our fiducial sample are directly comparable to the GRS. Both the full sample of COHRS clouds and the fiducial sub-sample are peaked around 5 kpc from the Galactic centre.

Figure \ref{F:cohrs_hist_props} shows that the effective radius of our fiducial sample has its median around $R_\mathrm{eff}\sim 9$\,pc, similar to the GRS clouds which have median sizes $R_\mathrm{eff}\sim 7.4$\,pc.  Nevertheless, our full sample has a median radius of $\sim1$\,pc  reflecting our ability to recover smaller-sized clouds. In terms of velocity dispersion, the COHRS fiducial sample has a median value of $\sigma_\mathrm{v} \sim 2.6 \mathrm{km~s}^{-1}$, while the GRS catalog is smaller: $\sigma_\mathrm{v}\sim1.3$\,km~s$^{-1}$. The COHRS object distributions for the fiducial sample are skewed towards larger $\sigma_\mathrm{v}$ values than those of \cite{rathborne2009}, and while this difference could be partially attributed to the relatively coarse spectral resolution of the COHRS data (1~km~s$^{-1}$) which is a factor 5 worse than that of the GRS (0.2~km~s$^{-1}$), the full catalog for COHRS is still able to recover a median $\sigma_\mathrm{v}\sim1.4$\,km~s$^{-1}$.
The main limitation, may in fact come from the fact that the GRS uses the $^{13}$CO(1-0) to observe the molecular gas, which is an optically thinner tracer than the $^{12}$CO(3-2) used in our study. In practice, this means that the $^{13}$CO(1-0) emission is able to trace higher density regions of the molecular clouds, that are not traceable with $^{12}$CO(3-2). As a result, the linewidths for the clouds measured from {$^{12}$CO} will be naturally larger than those of the GRS, particularly in clouds that contain high-density regions within them. Hence this optical depth effect affects more the fiducial sample, which naturally contains the most massive/dense star forming regions since they have associated compact continuum emission as detected with the BGPS.

The larger-than-expected linewidths will constitute one of the main biases of our study, and even with the deconvolution shown in Section~\ref{SS:props_exdc} we are not able to compensate for this effect.
This will potentially influence our measured virial masses and virial parameters. The average virial mass and virial parameters of the COHRS fiducial sample ($\sim6.4\times10^5$\,M$_{\odot}$, and $\alpha\sim2$) are a factor 4 larger than the respective GRS median values ($\sim1.4\times10^5$\,M$_{\odot}$, and $\alpha\sim0.5$). 

The average masses derived from CO luminosity and surface densities are fairly consistent between the two catalogs: $\langle M_{\rm lum}\rangle \sim3.2\times10^5$\,M$_{\odot}$ (COHRS fiducial) vs.~$\langle M_{\rm lum}\sim2.5\times10^5$\,M$_{\odot}\rangle$ (GRS). Nevertheless, the fiducial sub-sample contains clouds more massive than GRS. Similarly, the average cloud mass surface densities between the GRS catalog and the COHRS fiducial sample are approximately the same ($\Sigma_{\rm mol}\sim130$\,M$_{\rm \odot}$\,pc$^{-2}$ and $\Sigma_{\rm mol}\sim140$\,M$_{\rm \odot}$\,pc$^{-2}$, respectively), but we have clouds with larger surface densities in our sample.

\begin{figure}
\centering
\includegraphics[width=0.4\textwidth]{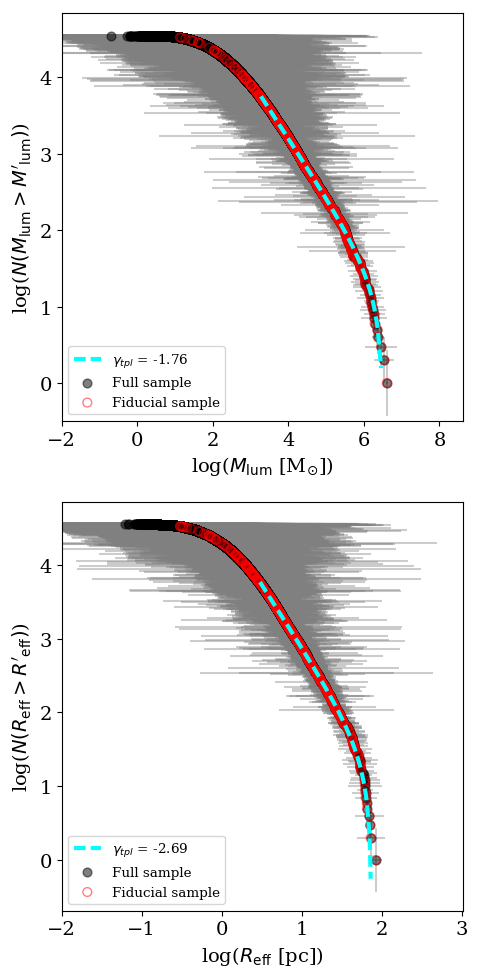}
\caption{COHRS mass (top) and size (bottom) spectra from the $M_{\rm lum}$ and $R_{\rm eff}$ of all the clouds in the catalog (black circle). The fiducial sample is indicated with red empty circles. Cyan dashed lines indicate the truncated power-law fit of the spectra above the estimated completeness limits. Gray lines indicate the error bars. For the $y-$axis the uncertainty is given by the counting error: $\sqrt{N}$.}
\label{F:cohrs_mrspect}
\end{figure}

\subsection{Cumulative distributions of cloud masses and sizes}
\label{SS:cum_dists}

Fitting cloud mass and size distributions can provide basic information about the cloud population and the molecular ISM itself. In this section we use the cumulative distributions of cloud $M_{\rm lum}$ and $R_{\rm eff}$  that we model as a truncated power-law distribution \citep{williams_mckee1997}:

\begin{equation}\label{E:cumdist}
N(X > X') = N_0\left[\left(\frac{X}{X_0}\right)^{\gamma+1}-1\right].
\end{equation}
Here, $X_0$ will correspond to $M_0$ for the mass distribution and to $R_0$ for the effective radius distribution, and represents the maximum value of the distribution, while $N_0$ is the number of clouds greater than $2^{1/(\gamma+1)}X_0$, i.e., where the distribution deviates from a power law. If $N_0\gg 1$ there is strong evidence for a truncation in the power law which indicates that physical effects are at work to limit the maximum value of a given cloud property. The truncation in the cloud mass distribution has been observed in several occasions  \citep{williams_mckee1997, rosolowsky05, freeman17, jeffreson18}. The ``cumulative'' form allows to fit the distributions considering the uncertainties on the cloud properties, and it is not influenced by the choice of the bin size that can bias binned distributions (\citealt{rosolowsky05}).  

Cumulative mass spectra are only well defined above the completeness limit of the survey. To estimate the mass completeness limit we consider the procedure illustrated in \citet[][equations 2 to 4]{heyer2001}. For their outer Galaxy cloud catalog the authors suggest a minimum CO luminosity given by: 

\begin{equation}\label{E:Lco_min}
L_{\rm CO}^{\rm min}(d)\mathrm{ [K\,km/s\,pc^2]} = N_p N_c T_{\rm th} \Delta v \Omega_{\rm mb} d^2;
\end{equation}

where $N_p$ is the minimum number of pixel per object, $N_c$ the minimum number of velocity channels, $T_{\rm th}$ the main-beam antenna temperature threshold, $\Delta v$ the dataset channel width, $\Omega_{\rm mb}$ beam solid angle, and $d$ the distance to the cloud. As explained by \cite{heyer2001}, the completeness limit is evaluated at 5$\sigma$ confidence limit:

\begin{equation}\label{E:Lco_c}
L_{\rm CO}^{\rm c} = L_{\rm CO}^{\rm min} + 5\sigma(L_{\rm CO});
\end{equation}

where:

\begin{equation}\label{E:sig_Lco}
\sigma_{L_{\rm CO}}\mathrm{ [K\,km/s\,pc^2]} = \sigma_{\rm RMS}\sqrt{N_p N_c} \Delta v \Omega_{\rm mb} d^2;
\end{equation}

and $\sigma_{\rm RMS}$ is the median RMS noise value across the full survey.

In our case we assume that the minimal object contains $N_p=18$ pixels (3 beams $\times$ 6 pixels per beam), $N_c=2$ channels (required by the dendrogram generation), minimum brightness $T_{\rm th}=\mathrm{SNR}\sigma_\mathrm{RMS}$ (where we assume $\sigma_\mathrm{RMS}=1$\,K as an conservative value across the COHRS fields, see \citealt{dempsey2013}; and SNR=3 as imposed by our masking method), channel width $\Delta v=1$\,km~s$^{-1}$ (COHRS data cube channel width), beam size of $\theta_\mathrm{FWHM}=17"$ (COHRS data beam). At a distance of $\sim15$\,kpc,  the largest distance in our catalog, we calculate a luminosity mass completeness of $\sim2\times10^3$\,M$_{\odot}$ by assuming our $\alpha_{\rm CO}=8.7$\,M$_{\odot}$\,(K\,km~s$^{-1}$\,pc$^2$)$^{-1}$. 

For the cloud size, considering the COHRS beam of $\theta_\mathrm{FWHM}=17''$ and that a cloud must span at least 3 beams to be regarded as an independent structure in the dendrogram, we get a effective radius completeness of 3\,pc.

These estimated completeness limits are conservative estimates for $M_{\rm lum}$ and $R_{\rm eff}$ at 15\,kpc. SCIMES does not extract objects at a fixed $T_{\rm mb}$ threshold. Instead, the masking level is set by the local noise properties and the SNR=3 threshold from the masking and dendrogram generation parameters.  Moreover, we use extrapolation and deconvolution which renders the measurement of the radius distribution more complex.  Thus, it is possible to find several objects at 15\,kpc with masses and effective radii below $\sim 2\times 10^3$\,M$_{\odot}$ and 3\,pc, respectively.

We fit equation~\ref{E:cumdist} to our spectra above these completeness limits using Orthogonal Distance Regression (ODR) as implemented in {\sc scipy} \footnote{\url{https://docs.scipy.org/doc/scipy/reference/odr.html}}, which takes into account the uncertainties on both dependent and independent quantities. Fig~\ref{F:cohrs_mrspect} shows the result of this experiment for both mass (top) and size (bottom) distributions. 

For the mass distribution of the full catalog, we find a power-law slope of $\gamma=-1.76\pm0.01$, $N_0=20.2\pm0.1$, $M_0=(3.14\pm0.03)\times10^6$\,M$_{\odot}$. The truncation indicates that clouds with mass above $M_0\sim 3\times10^6$\,M$_{\odot}$ are significantly absent in the region of the Milky Way surveyed by COHRS. A truncation around $10^6$\,M$_{\odot}$ has been observed for the first Galactic quadrant by other studies which use the cumulative representation of the mass spectrum (e.g.~\citealt{rosolowsky05}, \citealt{rice2016}).  Such a truncation mass is a critical feature for testing GMC evolution theories in the context of galaxy environment.  \citet{reina_campos2017} developed a model for the maximum mass scale of GMCs in galaxies as a function of local environment, finding a near constant upper limit mass of $\sim 10^6~M_{\odot}$ over the galactocentric radii of 4-8 kpc.  The physical effects that govern this mass scale are gravitational collapse on scales allowed by the Toomre stability criterion.  Our mass distributions, when separated into bins of galactocentric radius show a nearly constant truncation mass at all radii, consistent with those models. 

A spectral index $\gamma\sim-1.76$ indicates most of the molecular gas mass is contained in large objects. Our measurement is largely consistent to the $\gamma\sim-1.7$ observed for the inner Milky Way (e.g. \citealt{roman_duval2010}, \citealt{heyer_dame2015}, \citealt{rice2016}).  Our slightly steeper index can arise from the significantly higher resolution of the survey compared to the data supporting previous catalogs.  With coarse resolution, small clouds will be blended with large clouds, suppressing the number of small clouds recovered and increasing the mass of the larger clouds, thereby biasing the index to shallower values.

For the size distribution we get $\gamma=-2.70\pm0.01$, $N_0=26.4\pm0.2$, $R_0=73.6\pm 0.2$\,pc. The latter two quantities indicate that the $R_{\rm eff}$ distribution shows a truncation as well: the size of the clouds also reaches a maximum value in the inner Galaxy. In contrast to the mass spectrum, our size distribution appears significantly shallower than the one observed in the same region through $^{13}$CO observation \citep[$\gamma=-3.9$,][]{roman_duval2010}. This difference can be attributed to the action of the {\sc SCIMES} algorithm, which can recover objects significantly larger than the resolution element.  The GRS catalog is largely established by the size of the smoothing element, leading to a sharp cutoff in object sizes larger than 8 pc (i.e., the $6'$ smoothing kernel projected to 5 kpc). This is supported by Figure \ref{F:cohrs_hist_props}, which shows that our COHRS catalog recovers a tail of larger objects than those in the GRS catalog. 

\begin{figure*}
\centering
\includegraphics[width=0.7\textwidth]{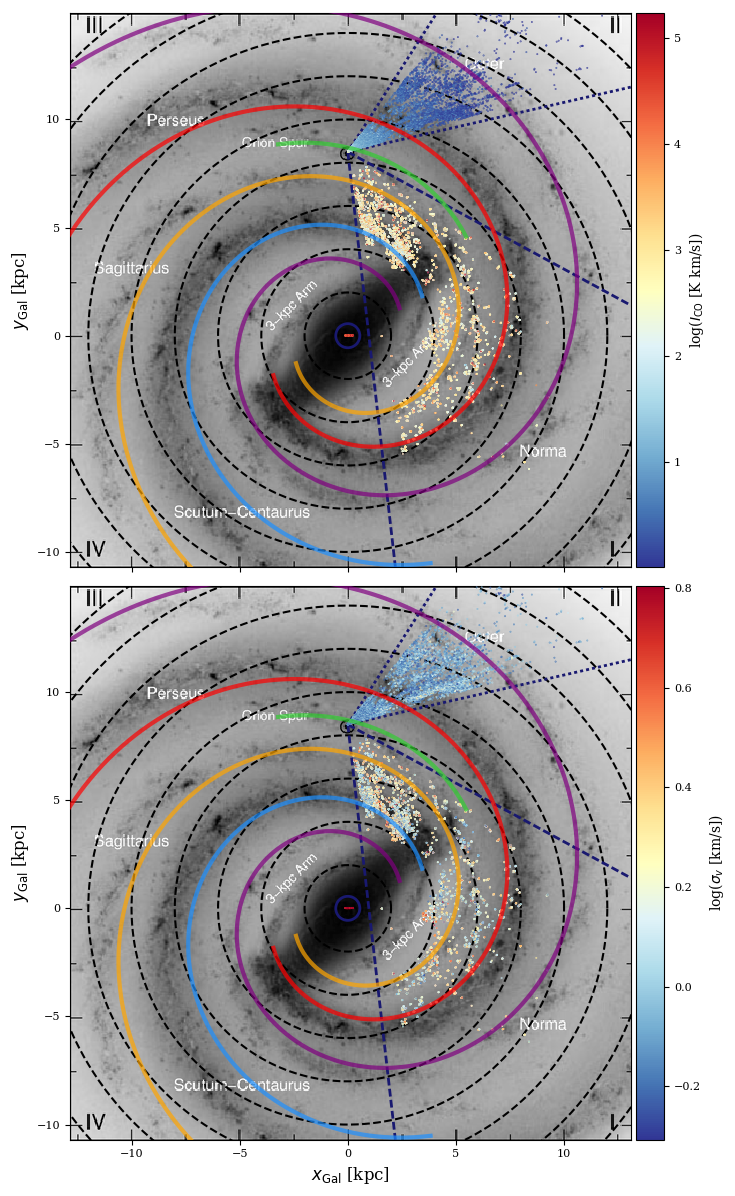}
\caption{COHRS, outer Galaxy, and Galactic centre cloud positions with respect to the Milky Way artistic impression (courtesy of NASA/JPL-Caltech/R. Hurt (SSC/Caltech)). Markers are color-encoded with a given properties: CO integrated intensity (top) and velocity dispersion (bottom). Those properties are independent from distance. The dark blue circle, dashed and dotted lines indicate, approximatively, the Milky Way regions observed by \citealt{oka2001}, \citealt{dempsey2013}, and \citealt{heyer2001} surveys, respectively. Note that the data present the \citet{oka2001} catalog of the Galactic centre are assumed to all be at a common $y_{\rm Gal}$ position in the Galactic centre. Galactocentric dashed black circles are placed 2\,kpc apart. The spiral arm positions are drawn from the results of Vallee 2017 (see Section~\ref{SS:mw_distribution_sparm} for further details) Scutum (blue), Sagittarium (yellow), Perseus (red), Local (green) and Norma/outer (purple).}
\label{F:cohrs_mw}
\end{figure*}

\section{COHRS clouds in the context of the Milky Way}
\label{S:mw_distribution}

In the previous section, we have looked at the global distribution of cloud properties using the COHRS dataset. In this section, we shall investigate if these properties change as a function of Galactic environment.

\subsection{Comparison between Galactic centre, inner Galaxy, and outer Galaxy clouds}
\label{SS:mw_distribution_comp}

In order to have a first glance at how the Galactic environment might be affecting cloud properties, we compare our clouds to those seen in other Galactic regions, where the environment is potentially different from the inner Galaxy. In particular we analyze our data with respect to the catalogs built for the outer Galaxy (\citealt{heyer2001}) and the Galactic centre (\citealt{oka2001}), which
have been constructed starting from $^{12}$CO observations\footnote{Note, however, that Heyer et al. 2001 and Oka et al. 2001 uses $^{12}$CO(1-0) which has different density sensitivity with respect to $^{12}$CO(3-2) ($\sim1.4\times10^3$\,cm$^{-3}$ versus $\sim7\times10^4$\,cm$^{-3}$), respectively. Therefore, $^{12}$CO(1-0) would potentially trace more extended structures with respect to $^{12}$CO(3-2), and give broader line-widths.} that are similar in term of spatial and spectral resolution to COHRS (however bias effects introduced by the segmentation methods can apply, see Section~\ref{S:surveys}). 

\begin{figure}
\centering
\includegraphics[width=0.5\textwidth]{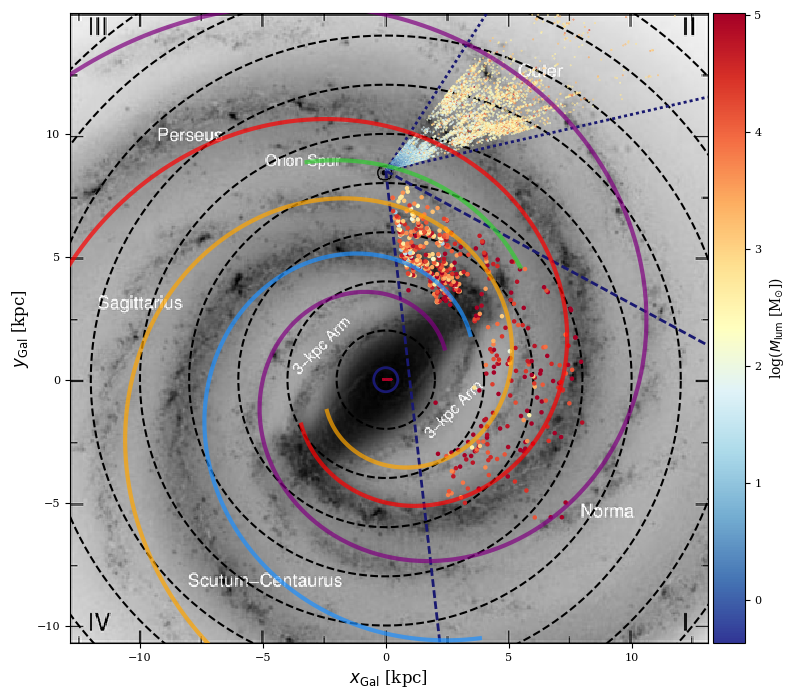}
\caption{COHRS, outer Galaxy, and Galactic centre cloud positions with respect to the Milky Way artistic impression (courtesy of NASA/JPL-Caltech/R. Hurt (SSC/Caltech)). Markers are color-encoded with a the cloud mass from CO luminosity. For the COHRS data only the fiducial sub-sample is indicated. Other symbol notations follow Fig~\ref{F:cohrs_mw}.}
\label{F:cohrs_mw_mass}
\end{figure}

In Fig.~\ref{F:cohrs_mw} we compare the clouds in the three surveys using two distance-independent properties: CO integrated intensity ($I_{\rm CO}$, upper panel) and velocity dispersion ($\sigma_{\rm v}$, lower panel). The cloud data are color-encoded by their respective properties and plotted on the face-on view of the Milky Way. In terms of $I_{\rm CO}$, the clouds in the three Galactic regions are starkly different: objects in the outer Galaxy reach maximum integrated intensities around $5\times10^3$\,K\,km~s$^{-1}$, while clouds in the Galactic Center have $3\times10^3<I_{\rm CO}<6\times10^6$\,K\,km~s$^{-1}$. COHRS objects show values of CO integrated intensity in between these two extremes. Similar conclusions can be drawn from the velocity dispersion comparison, even if the difference is less sharp. In Fig.~\ref{F:cohrs_mw_mass} we show the data points color-encoded by their mass from CO luminosity. In this case, for COHRS we plot only the clouds from the fiducial sample, for which we have good estimations of their distances. This experiment highlights the fact that the fiducial sample is mostly constituted by massive objects, and their masses appear to be similar to the clouds identified in the Galactic centre. 

\begin{figure*}
\centering
\includegraphics[width=1\textwidth]{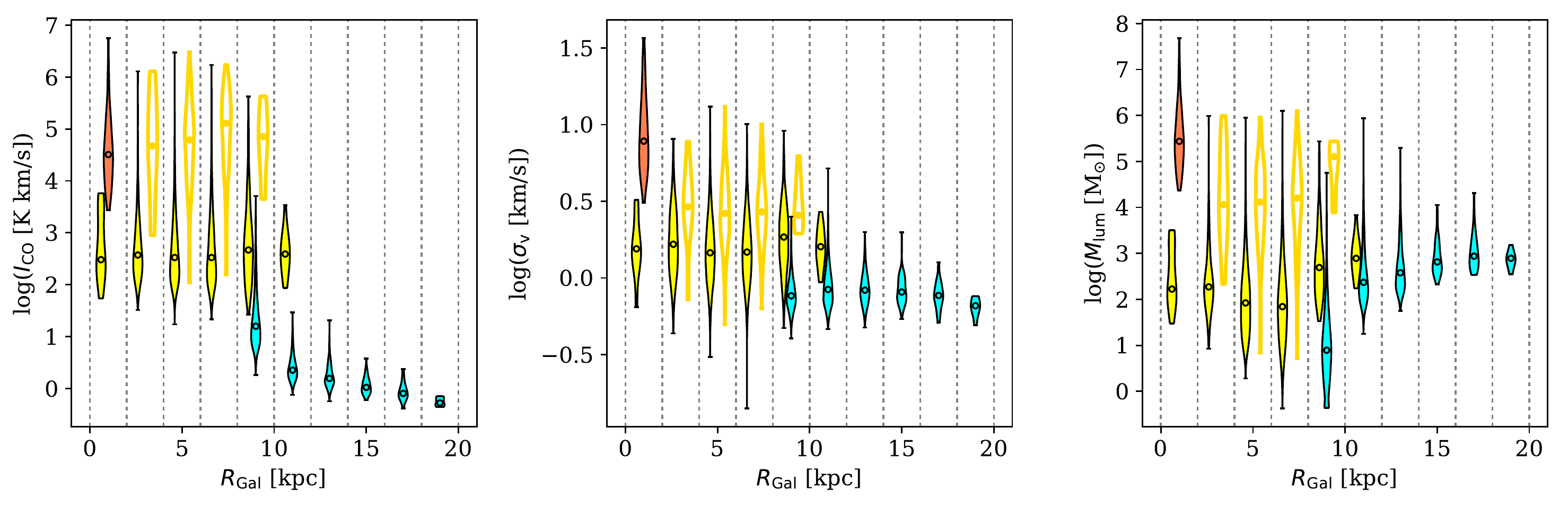}
\caption{Violin plot representations of CO integrated intensity (left), velocity dispersion (middle) and luminosity mass (right) for the clouds in the outer Galaxy (cyan, Heyer et al. 2001 catalog), Galactic Center (red, Oka et al. 2001 catalog), COHRS full catalog (yellow full violins) and fiducial catalog (yellow empty violins). Violin plots are histograms, where the width along the x-axis indicates the normalized fraction of data at the corresponding y-axis value. The violins extend from the minimum to the maximum of the distributions. Circles indicate the median of the property distributions within each Galactocentric radial bins of 2\,kpc.}
\label{F:cohrs_raddists}
\end{figure*}

While drawing the clouds on a face-on view of the Milky Way is useful to visualize their locations across the Galactic disk, the large overlap between the data points does not allow to derive firm conclusions about the radial gradient of the cloud properties. Therefore, in Fig.~\ref{F:cohrs_raddists} we display violin plots of CO integrated intensity, velocity dispersion, and luminosity mass within bins of 2\,kpc Galactocentric radius for the cloud within the three surveys. This representation confirms that, on average, clouds segmented out from COHRS data have properties intermediate between Galactic centre and outer Galaxy objects. Nevertheless, intrinsic biases of the different segmentation methods might play a role here (see Section~\ref{S:surveys}). 
For instance, outer Galaxy cloud masses in the most external bin reach estimates closer to the ones of COHRS objects, but the monotonic increment of the average cloud mass in the outer Galaxy suggests that the extracted properties are scaling with the distance, as would be the case when the segmentation method extracts clouds around a specific angular-scale, rather than to the actual physics involved in the different Galactic regions.

\begin{figure*}
\centering
\includegraphics[width=1\textwidth]{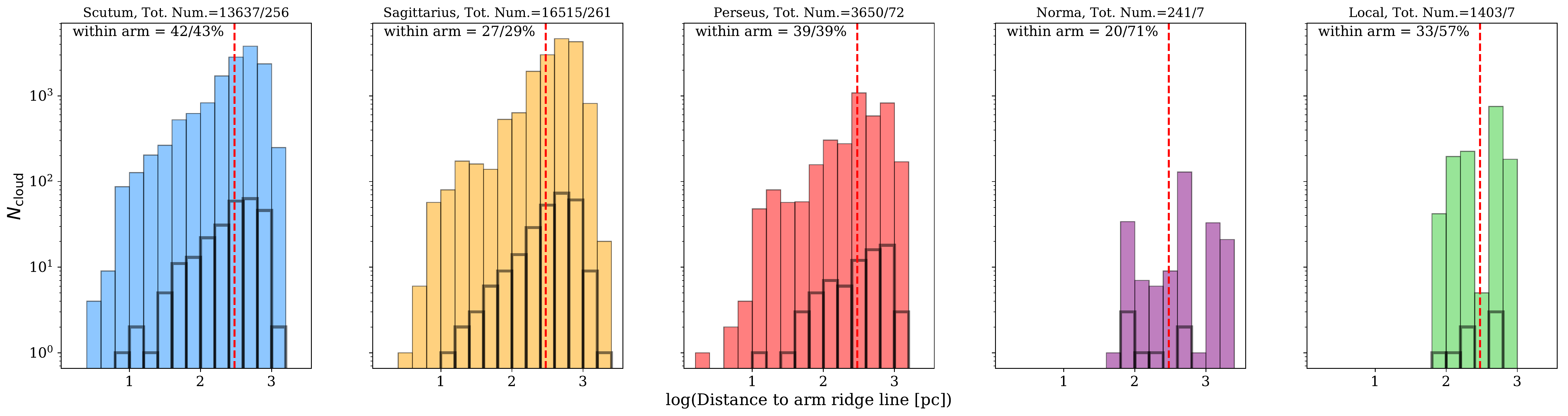}
\caption{Histograms of the number of clouds from the ``full'' sample associated with (from left to right) Scutum, Sagittarius,  Perseus, Norma/outer and Local arms described as log-normal spirals as drawn in Fig.~\ref{F:cohrs_mw}-\ref{F:cohrs_mw_mass}. Black transparent histograms highlight the clouds in the ``fiducial'' sample. The total number of full and fiducial sample clouds associated to a specific arm is indicated in the title of each panel. The percentage of full and fiducial sample clouds located within the spiral arms is given in the upper left side of each panel. The vertical, red, dashed lines mark the average spiral arm half width (300\,pc; Vallee 2017).}
\label{F:cohrs_sparm}
\end{figure*}

\subsection{Galactocentric dependency of properties in the COHRS sample}
\label{SS:galcentre_prop}

Given that interpreting the environmental dependencies of cloud properties with the intercomparison of different surveys is not straightforward, we explore if we can detect any trends within the COHRS sample alone. To do so, we have divided the full sample into four bins according to the galactocentric distance of the clouds, each containing an equal number of clouds, and estimated their individual cummulative distributions of cloud masses and radius (see Figure~\ref{F:cum_arm_interam}, top row). All bins show mass spectrum slopes consistent with the full sample one ($\gamma\sim-1.75$). This indicates that  most of the molecular gas is contained in massive clouds, for all radial bins. Nevertheless, the innermost annulus (where $1.6<R_{\rm Gal}<4.8$\,kpc) exhibits a distribution shifted towards higher masses than the other bins. This annulus contains the most massive clouds of the sample, corresponding to rich reservoir of molecular gas in the Galactic ring. Furthermore, the dynamics likely favor the formation of large cloud complexes at the end of the Milky Way's stellar bar. The two outer annuli we consider ($5.5<R_{\rm Gal}<6.3$\,kpc and $6.3<R_{\rm Gal}<10.4$\,kpc) show  similar mass distributions. However, the region between $4.8<R_{\rm Gal}<5.5$\,kpc appears to contain less massive clouds compared to other Galactocetric annuli. 

The radial variations in the size distributions constructed within the same annuli do not always reflect the radial changes in the mass spectra. The innermost ring and the two outermost annuli are almost indistinguishable. However, the spectrum of the ring between $4.8<R_{\rm Gal}<5.5$\,kpc appears bended towards lower effective radii, reflecting the behavior of the corresponding mass distribution.

From this analysis, it appears that various environmental effects are at work in the surveyed Milky Way region that contribute to create mass and size distributions with different shapes. \cite{zetterlund2018} find a significant steepening of the mass distribution of dense gas clumps in the the range $4.6 < d_\mathrm{Gal} < 6.3$ kpc. We note that variations in the mass spectra remain when examined in a distance-limited sample, and are not likely to be attributable to distance scalings alone. 

\subsection{Spiral arm versus Inter-arm clouds in the COHRS sample}
\label{SS:mw_distribution_sparm}

From Fig~\ref{F:cohrs_mw}-\ref{F:cohrs_mw_mass} it appears that a significant number of clouds in the COHRS sample might be associated to the inter-arm regions of the Milky Way. It is interesting now to verify this rudimentary visual impression with a more rigorous test. For this experiment and to draw the arms in Fig~\ref{F:cohrs_mw}-\ref{F:cohrs_mw_mass} we use the spiral arm models defined by the log-normal spiral:

\begin{equation}\label{E:sparm}
\log(R/R_{\rm ref}) = -(\beta - \beta_{\rm ref})\tan(\psi)
\end{equation}

where the reference Galactocentric radius ($R_{\rm ref}$), Galactocentric azimuth ($\beta_{\rm ref}$), and pitch angle ($\psi$) are taken from the recent update of \cite{vallee2017}. In the model of \cite{vallee2017}, the pitch angle of each arm is kept constant ($\psi=13^{\circ}$). We assume that the four main Milky Way arms originate from the tip of the long bar which have a semi-major axis of 5\,kpc \citep{wegg2015}, therefore $R_{\rm ref}=5$\,kpc for each arm. The values of $\beta_{\rm ref}$ are assumed to be 0$^{\circ}$, 90$^{\circ}$, 180$^{\circ}$, 270$^{\circ}$ for Scutum, Sagittarius, Perseus, Norma arms, respectively. Given the average inclination of the bar with respect to the line that connects Sun and Galactic centre, $a=30^{\circ}$ \citep{wegg2015}, the starting Galactocentric azimuth of Scutum and Norma arms are fixed to $\beta=60^{\circ}$, while for Sagittarius and Perseus arms $\beta=240^{\circ}$. The parameters of the Local spurs have been precisely defined by the VLBI maser parallaxes measurements of \cite{reid2014} (see their Table 2): $R_{\rm ref}=8.4$\,kpc, $\beta_{\rm ref}=8.9^{\circ}$, $\psi=12.8^{\circ}$.

To quantity the number of clouds associated with a spiral arm we calculate the Euclidean distance between the cloud centroid position in Galactocentric coordinates ($x_{\rm Gal}$ and $y_{\rm Gal}$ in the catalog) and the closest point of each spiral arm ridge line described by the model.

The result of the analysis is reported in Fig.~\ref{F:cohrs_sparm}. Most of the objects appear almost equally distributed between Sagittarius and Scutum arms, followed by Perseus, Local, and Norma arms. Similarly to the visual impression obtained by looking at Fig.~\ref{F:cohrs_mw}-\ref{F:cohrs_mw_mass}, a small fraction of the clouds ($\sim35\%$) are found within the spiral arms, if we consider an average arm width of 600\,pc \citep{vallee2017}. By assuming a larger average width, 800\,pc as calculated by the same author in an earlier paper \citep{vallee2014} almost 50\% of the clouds appear to be enclosed within spiral arms. Clouds within the spiral arms encompass almost 50\% of the cloud flux in the catalog (considering also the small objects excluded from most of the analysis in the paper). These fractions are very similar to the findings in nearby galaxies (for M51 only $\sim60\%$ of the flux comes from spiral arms, \citealt{colombo14a}).

\begin{figure*}
\centering
\includegraphics[width=0.75\textwidth]{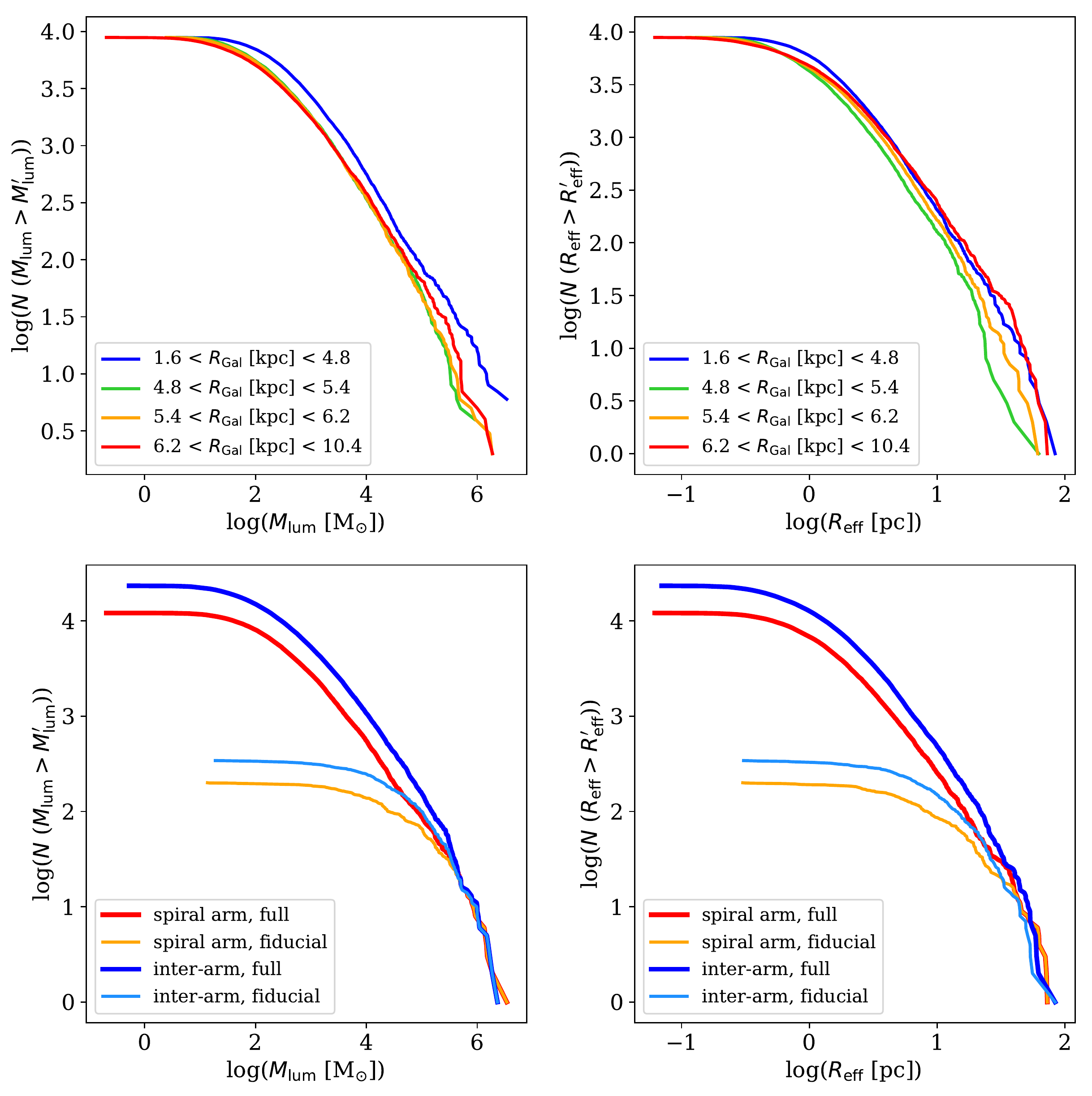}
\caption{Cumulative mass (left) and size (right) distributions for the clouds in COHRS as function of the environment. \emph{Top:} colored spectra are calculated for the full sample within various Galactocentric radii that contain the same number of clouds. \emph{Bottom:} distributions of cloud mass and size within the spiral arms and in the inter-arm regions for the full and fiducial samples separately.}
\label{F:cum_arm_interam}
\end{figure*}

Using these sub-samples, we explored whether there are any noticeable differences between the cloud properties in either sub-sample. The cumulative distributions of the mass and radius are shown in Fig.~\ref{F:cum_arm_interam} (bottom row). From there, we can see that the distributions look overall very similar, independently of the arm/interarm assignment, particularly when looking at the slope of the distributions. A two-sided KS test on the full sample has low p-values, which would suggest that the two are not drawn from the same distribution. However, the KS test on the fiducial sample shows that the distributions are very similar. Interestingly, despite the lower numbers of clouds in the spiral arm regions, we find that both distributions reach similar maximum sizes and masses.

We note, however, that while it is interesting to see that we do not recover a strikingly different cumulative mass or size distribution in these two different environments \citep[as opposed to what has been observed in, e.g., M51][]{colombo14a}, there are a couple of caveats to this study that we should bear in mind. 

Firstly, as noted in Section~\ref{SS:properties}, optical depth effects could play a key role, particularly in our ability to trace the higher-density regions within molecular clouds, which could potentially underestimate the masses of our clouds, especially for the most massive complexes where more of the mass is enclosed in high-density regions. This effect could potentially alter the shape of the tails of the distribution, which is where we would expect to see a different behavior between the two distributions \citep[e.g.][]{koda2009,duarte-cabral2016}

Secondly, the specific assignment of each cloud into arms or interams, is highly uncertain, partly due to our limited knowledge of the position and extent of each arm in the Milky Way. Indeed, \cite{reid2014} calculated a variable width for the spiral arms of the Milky Way: 170\,pc for Scutum, 260\,pc for Sagittarius, 380\,pc for Perseus , 630\,pc for Norma, and 330\,pc for the Local arms (see their Table 2). Using this width the number of clouds and the flux within the spiral arms appear strongly reduced: only $12\%$ of the clouds are found within the spiral arms, carrying a similar percentage of fluxes with respect to the total cataloged cloud fluxes. Nevertheless, the derivation of \cite{reid2014} are valid for the 2$^{\rm nd}$ Galactic quadrant. Moreover, the authors noticed that the arm width tends to increase with Galactocentric radius. Therefore, the spiral arm widths might be different than the assumed ones in the region surveyed by COHRS. But even if the model derived by \cite{vallee2017} appears to be the more probable given the variety of tracers analyzed and summarized by the author, our position within the Galactic plane makes difficult to define the spiral arm clouds with good precision. For instance, we have assumed that a cloud whose centroid falls within 300\,pc from the assigned spiral arm is fully contained within it, but the clouds' extension and morphology is not taken into account. Some of the clouds might straddle the arm and inter-arm regions (as in the case of feather or spurs observed in nearby grand-designed spirals, e.g. \citealt{schinnerer2017}), and this is not accounted for. 

Lastly, the kinematic distances used to place the COHRS clouds in a face-on view of the Galaxy, are affected by large uncertainties, especially when considering clouds that are in or close to a spiral arm, where the velocities deviate from the circular motions assumed for the gas when using a model of the rotation curve \citep[e.g.][]{ramon-fox2018,duarte-cabral2015}. This effect can place clouds in an arm while they should be inter-arm clouds, and vice versa, making this exercise non-trivial.

\section{Scaling relations between cloud properties}
\label{S:scaling_relations}

So far we have only analyzed each cloud property on its own. However, the correlations between  different properties have been the primary channel by which we understand the dynamical state and evolution of the molecular ISM. The seminal work in this area is from \citet{larson81} who identified three fundamental scaling relations between the molecular cloud properties, often called ``Larson's laws''. \citet{larson81} measured a correlation between velocity dispersion and size among molecular objects from sub-parsec to few hundreds parsecs size (the size-linewidth relation) as $\sigma_v=1.10L^{0.38}$. The author interpreted this relation as the manifestation of Kolomogorov's incompressible turbulence, which was proposed to be the main agent for creating molecular overdensities.  The second relation measured by \cite{larson81} was between velocity dispersion and mass ($\sigma_v=0.42M^{0.20}$) which implies that the clouds are in approximate virial equilibrium.
Consistent with this hypothesis, \citet{larson81} demonstrated there was no discernible relationship between the virial parameter, $\alpha$, and cloud size. An anti-correlation between cloud mean density and size ($n=3400L^{-1.10}$) was consistent with clouds having a constant surface density. \cite{solomon87} then used a homogeneous cloud sample in the inner Galaxy to remeasure the Larson's relations, and found a constant mass surface density $\Sigma_\mathrm{mol}=170$\,$M_\odot$\,pc$^{-2}$.  

However, these scalings are not independent (\citealt{heyer2009}, \citealt{wong2011}). The second Larson's relation implies that the clouds are in virial equilibrium; in terms of the virial parameter, $\alpha\sim1$, or $M_\mathrm{lum} \sim 5\sigma_v^2 R_\mathrm{eff}/G$. Together with the definition of cloud mass surface density, $\Sigma_\mathrm{mol} = M_\mathrm{lum}/(\pi R_\mathrm{eff}^2)$, the cloud velocity dispersion can be expressed as:
\begin{equation}\label{E:sigv_lars}
\sigma_v = \sqrt{\frac{\pi}{5} G \Sigma_\mathrm{mol} R_\mathrm{eff}}.
\end{equation} 
For a $\Sigma_\mathrm{mol}\sim200$\,$M_\odot$\,pc$^{-2}$ which is typical for the inner Galaxy (e.g., \citealt{heyer_dame2015}), we find a formulation of the first Larson's relation:
\begin{equation}\label{E:lars1}
\sigma_v=0.74R_\mathrm{eff}^{0.50}, 
\end{equation}
similar to the one observed by \cite{solomon87}. In this aspect the scaling between size and linewidth of the cloud 
\begin{equation}
\sigma_0\equiv \sigma_v \left( \frac{R_\mathrm{eff}}{\mathrm{1~pc}}\right)^{-1/2}
\end{equation} 
should be constant. \citet{heyer2009} reanalyzed \cite{solomon87} clouds using $^{13}$CO data drawn from the GRS to independently derive the cloud mass through an LTE analysis and found a scaling between $\sigma_0$ and the mass surface density of the clouds. This implies that the clouds in the first Galactic quadrant do not follow the original Larson relations: they cannot be defined by a single scaling between $\sigma_v$ and $R_\mathrm{eff}$, as they do not have constant $\Sigma_\mathrm{mol}$, and they are not necessarily gravitationally bound.\\ 

\begin{table*}
\begin{tabular}{r|cccc|cccc}
\hline
Relation & \multicolumn{4}{c}{Fiducial} & \multicolumn{4}{c}{Full} \\
& $\beta$ & $\gamma$ & Scatter [dex] & $r$ & $\beta$ & $\gamma$ & Scatter [dex] & $r$ \\
\hline
$\sigma_v$ [km/s] = $\beta$($R_\mathrm{eff}$ [pc])$^{\gamma}$ & 1.79 $\pm$ 1.05 & 0.24 $\pm$ 0.02 & 0.41 & 0.40 & 1.91 $\pm$ 1.00 & 0.11 $\pm$ 0.01 & 0.41 & 0.16 \\
$\sigma_v$ [km/s] = $\beta$($M_\mathrm{lum}$ [M$_{\odot}$])$^{\gamma}$ & 1.05 $\pm$ 1.10 & 0.10 $\pm$ 0.01 & 0.41 & 0.42 & 1.45 $\pm$ 1.01 & 0.05 $\pm$ 0.01 & 0.40 & 0.21 \\
$\sigma_v$ [km/s] = $\beta$($\Sigma_\mathrm{mol} R_\mathrm{eff}$ [M$_\odot$ pc$^{-1}$])$^{\gamma}$ & 0.80 $\pm$ 1.10 & 0.19 $\pm$ 0.01 & 0.41 & 0.40 & 1.24 $\pm$ 1.01 & 0.11 $\pm$ 0.01 & 0.40 & 0.23 \\
$\alpha$ = $\beta$(M$_\mathrm{lum}$ [M$_\odot$])$^{\gamma}$ & 237.86 $\pm$ 1.28 & -0.43 $\pm$ 0.02 & 0.82 & -0.62 & 512.86 $\pm$ 1.02 & -0.58 $\pm$ 0.01 & 0.75 & -0.69 \\
$L_\mathrm{CO}$ [K km/s pc$^2$] = $\beta$($R_\mathrm{eff}$ [pc])$^{\gamma}$ & 32.47 $\pm$ 2.30 & 2.17 $\pm$ 0.38 & 0.28 & 0.95 & 19.05 $\pm$ 1.02 & 2.26 $\pm$ 0.06 & 0.29 & 0.92 \\
M$_\mathrm{vir}$ [M$_\odot$] = $\beta$($L_\mathrm{CO}$ [K km/s pc$^2$])$^{\gamma}$ & 188.19 $\pm$ 2.12 & 0.75 $\pm$ 0.09 & 0.73 & 0.82 & 544.99 $\pm$ 1.06 & 0.65 $\pm$ 0.02 & 0.73 & 0.73 \\
\hline
\end{tabular}
\caption{Summary of the Larson's laws fitting performed using PCA. $r$ indicates the Pearson's correlation coefficient value \label{T:larslaws}}.
\end{table*}

In this Section we analyze the correlations between cloud integrated properties using the Principal Component Analysis technique (PCA, \citealt{pearson1901}) applied to the bivariate relationships between cloud properties (see Appendix~\ref{A:PCA_details} for more details).

\subsection{Scaling relations from the COHRS dataset}
\label{SS:scaling_cohrs}

\begin{figure*}
\centering
\includegraphics[width=0.75\textwidth]{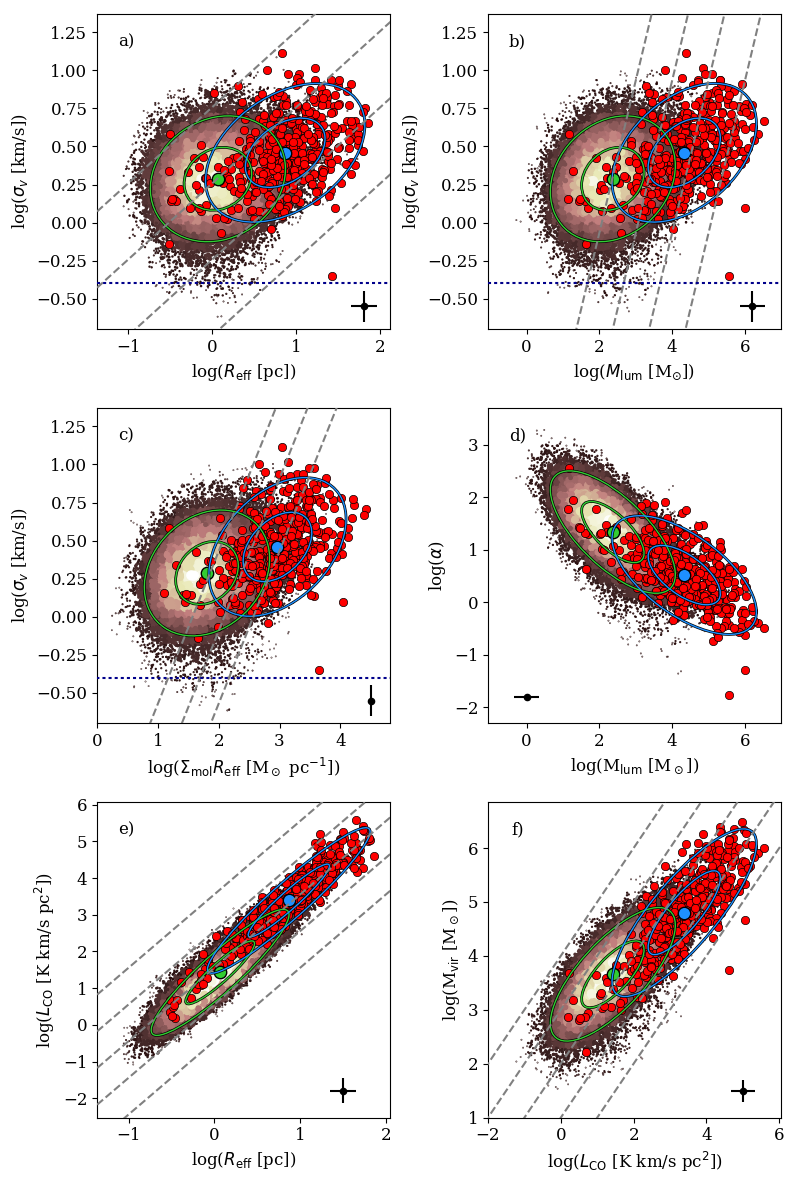}
\caption{COHRS cloud scaling relations. The scaling relations are shown as bi-dimensional histograms, where the colors get brighter proportional to the number of data. The fiducial data are highlighted with red symbols. Green and cyan ellipses represent the PCA of the full and fiducial samples, respectively. Green and cyan points within the inner ellipses show the mean of the properties for full and fiducial sample, respectively.  The ellipses contain $\sim68$\% and $\sim 95\%$ of the data. The horizontal, dotted line marks the spectral resolution of the COHRS data. The straight, dashed, gray lines indicate (from top to bottom, left to right): ($a$) virialized clouds ($\alpha=1$) with $\Sigma_\mathrm{mol}=10^4, 10^3, 10^2, 10$\,$M_\odot$\,pc$^{-2}$; ($b$) virialized clouds ($\alpha=1$) with $R_{\rm eff}=0.1,1,10,100$\,pc for a constant $\alpha_{\rm CO}=8.7$\,$M_{\odot}$/(K\,km~s$^{-1}$\,pc$^2$); ($c$)
$\alpha=10,3,1$; ($e$) $\Sigma_\mathrm{mol}=10^4, 10^3, 10^2, 10, 1$\,$M_\odot$\,pc$^{-2}$; ($f$) $\alpha_{\rm CO} = 10^4, 10^3, 10^2, 10, 1$\,$M_\odot$/(K\,km~s$^{-1}$\,pc$^2$)$^{-1}$ assuming virialized clouds. In the bottom of each panel the typical error bars are given.}
\label{F:cohrs_larslaws}
\end{figure*}

We investigate the various scaling relations within the COHRS sample, in Fig.~\ref{F:cohrs_larslaws}, where the results of our PCA analysis are visualized through ellipses. We plot $1\sigma$ and $2\sigma$ confidence ellipses, which contain approximately 68\% and 95\% of the data points respectively, and we consider the full COHRS cloud sample and the fiducial sample separately. Dashed lines in Fig.~\ref{F:cohrs_larslaws} indicate loci where a given set of parameters are constant around a certain estimate, as would be expected from the Larson's relations.

Fig.~\ref{F:cohrs_larslaws}$a$ shows the relationship between cloud size and linewidth, which are the only two independent properties we can measure for the clouds. For the fiducial sample we obtain a formulation of the size linewidth relation close to the original work of \citet{larson81}, but the two quantities are only moderately correlated. For the full sample the two quantities are very weakly correlated. As observed in Section~\ref{SS:properties}, the coarse spectral resolution of our dataset and the optically thick CO tracer used by COHRS may play a role in rendering the relationship between $\sigma_{\rm v}$ and $R_{\rm eff}$ shallower than predicted by equation~\ref{E:lars1}.
Nevertheless, the relationship between $\sigma_{\rm v}$ and $R_{\rm eff}$ shows a large amount of scatter from both full and fiducial catalogs, which suggests that the clouds in our catalog cannot be described as simple virialized objects with a single value of molecular gas mass surface density. A similar conclusion can be drawn by the second Larson's law relation which connects $\sigma_{\rm v}$ and cloud mass from CO luminosity.

The scatter in the first Larson's relation, as well as the large velocity dispersions measured for the clouds, could be driven by surface density variations (\citealt{heyer2009}). In that case, $\sigma_{\rm v}$ and $\Sigma_{\rm mol}R_{\rm eff}$ should show a larger degree of correlation than other relations that involve the velocity dispersion \citep[e.g.]{miville_deschenes2017}. For our catalog, however, this is not the case (see Fig.~\ref{F:cohrs_larslaws}$c$), but we also measure a much shallower correlation than the one imposed by self-gravitation, as shown by the lines of constant $\alpha$ in Fig.~\ref{F:cohrs_larslaws}$c$. The clouds in the fiducial catalog are globally closer to the virial equilibrium $\alpha\sim1-2$, but the bulk of objects in the full sample seem to have $\alpha\gg1$ (as observed in Section~\ref{S:props}). The clouds in our fiducial sample are generally the most massive ones in the catalog, so it is not surprising that those are closer to virialization, given the covariance between $\alpha$ and $M_{\rm lum}$ (Fig.~\ref{F:cohrs_larslaws}$d$).

Plotting the cloud luminosities from CO against their effective radii (in Fig.~\ref{F:cohrs_larslaws}$e$) provides an additional diagnostic for the surface density of the clouds. Both measurements from fiducial and full samples appear tightly clustered along the $\Sigma_{\rm mol}=100$\,M$_{\odot}$\,pc$^2$, however the average for the fiducial catalog clouds is slightly above this line, while the average from the full catalog is slightly below. The two quantities are covariant and result strongly correlated. 

The virial parameter and cloud mass surface density that somehow shape the appearance of the Larson's relations depend upon $M_{\rm mol}$ which in turn depends on the global value of the CO-to-H$_2$ conversion factor, $\alpha_{\rm CO}$. In Section~\ref{S:r31} we estimated an $\alpha_{\rm ^{12}CO(3-2)}=8.7$\,M$_{\odot}$\,(K\,km~s$^{-1}$\,pc$^2$)$^{-1}$ by correcting the standard Galactic $\alpha_{\rm ^{12}CO(1-0)}$ by the $^{12}$CO(3-2)/$^{12}$CO(1-0) ratio, $R_{31}$. $R_{31}$ looks highly variable across our survey area and the $\alpha_{\rm ^{12}CO(3-2)}$ value we calculated is an approximation spanning both bright and faint emission. The $M_{\rm vir}-L_{\rm CO}$ diagram can be a useful diagnostic to test both cloud dynamical state and variations in $\alpha_{\rm ^{12}CO(3-2)}$. The correlation between these two quantities appears significant considering that both fiducial and full sample register quite high values of Pearson's correlation coefficients ($r\sim0.8$).
If the clouds are virialized, the relationship between $M_{\rm vir}$ and $L_{\rm CO}$ appears to be clustered across the $\alpha_{\rm ^{12}CO(3-2)}=10$\,M$_{\odot}$\,(K\,km~s$^{-1}$\,pc$^2$)$^{-1}$ constant line, which is close the value of $\alpha_{\rm ^{12}CO(3-2)}$ we estimated independently.  This set of observations argues that the clouds being near virial equilibrium is consistent with the values of $\alpha_{\rm ^{12}CO(3-2)}$ that we argued for previously.  This relationship is clearest for the massive clouds in the fiducial sample. Taking the full sample as virialized would imply the global average $\alpha_{\rm ^{12}CO(3-2)}$ is much closer to $100$\,M$_{\odot}$\,(K\,km~s$^{-1}$\,pc$^2$)$^{-1}$ and the slope of the correlation is shallower than linear ($\gamma\sim0.65$). Given the small scale of these objects, we are inclined to believe these small objects are unbound molecular clouds rather than our $\alpha_{\rm ^{12}CO(3-2)}$ being grossly inappropriate \citep{heyer01}.

The scaling relations we observe for COHRS clouds all have a significant amount of scatter ($\sim$0.5\,dex) and are shallower than typically measured in the inner Galaxy despite our significantly larger sample sizes. This is particularly true if we consider the full catalog instead of the fiducial sub-sample. Part of the reason might be attributed to the coarse spectral resolution of our dataset and the optically thick tracer we use, but we can also interpret this as the fact that we are genuinely sensitive to a cloud population that show large linewidths (with respect to their sizes) and that are truly in a gravitationally unbound state.

\subsection{COHRS cloud scaling relations in the context of other surveys}
\label{S:surveys}
\begin{figure*}
\centering
\includegraphics[width=1\textwidth]{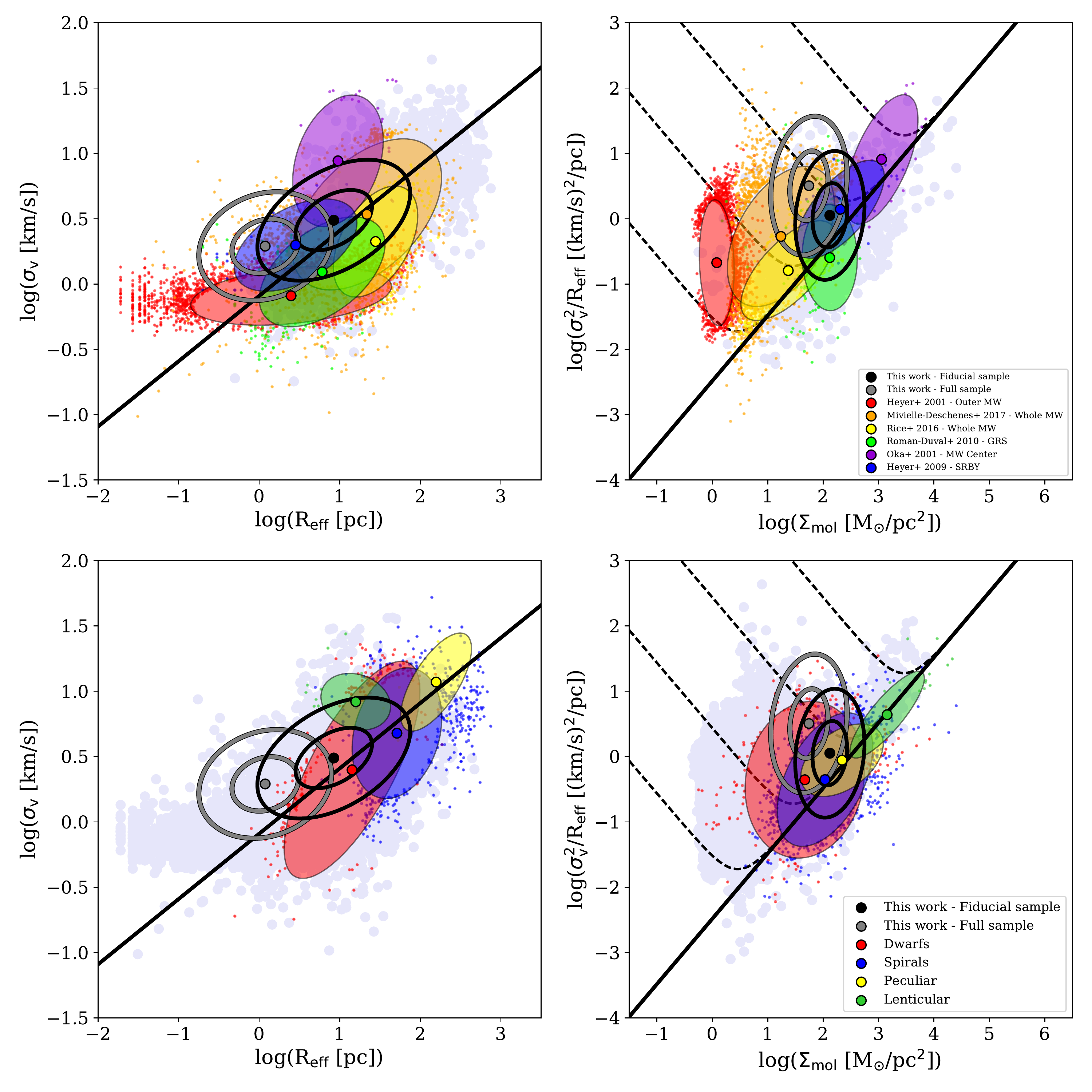}
\caption{Comparison of scaling relationships between cloud properties from different catalogs. The coloured ellipses show the density distribution of the data approximated with PCA and contain $\sim 95\%$ of the data. For the COHRS catalog, the inner ellipse contains $\sim68$\% of the data. The central dots show the mean of the cloud property distributions. Small colored markers illustrate data outside the 95\% confidence ellipses, except for the COHRS clouds. The upper panels include Milky Way cloud catalogs as PCA ellipses, while gray markers indicates the cloud properties from the collection of nearby galaxy catalogs. The opposite applies for the lower panels. Galaxy groups and references are summarized in Tables~\ref{T:surveys_props} and \ref{T:surveys_clouds}. In the right panels, the black solid line shows equation~\ref{E:sigv_lars} for $\Sigma_{\rm mol}=200$\,M$_{\odot}$\,pc$^{-2}$. In the left panels, the black solid line represents equation~\ref{E:sigv_press} with $P_e=0$. Dashed black lines show equation~\ref{E:sigv_press} where the ambient medium pressure is included, from bottom to top: $P_e/k_B=10^2,10^4,10^6,10^8$\,K\,cm$^{-3}$.}
\label{F:cohrs_lcomp}
\end{figure*}

In this section we have gathered data from a variety of CO surveys in order to compare the scaling relations from the clouds in the COHRS sample to those derived from other Milky Way regions and nearby galaxies, as listed in Table~\ref{T:surveys_props}\footnote{Some of this data was obtained via the Cloud Archive for MEtadata, Library and Online Toolkit, CAMELOT \citep[][\url{http://camelot-project.herokuapp.com/}]{camelot2016}}. All of these studies define the identified objects as ``molecular clouds'' (or depending on their sizes and masses, to ``clumps'' or ``Giant Molecular Clouds'') and they all use CO transition or isotopogues for the calculation of cloud properties. The dataset parameters from where those clouds have been extracted are illustrated in Table~\ref{T:surveys_props}. Average properties of the cloud ensembles are collected in Table~\ref{T:surveys_clouds}. Figure \ref{F:cohrs_lcomp} illustrates the differences between cloud samples drawn from various galaxies in scaling relationship diagrams. As per Section~\ref{S:scaling_relations}, we use PCA to describe the relationship between the cloud properties in the different surveys and to illustrate where these objects are found in the various parameter spaces.

\begin{table*}
\begin{tabular}{cccccc}
\hline
Region & Tracer & $\theta_\mathrm{FWHM}$ & $\Delta v$ & Reference \\
\hline
&&& km~s$^{-1}$ & & \\
(1) & (2) & (3) & (4) & (5) \\
\hline
\multicolumn{5}{c}{Milky Way} \\
\hline
MW - 1$^{\rm st}$ quadrant & $^{12}$CO(3-2) & $16.6''$ & 1 & Dempsey et al. 2013 \\
MW Center & $^{12}$CO(1-0) & $34''$ & 0.65 & Oka et al. 1998 \\
Outer MW & $^{12}$CO(1-0) & $50''$ & 0.98 & Heyer et al. 1998 \\
MW - 1$^{\rm st}$ quadrant & $^{13}$CO(1-0) & $45''$ & 1 & Sanders et al. 1986 \\
MW - 1$^{\rm st}$ quadrant & $^{12}$CO/$^{13}$CO(1-0) & $46''$ & 0.212 & Jackson et al. 2006 \\
Whole MW & $^{12}$CO(1-0) & 0.125$^{\circ}$/0.25$^{\circ}$ & 0.65/1.3 & Dame et al. 2001 \\
\hline
\multicolumn{5}{c}{Dwarfs} \\
\hline
Nearby Dwarfs & $^{12}$CO(1-0)/(2-1) & 37 pc & 1.8 & Bolatto et al. 2008 \\
LMC & $^{12}$CO(1-0) & 8 pc & 0.53 & Hughes et al. 2010 \\
NGC300 & $^{12}$CO(2-1) & 37 pc & 1.056 & Faesi et al. 2016 \\
NGC6822 & $^{12}$CO(2-1) & 2 pc & 5 & Schruba et al. 2017 \\
\hline
\multicolumn{5}{c}{Spirals} \\
\hline
M64 & $^{13}$CO(1-0) & 75 pc & 4.25 & Rosolowsky \& Blitz 2005 \\
M33 & $^{12}$CO(1-0) & 48 pc & 2.6 & Gratier et al. 2010 \\
Nearby spirals & $^{12}$CO(1-0) & <78 pc & 5.8 & Donovan-Meyer et al. 2013 \\
M51 & $^{12}$CO(1-0) & 40 pc & 5 & Colombo et al. 2014 \\
M100 & $^{12}$CO(1-0) & 216 pc & 5 & Pan et al. 2017 \\
NGC1068 & $^{13}$CO(1-0) & 98 pc & 1.5 & Tosaki et al. 2017 \\
NGC253 & $^{12}$CO(1-0) & 37 pc & 5 & Leroy et al. 2015 \\
M83 & $^{12}$CO(1-0) & 22.8 pc & 2.57 & Freeman et al. 2017 \\
\hline
\multicolumn{5}{c}{Peculiar} \\
\hline
Antennae & $^{12}$CO(2-1) & 160 pc & 4.9 & Wei et al. 2012 \\
\hline
\multicolumn{5}{c}{Lenticular} \\
\hline
NGC4526 & $^{12}$CO(2-1) & 18 pc & 10 & Utomo et al. 2015 \\
\hline
\end{tabular}
\caption{Summary of the observation from which the clouds in Section~\ref{S:surveys} comparison have been identified: (1), Galactic region or nearby galaxy related to the observation; (2), CO isotopologue and transition used for the observation; (3) spatial resolution of the observation in arcsecond, degrees, or pc; (4) spectral resolution of the observation in km~s$^{-1}$; (5) reference where the previous data sets have been obtained. Notes: Heyer et al. 2009 use a combination of $^{12}$CO(1-0) data from the University of Massachusetts-Stony Brook Galactic Plane Survey (Sanders et al. 1986) and the $^{13}$CO(1-0) data from the Boston University-FCRAO Galactic Ring Survey (Jackson et al. 2006) to redefine the clouds properties previously identified by Solomon et al. 1987. Data from nearby dwarfs (Bolatto et al. 2008) and nearby spirals (Donovan-Meyer et al. 2013) are obtained from a variety of datasets (see the cited works for further references and details); the values reported in the table are averages from the parameters of the different datasets involved in the paper.}
\label{T:surveys_props}
\end{table*}

The left panels of Fig.~\ref{F:cohrs_lcomp} show the ``size-linewidth'' relation expressed through the effective radius and velocity dispersion of the clouds, for the compilation of Galactic studies on the top, and extragalactic studies on the bottom. 

In Galactic studies (Fig.~\ref{F:cohrs_lcomp} top left), we can see that clouds identified in the Galactic Center (\citealt{oka2001}) and the outer Galaxy (\citealt{heyer2001}) show the most extreme variations when compared to the bulk of the MW (as already suggested from Section.~\ref{SS:mw_distribution_comp}). Our fiducial sample overlaps quite well with the ellipses defined by the clouds from the GRS by \cite{roman_duval2010}  (in green) and those from \cite{heyer2009} (in blue), all of which have been identified in a similar region of the Galaxy. Nevertheless, both our clouds and those of \cite{heyer2009} show slightly larger velocity dispersions than the GRS for a same effective radii. This behaviour can potentially be attributed to a difference in tracer, given that both our sample and that of \cite{heyer2009} clouds have been segmented from $^{12}$CO emission, whilst \citet{roman_duval2010} objects are drawn from $^{13}$CO data. Indeed, $^{12}$CO can suffer from high-optical depth effects which can artificially broaden the line emission, and can also suffer from severe blending due to their higher abundance and ability to trace lower density regimes \citep{hughes13b} - effects that are reduced for the less abundant $^{13}$CO. Another cause of the mismatch could be attributed to the lower spectral resolution of both our dataset and that of \cite{heyer2009}, compared to the GRS data, unabling us from detecting lower velocity dispersion objects. Finally, the segmentation method could also play a role, but the effects of the different decomposition methods are more difficult to quantify. Nevertheless, we note that the \cite{heyer2009} and \cite{roman_duval2010} catalogs are constructed by watershed segmentation methods which tend to decompose objects with sizes closer to the resolution element \citep[][see also Table~\ref{T:surveys_clouds}]{hughes13b}. We, instead, consider the full dynamical range of the survey. Therefore it is not surprising that our sampling in effective radius is larger, especially for the full catalog.
We can also see that the clouds identified in the Dame survey data by \citet{rice2016} and \cite{miville_deschenes2017}, which span the entire Galactic plane, sample larger effective radii that the other samples, but this may be a consequence of the coarser resolution of the Dame survey. Interestingly, the catalog construction techniques developed by the authors of those two works have been designed to overcome the data limitations\footnote{The two catalogs have been built following two different philosophies: \cite{rice2016} catalog collects only Milky Way GMCs, while \cite{miville_deschenes2017} work aims to assemble all of the flux into discrete structures of any mass, see Table~\ref{T:surveys_clouds}.}, but they produce quite dissimilar results.

Looking at the extragalactic works (Fig.~\ref{F:cohrs_lcomp} bottom left panel), clouds from the different galaxy groups occupy different parts of the size-linewidth plane. Objects from our fiducial sample span sizes that range from the smaller clouds in dwarf galaxies to average clouds in spirals. Velocity dispersions between those sub-samples are also comparable, however clouds in our fiducial and full catalogs show larger $\sigma_{\rm v}$  values with respect to similar sized objects in dwarf galaxies.
As for the Milky Way catalogs, this can be simply due to differences in spatial and/or velocity resolution \citep{hughes13b}.

The slope of the relationship between $\sigma_{\rm v}$ and $R_{\rm eff}$ calculated through PCA of our fiducial sample, the GRS catalog and full Milky Way catalog from \cite{miville_deschenes2017} are quite comparable ($\sim0.3-0.4$), while the slope for the outer Galaxy clouds is shallower ($\sim0.05$), and the Galactic centre clouds and the full Galaxy cloud catalog of \cite{rice2016} are steeper ($\sim0.7$).
The slope of the relationships from the various galaxy groups, deducible from the PCA ellipse orientation, are also quite different between each other: we obtain $\sim0.65$ for  ``Spirals'', $\sim0.8$ for ``Peculiar'', and $\sim0.1$ for ``Dwarfs'' and ``Lenticular'' types. Nevertheless, these orientations could also be set by the resolution biases as suggested by \cite{hughes13b}. Note, also, that PCA results can be different from the fit between $\sigma_{\rm v}$ and $R_{\rm eff}$ performed on the various papers where the data are taken from, due to different fitting techniques, sub-sample considered, etc. However, in a uniform analysis of the data, drawn without homogenization from different surveys, shows significant variation between the different catalogs. This highlights the fact that these scaling relations are rather sensitive to the segmentation method used as well as the specifications of each dataset, and should therefore be used with caution.\\

So far we have considered clouds as self-contained objects, however, the Galactic environment could potentially shape the properties of the clouds in different ways as we noticed in Section~\ref{S:mw_distribution}. This is particularly evident in nearby spirals (\citealt{colombo14a}, \citealt{leroy2017}, \citealt{sun2018}). Comparing the Milky Way catalogs on the $\sigma_{\rm v}^2/R_{\rm eff}-\Sigma_{\rm mol}$ plane provides a different interpretation of the cloud dynamical state and their interaction with the environment. Clouds bound simply by their self-gravity would tightly cluster across the solid black line in Fig.~\ref{F:cohrs_lcomp} (upper and lower right panels). However, clouds seem almost always shifted toward higher values of $\sigma_\mathrm{v}^2/R_\mathrm{eff}$ with respect to the line imposed by virial equilibrium. \cite{dobbs11} interpret this fact as evidence that clouds are simply gravitationally unbound entities \citep[even though they could still be confined by ram/thermal pressure, as later suggested by e.g.][]{duarte-cabral2017}. Indeed, \cite[][and references therein]{field2011} consider the case of clouds being confined by an external pressure ($P_e$) and suggest an alternative version of equation~\ref{E:sigv_lars} that takes into account this parameter:

\begin{equation}\label{E:sigv_press}
\sigma_v = \sqrt{\frac{1}{3}\left(\pi\Gamma G \Sigma_\mathrm{mol} + \frac{4P_e}{\Sigma_\mathrm{mol}}\right)R_\mathrm{eff}},
\end{equation} 

\noindent where $\Gamma=0.6$ for clouds with constant density. Constant pressure loci are indicated as dashed lines in Fig.~\ref{F:cohrs_lcomp} (upper and lower right panels). Looking at the upper right panel of Fig.~\ref{F:cohrs_lcomp} it appears that proceeding from the outer Galaxy to the Galactic centre passing from the inner Galaxy, clouds could be bound if subjected to a crescent ambient pressure. The lower right panel of Fig~\ref{F:cohrs_lcomp} shows that most of the clouds in the various groups appear bound by self-gravity; only objects in the ``Dwarfs'' group clearly require ambient pressure confinement to explain their linewidths if they are bound. Clouds have also similar values of molecular gas mass surface densities, on average quite close to $\Sigma_{\rm mol}\sim200$\,M$_{\odot}$\,pc$^{-2}$, except for the ``Lenticular'' type which has cloud surface densities an order of magnitude larger than the other groups. Again, drawing firm conclusions about the intrinsic physics described by these diagrams, is hampered by the various catalog generation techniques and survey designs which need to be taken with caution.\\

To summarize, it is challenging to derive firm conclusions from cloud catalogs built from heterogeneous datasets. Cloud definition and their derived properties can be strongly influenced by the combination of survey designs and cloud identification methods. Nevertheless across the scaling relation diagrams of Milky Way and nearby galaxies a few common features can be discerned. Most of the cloud samples show a sub-linear scaling between their velocity dispersions and effective radii which tend to virialization prescription. Small objects (of typically a pc in size), however, do not follow this description and can be considered simply gravitationally unbound or bound by ambient pressure. For the Milky Way, those clouds appear to have a little contribution to the total molecular gas budget, since mass spectra from COHRS clouds and from other surveys (\citealt{heyer01}, \citealt{roman_duval2010}, \citealt{rice2016}, \cite{heyer_dame2015}) all indicate that most of the molecular material is contained into few large GMC-like entities. The difficulties in the interpretation of the results of this analysis, in term of actual difference in cloud properties, underlines also the needs of having, at least for the Milky Way, a single cloud catalog constructed with a consistent extraction method and from surveys with similar designs.

\begin{table*}
\begin{tabular}{cccccccc}
\hline
Region & N. objects & $R_\mathrm{eff}$ & $\sigma_v$ & $\Sigma_\mathrm{mol}$ & $\sigma_0^2$ & Method & Reference \\
&& pc & km~s$^{-1}$ & M$_\mathrm{\odot}$ pc$^{-2}$ & km$^2$~s$^{-2}$ & & \\
(1) & (2) & (3) & (4) & (5) & (6) & (7) & (8)\\
\hline
\multicolumn{8}{c}{Milky Way} \\
\hline
MW - 1$^{\rm st}$ quadrant & 542 & $ 8.93^{19.19}_{ 4.51}$ & $ 2.92^{ 4.39}_{ 2.17}$ & $129.27^{216.48}_{79.86}$ & $ 1.06^{ 2.36}_{ 0.51}$ & SCIMES & This work - fiducial sample \\
MW - 1$^{\rm st}$ quadrant & 35446 & $ 1.13^{ 2.16}_{ 0.60}$ & $ 1.94^{ 2.66}_{ 1.44}$ & $53.30^{92.15}_{32.10}$ & $ 3.23^{ 7.33}_{ 1.41}$ & SCIMES & This work - full sample \\
MW Center & 165 & $ 9.10^{14.60}_{ 5.70}$ & $ 7.80^{12.70}_{ 5.40}$ & $1171.95^{1922.26}_{690.73}$ & $ 8.12^{17.21}_{ 3.38}$ & Sigma clipping & Oka et al. 2001 \\
Outer MW & 10156 & $ 2.69^{ 7.39}_{ 1.04}$ & $ 0.79^{ 0.94}_{ 0.68}$ & $ 1.14^{ 1.45}_{ 0.94}$ & $ 0.23^{ 0.60}_{ 0.09}$ & Sigma clipping & Heyer et al. 2001 \\
MW - 1$^{\rm st}$ quadrant & 158 & $ 3.20^{ 5.47}_{ 1.63}$ & $ 2.00^{ 2.60}_{ 1.60}$ & $189.54^{372.64}_{109.29}$ & $ 1.34^{ 2.48}_{ 0.73}$ & Sigma clipping & Heyer et al. 2009 \\
MW - 1$^{\rm st}$ quadrant & 749 & $ 7.40^{11.90}_{ 3.40}$ & $ 1.32^{ 1.70}_{ 0.98}$ & $144.00^{184.60}_{103.60}$ & $ 0.23^{ 0.44}_{ 0.14}$ & ClumpFind & Roman-Duval et al. 2010 \\
Whole MW & 8107 & $25.07^{44.70}_{12.81}$ & $ 3.63^{ 5.50}_{ 2.25}$ & $16.50^{42.57}_{ 6.62}$ & $ 0.57^{ 1.19}_{ 0.26}$ & Own & Mivielle-Deschenes et al. 2017 \\
Whole MW & 1038 & $27.07^{41.48}_{17.70}$ & $ 2.08^{ 2.96}_{ 1.47}$ & $21.22^{50.87}_{10.99}$ & $ 0.17^{ 0.29}_{ 0.09}$ & Dendrogram & Rice et al. 2016 \\
\hline
\multicolumn{8}{c}{Dwarfs} \\
\hline
Nearby Dwarfs & 110 & $27.54^{41.69}_{18.73}$ & $ 3.39^{ 4.65}_{ 2.63}$ & $74.64^{132.69}_{37.40}$ & $ 0.40^{ 0.69}_{ 0.23}$ & CPROPS & Bolatto et al. 2008 \\
LMC & 524 & $12.05^{15.80}_{ 8.57}$ & $ 1.40^{ 1.79}_{ 1.07}$ & $60.78^{96.03}_{38.19}$ & $ 0.17^{ 0.28}_{ 0.10}$ & CPROPS & Wong et al. 2011 \\
NGC300 & 45 & $ 4.15^{ 4.58}_{ 3.69}$ & $ 1.83^{ 2.38}_{ 1.49}$ & $686.59^{1303.61}_{352.29}$ & $ 0.82^{ 1.27}_{ 0.61}$ & CPROPS & Faesi et al. 2016 \\
NGC6822 & 111 & $ 2.60^{ 3.30}_{ 2.05}$ & $ 1.15^{ 1.41}_{ 0.80}$ & $ 7.47^{11.93}_{ 3.80}$ & $ 0.45^{ 0.73}_{ 0.27}$ & CPROPStoo & Schruba et al. 2017 \\
\hline
\multicolumn{8}{c}{Spirals} \\
\hline
M64 & 25 & $86.00^{115.00}_{62.00}$ & $ 9.79^{14.89}_{ 7.66}$ & $57.14^{69.89}_{43.20}$ & $ 1.14^{ 2.22}_{ 0.73}$ & ClumpFind & Rosolowsky \& Blitz 2005 \\
M33 & 308 & $37.00^{56.00}_{24.75}$ & $ 8.85^{11.00}_{ 7.20}$ & $31.66^{64.39}_{16.23}$ & $ 2.10^{ 3.57}_{ 1.29}$ & CPROPS & Gratier et al. 2012 \\
Nearby spirals & 96 & $87.90^{108.75}_{61.33}$ & $ 4.77^{ 7.84}_{ 3.79}$ & $188.05^{272.61}_{138.50}$ & $ 0.32^{ 0.54}_{ 0.19}$ & ClumpFind & Donovan-Meyer et al. 2013 \\
M51 & 1506 & $35.00^{55.00}_{32.00}$ & $ 5.90^{ 8.00}_{ 4.20}$ & $139.38^{258.87}_{82.87}$ & $ 0.80^{ 1.58}_{ 0.43}$ & CPROPS & Colombo et al. 2014 \\
NGC253 & 10 & $30.50^{40.75}_{24.25}$ & $ 9.36^{10.53}_{ 7.55}$ & $4760.04^{9018.37}_{2218.98}$ & $ 2.61^{ 3.00}_{ 2.34}$ & CPROPStoo & Leroy et al. 2015 \\
M100 & 165 & $301.70^{403.90}_{229.40}$ & $ 7.30^{ 9.90}_{ 5.20}$ & $32.29^{52.75}_{17.53}$ & $ 0.17^{ 0.36}_{ 0.09}$ & CPROPS & Pan et al. 2017 \\
NGC1068 & 187 & $100.00^{133.00}_{74.50}$ & $ 4.12^{ 5.33}_{ 3.08}$ & $26.12^{59.07}_{15.21}$ & $ 0.18^{ 0.26}_{ 0.11}$ & ClumpFind & Tosaki et al. 2017 \\
M83 & 873 & $46.10^{56.89}_{36.68}$ & $ 3.24^{ 4.61}_{ 2.36}$ & $79.53^{146.65}_{54.30}$ & $ 0.22^{ 0.45}_{ 0.12}$ & CPROPStoo & Freeman et al. 2017 \\
\hline
\multicolumn{8}{c}{Peculiar} \\
\hline
Antennae & 67 & $168.00^{233.50}_{118.00}$ & $11.90^{16.25}_{ 9.30}$ & $224.78^{387.46}_{112.18}$ & $ 0.88^{ 1.31}_{ 0.59}$ & ClumpFind & Wei et al. 2012 \\
\hline
\multicolumn{8}{c}{Lenticular} \\
\hline
NGC4526 & 103 & $16.06^{22.53}_{11.86}$ & $ 8.20^{ 9.14}_{ 7.24}$ & $1237.36^{1740.62}_{907.78}$ & $ 3.95^{ 6.15}_{ 2.58}$ & CPROPS & Utomo et al. 2015 \\
\hline
\end{tabular}
\caption{Summary of the cloud properties used for the comparison in Section~\ref{S:surveys}: (1), Galactic region or nearby galaxy related to the observation; (2), number of objects identified; (3), effective radius; (4), velocity dispersion; (5), molecular gas surface density; (6), scaling parameter; (7), method used to identify clouds: GaussClump (Stutzki \& Guesten 1990), ClumpFind (Williams et al. 1994), CPROPS (Rosolowsky \& Leroy 2006), dendrogram (Rosolowsky et al. 2008), SCIMES (Colombo et al. 2015), CPROPStoo (Leroy et al. 2015); (8), catalog reference. All properties are presented as median, 25$^\mathrm{th}$ and 75$^\mathrm{th}$ percentiles of their distributions.}
\label{T:surveys_clouds}
\end{table*}

\section{Summary and Outlook}
\label{S:summary} 

We presented an analysis of the molecular gas properties imaged by the JCMT $^{12}$CO(3-2) High Resolution Survey through the cloud identification. We applied the SCIMES algorithm to obtain a catalog of integrated properties from more than 85\,000 clouds in the first Galactic quadrant. We corrected cloud properties for instrumental biases by applying the techniques described by \cite{rl06}. Our main results can be summarized as follows.

\begin{enumerate}

\item By comparing the the University of Massachusetts Stony Brook Survey $^{12}$CO(1-0) data with the smoothed and reprojected COHRS data we obtain an average $^{12}$CO(3-2) over $^{12}$CO(1-0) flux ratio, $R_{31}=0.5$. This translates into a $^{12}$CO(3-2)-to-H$_2$ conversion factor: $X_{^{12}\mathrm{CO(3-2)}} = 4\times10^{20}\,{\rm (cm^2\,K\,km\,s^{-1}})^{-1}$ or $\alpha_{^{12}\mathrm{CO(3-2)}} = 8.7\,{\rm M_{\odot}\,( K\,km\,s^{-1}\,pc^{2})^{-1}}$. 

\item We calculated the distance to the clouds using the distance measurements from \cite{zetterlund2018} which applied the methods of \cite{ellsworth_bowers2013} starting from the Bolocam Galactic Plane Survey. This establishes a grid of 2202 pixels in the position-position-velocity coordinate system of COHRS dataset. In this way we obtained 406 clouds with a distance pixel within them, 41\,896 clouds where a parental structure of the dendrogram have a distance point within it, and 42\,718 isolated clouds for which we choose the distance pixel closest to the outer boarder of the object.

\item We separated the analysis between two sub-samples: $\sim35500$ well resolved objects, defined ``full'' sample, and a ``fiducial'' sample constituted by 542 clouds with well defined distances. The latter shows properties quite similar to Galactic Ring Survey catalog of \cite{roman_duval2010}. Nevertheless, velocity dispersion and the properties that depend on it (as virial mass and virial parameter) are shifted towards higher values with respect to GRS.  We attribute this to: 1) the coarse velocity resolution of the COHRS data relative to the structures that COHRS can resolve spatially and 2) to the different tracers used to image the molecular medium by COHRS and GRS.

\item We analyzed the cumulative distributions of mass derived from CO luminosity and effective radius in detail. We found a $\gamma\sim-1.75$ and $\gamma\sim-2.80$ for the spectral index of mass and size spectra, respectively. Both distributions show high-end truncations at $M_{\rm lum}\sim3\times 10^6~M_{\odot}$ and $R_{\rm eff}\sim70$\,pc, respectively. Mass and size spectra show subtle differences when calculated within different Galactocentric annuli.

\item COHRS clouds show CO integrated intensity and velocity dispersion values intermediate between the Galactic centre and the outer Galaxy. In particular we observed that clouds in the fiducial sample have masses similar to Galactic centre clouds.

\item Approximately 35\% of the clouds and the cataloged flux look embedded into spiral arms, by considering a model that foresees 4 main spiral arms and the Local spurs, where arms have a fixed width of 600\,pc.

\item We used principal component analysis to study the scaling relations from ``full'' and ``fiducial'' samples separately. We found mainly moderate correlations (Pearson $r\sim0.5$) between most of the properties only when the fiducial sample clouds are considered. Only the $L_{\rm CO}-R_{\rm eff}$ and $M_{\rm vir}-L_{\rm CO}$ show high levels of correlation ($r>0.8$ from both subsamples). However, these quantities are expected to be intrinsically correlated given the observables that contribute to them. For the fiducial sample, we observed a size-linewidth relation shallower (slope$\sim0.3$) and weaker (Pearson $r\sim0.4$) when compared to previous studies in the same Galactic region. We attribute this disparity to the different tracers (which have different optical depths), and the different velocity resolutions of the surveys.

\item By collecting catalogs from various, heterogeneous Galactic and nearby galaxy CO surveys we found that, on average, the scaling between cloud effective radii and velocity dispersions largely scatter around a power law of slope 0.5.
Most of the cloud linewidths would be consistent with a state of virialization if the clouds are bound by an over-pressurized environment.
We warn that physical interpretations can be strongly shaped by methodological choices in term of survey designs, cloud identification methods, and the CO isotopologues observed.

\end{enumerate}

In this work we have studied the properties of the clouds identified in the COHRS data through the approach of using integrated cloud properties. These properties reduce all the available information from the mapping of the clouds down to single numbers. These measurements can also be achieved when the objects are point sources, i.e. a few beams apart, as in the case of nearby galaxy observations. Even with the advent of ALMA the smallest scale resolvable outside the Local Group remains the size of a GMC ($\sim20$\,pc). However, in recent years, the Galactic plane has been explored by a wealth of high resolution spectroscopic large scale, blind surveys able to image down to the size of clumps (e.g., COHRS, CHIMPS, SEDIGISM, FUGIN; and the ongoing FQS, \citealt{benedettini2017}, and OGHReS, PI C. K\"onig). The application of SCIMES on these surveys will allow us to obtain thousands of clouds with a well resolved internal structure and defined outer edges. This will also give the opportunity to study the ``resolved'' properties of the clouds as probability distribution functions (PDFs), turbulence, morphology (elongation versus roundness), kinematics (through moment maps), clump formation efficiency (CFE, \citealt{eden2013}) and star formation efficiencies (in combination with sub-mm surveys such as ATLASGAL, \citealt{schuller2009}, \citealt{csengeri2014}, \citealt{urquhart2014}); Hi-GAL, (\citealt{molinari2016}, \citealt{elia2017}); BGPS, \citealt{aguirre2011}, \citealt{ginsburg2013}) in a truly statistical fashion.


\section*{Acknowledgements}
We thank the anonymous referee for carefully reading our manuscript. DC acknowledges support by the Deutsche Forschungsgemeinschaft, DFG through project number SFB956C. DC thanks Axel Wei{\ss}, Karl Menten, Friedrich Wyrowski, and Carsten K\"onig for the useful discussions. ER acknowledges the support of the Natural Sciences and Engineering Research Council of Canada (NSERC), funding reference number RGPIN-2017-03987. ADC acknowledges the support of the UK Science and Technology Facilities Council consolidated grants ST/N000706/1. This research made use of {\sc Astropy}, a community-developed core Python package for Astronomy (\citealt{astropy2013}).
\footnotesize{
\bibliographystyle{mn2e_new}
\bibliography{cold}
}

\appendix

\section{New SCIMES version}
\label{A:new_scimes}
Together with this paper we release a new version of SCIMES (v.0.3.2) on the github page of the project\footnote{\url{https://github.com/Astroua/SCIMES}}, which includes several major upgrades. The new version is approximately 30 times faster than the version previously released online. A few new features have been added. In the final catalog, the code can now retain dendrogram branches and leaves that cannot be uniquely associated with the identified clusters. The affinity matrix scaling parameters (see \citealt{colombo15} for further details) can be restricted to only searching above a specified signal-to-noise ratio (default: SNR$>3$). The ``SNR affinity matrix'' is built by inputing the temperature peak value of each structure considered in the matrix construction divided by a fixed (user-defined) noise estimation. The temperature peak value is collected in the attribute \verb"height" of the \verb"astrodendro.structure.Structure" class. The scaling parameter is now set by the largest gap within the affinity matrices above the given SNR. This increases the code stability and avoids considering large and incoherent objects. The algorithm updates information about the cluster type within the dendrogram: ``T'' for trunks (i.e. dendrogram branches without parent); ``B'' for branches (structures with parent and children); ``L'' for leaves (structures without children). The code now also reports the number of leaves and unique identifiers for the ID of the structure parent and ancestor. Finally, the SCIMES main class also provides mask cubes for the identified clusters, as well as for all leaves and trunks in the dendrogram.

\section{Comparison between assumed and kinematic distances to the clouds}
\label{A:kda}

The kinematic distance method consists of the derivation of galactocentric radius and distance to a clump of gas through the knowledge of its spectroscopic radial velocity and the assumption of a given rotation curve (e.g. \citealt{roman_duval2009}). The galactocentric radius can be uniquely determined through the following equation for a given longitude ($l$) and radial velocity ($v_r$):

\begin{equation}\label{E:r_kda}
r = R_{\odot}\sin(l)\frac{v(r)}{v_r + v_{\odot}\sin(l)}
\end{equation}

where $R_{\odot}$ and $v_{\odot}$ are the galactocentric radius and orbital velocity of the Sun, respectively; $v(r)$ indicates the rotation curve, and $v_r$ is the spectroscopical radial velocity of the cloud. 

In the inner Galaxy ($r<R_{\odot}$) the distance calculated through the kinematic method is not unique. Each galactocentric radius corresponds to two distances
along the line of sight, called \emph{near} and the \emph{far} kinematic distances, which are located on either side of the \emph{tangent point}. The tangent point is the region where the velocity vector of the cloud is aligned with the line-of-sight. The near and far distances can be calculated via: 

\begin{equation}\label{E:d_kda}
d = R_{\odot}\cos(l)\pm\sqrt{r^2 - R_{\odot}^2\sin(l)^2}.
\end{equation}

The duality of the distance calculated within the inner Galaxy is referred to as the \emph{kinematic distance ambiguity} which is challenging to break without additional information.

\begin{figure}
\centering
\includegraphics[width=0.45\textwidth]{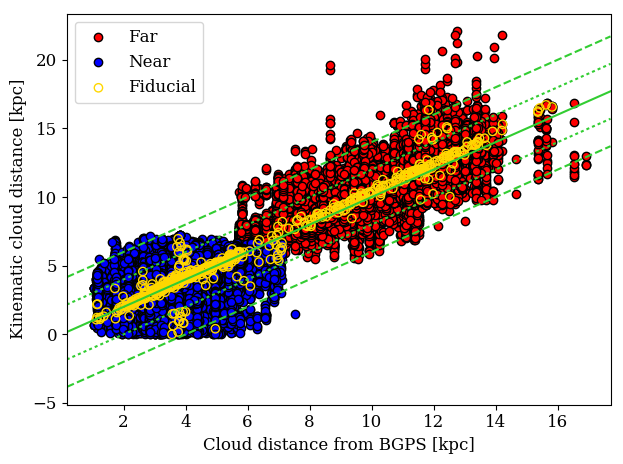}
\caption{Comparison between cloud assumed distance from BGPS sources and calculated kinematic distance. Blue points indicate near cloud distance, while red points the far cloud distances. The fiducial sample is highlighted with yellow circles. The green full line represent unity, while the green dotted and dashed line a factor 2 and 4 scatter, respectively.}
\label{F:cohrs_kda}
\end{figure}

Fig.~\ref{F:cohrs_kda} shows the comparison between the distance we assumed for the clouds via association with BGPS sources (as described in Section~\ref{S:cohrs_distance}) and the kinematic distance calculated to the clouds through equations~\ref{E:r_kda} and \ref{E:d_kda}. The kinematic distance attributed to each cloud is given by the longitude and radial velocity of its centroid position (\texttt{glon\_deg} and \texttt{vlsr\_kms} parameters in the catalog, respectively). As Galactic rotation curve we use the model defined by \cite{brand_blitz1993} with the parameter calculated by \citealt{reid2014}:

\begin{equation}\label{E:vc_kda}
v(r) [\mathrm{km\,s}^{-1}]= (240\pm9) (r/R_{\odot})^{0.00\pm0.02}.
\end{equation}

For a given cloud the near or the far distance is plotted considering the one which is closer in value to the assumed one. Around 60\% of the sample analyzed in the paper can be attributed to the far distance, while the remaining 40\% to near distances. Nevertheless, only 30\% of the sources in the BGPS catalog are at  far distances. Indeed, most of the COHRS clouds at the far distances have  ``broadcasted'' or ``closest'' distance attribution. Indeed, the exact values of the kinematic distances can be quite dissimilar to the assumed ones for the full sample. Nevertheless, the distances assumed for the clouds in the ``fiducial'' sample are almost indistinguishable to the ones derived using the kinematic method, with only few exceptions. Moreover, 60\% of ``fiducial'' clouds is attributed to the near distances, the remaining 40\% to the far distances. 

\section{Principal Component Analysis}
\label{A:PCA_details}

In Section \ref{S:scaling_relations}, we analyze the correlations between cloud integrated properties using the Principal Component Analysis technique (PCA, \citealt{pearson1901}) applied to the bivariate relationships between cloud properties. While PCA is a dimensionality reduction technique, we use it in the place of regression to identify the directions of maximal variance of the data and assume that this direction defines the scaling relationship.  The perpendicular directions define the scatter within the data. This approach is particularly useful for our analysis since the underlying correlations have large intrinsic scatter.

The PCA algorithm constructs the covariance matrix between quantities under analysis, calculating the largest and smallest eigenvectors of this matrix. The orientation of the largest eigenvector with respect to the $x$-axis defines the slope of the relation (i.e., the tangent of the angle between the largest eigenvector and the positive x-axis directions). By construction, the smaller eigenvector is perpendicular to the larger eigenvector. The scatter of the relation is provided by PCA as $\sigma_{\rm min}=2\sqrt{\lambda_{\rm min}}$, where $\lambda_{\rm min}$ is the smaller eigenvalue. Generally, two quantities can be considered as correlated if $\lambda_{\rm maj}\gg\lambda_{\rm min}$, where $\lambda_{\rm maj}$ is the largest eigenvalue. 

To obtain the uncertainty on the relationship slopes we generate 1000 random, synthetic datasets by redistributing the data within the $x$- and $y$-axis  error bars. PCA is applied at each iteration, the slope uncertainty is calculated as standard deviation of the resulting slope distributions. Table~\ref{T:larslaws} summarizes the results from PCA fitting of the scaling relations analyzed here.

\section{Extrapolation and deconvolution effects on cloud properties}
\label{A:props_exdec}

\begin{figure*}
\centering
\includegraphics[width=0.8\textwidth]{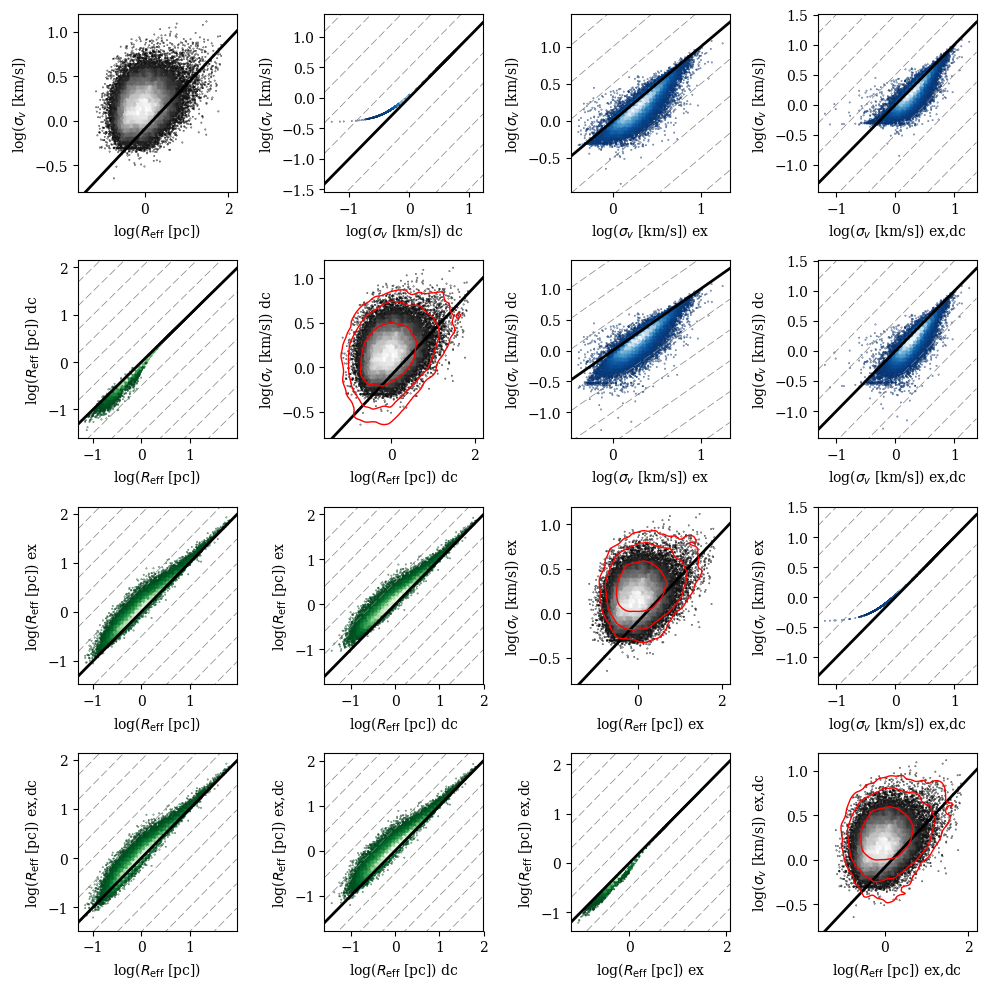}
\caption{Comparison between extrapolated (``ex''), deconvolved (``dc''), extrapolated and deconvolved (``ex\_dc''), and uncorrected $R_{\rm eff}$ and $\sigma_{\rm v}$ (no suffix), and their effects on the size-linewidth relation (along the figure main diagonal). Across the main diagonal the black full line indicates equation~\ref{E:lars1}, while in the panels off diagonal the 1:1 relation. Across the main diagonal uncorrected properties are always indicated with gray markers, while the red markers indicates the corrected properties. Dashed gray lines are spaced of 0.5\,dex.}
\label{F:cohrs_lars1_ex_dc}
\end{figure*}

\begin{figure*}
\centering
\includegraphics[width=0.8\textwidth]{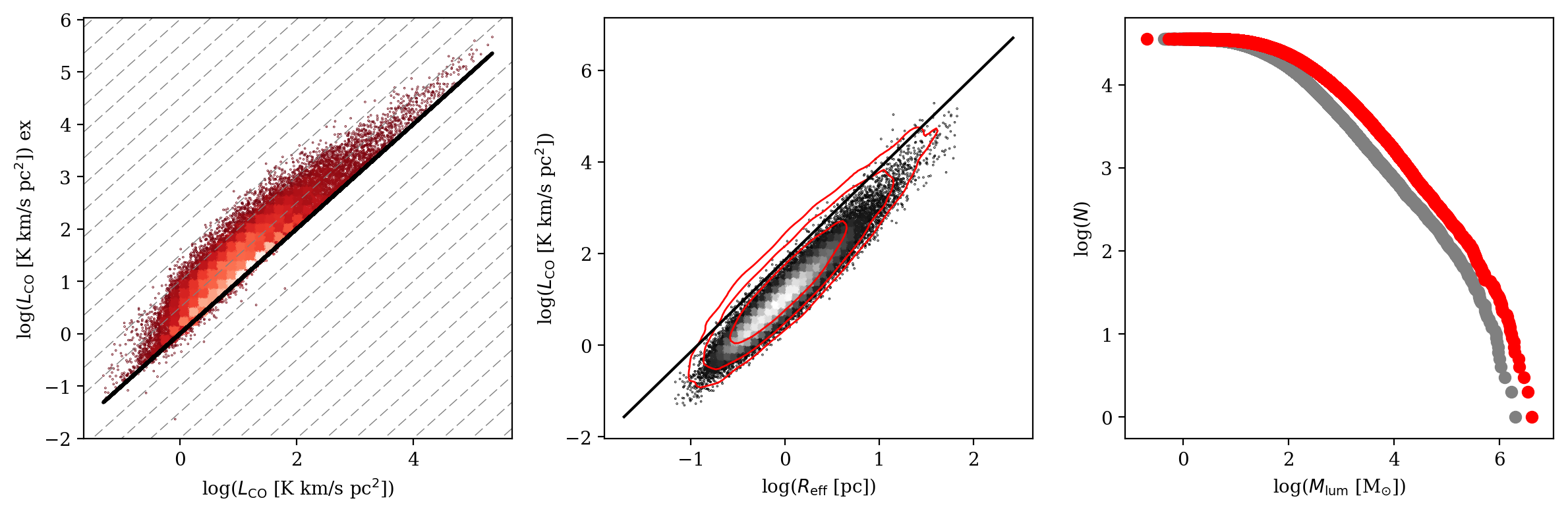}
\caption{Comparison between extrapolated (``ex'') and uncorrected (no suffix) CO luminosity (left). Effect of luminosity extrapolation on the $L_{\rm CO}$-$R_{\rm eff}$ relation (middle) and the cumulative mass spectrum (right). In the left panel the black line represents the 1:1 relation while in the middle one, the locus defined by a constant $\Sigma_{\rm mol}=200$\,M$_{\odot}$\,pc$^{-2}$ and $\alpha_{\rm CO}=8.7$\,M$_{\odot}$\,(K\,km~s$^{-1}$\,pc$^2$)$^{-1}$. In the last two panels, corrected properties (extrapolated for $L_{\rm CO}$, extrapolated and deconvolved for $R_{\rm eff}$) are indicated with red markers, while uncorrected ones with gray markers. Dashed gray lines in the first panel are spaced of 0.5\,dex.}
\label{F:cohrs_lum_ex_dc}
\end{figure*}

In this paper, we have used extrapolation and deconvolution to correct the measured cloud properties. Nevertheless, this paradigm has been mostly used for nearby galaxy studies and only rarely in the Milky Way \citep[e.g.,][]{blitz80}. Here we explore the differences between properties directly calculated by {\sc astrodendro} and those corrected for instrumental biases. 

Fig.~\ref{F:cohrs_lars1_ex_dc} summarizes the differences between the various approaches for calculating the cloud properties in term of $R_{\rm eff}$ and $\sigma_{\rm v}$. Generally, the deconvolution effectively reduces the cloud velocity dispersion starting from both the uncorrected and extrapolated $\sigma_{\rm v}$. Large values of $\sigma_{\rm v}$ are not influenced by this correction. In contrast, the extrapolation almost always shifts the velocity dispersion toward values $\sim0.5$\,dex higher with respect to the uncorrected velocity dispersion. The effect is similar for the cloud size. Nevertheless, extrapolated and deconvolved $R_{\rm eff}$ appears more equally distributed across the 1:1 line with respect to uncorrected and deconvolved only effective radii. The deconvolution alone does not significantly change the form of the size-linewidth relation we measured in the paper, but adds several points at low $\sigma_{\rm v}$. The same is true for the extrapolation and deconvolution combined. 

The extrapolation produces a more significant boost of the properties (up to 1\,dex) when it is applied to the CO flux. However, this boost closes the gap between the emission in the cataloged structure and the total flux obtained by averaging over the entire data cube. 
Figure~\ref{F:cohrs_lum_ex_dc} (left panel) shows that extrapolated cloud $L_{\rm CO}$ can be even 1\,dex higher than uncorrected CO luminosity. This effect is small for clouds with large $L_{\rm CO}$ values in the catalog.  

The instrumental correction on both $L_{\rm CO}$ and $R_{\rm eff}$ provides a relationship between the two quantities that is closer to the theoretical model described by constants $\Sigma_{\rm mol}=200$\,M$_{\odot}$\,pc$^{-2}$ and $\alpha_{\rm CO}=8.7$\,M$_{\odot}$\,(K\,km~s$^{-1}$\,pc$^2$)$^{-1}$ (Fig.~\ref{F:cohrs_lum_ex_dc}, middle). At the same time, the shape of the mass spectrum (Fig.~\ref{F:cohrs_lum_ex_dc}, right) from uncorrected and extrapolated values of $M_{\rm lum}$ is similar. The fit of the mass spectrum from the uncorrected $M_{\rm lum}$ has a spectral index $\gamma=-1.7$, consistent with the one measured from the extrapolated mass reported in the Section~\ref{SS:cum_dists}. Nevertheless, the truncation mass given by the uncorrected property spectra is a factor 2 lower $M_0\sim1.7\times10^6$\,M$_{\odot}$.

In conclusion, we regard these corrections for emission at low significance and for the effects of instrumental resolution as significant.  While the models used for correction cannot completely overcome the limitations of the observations, the combination of corrections adopted and the uncertainties established by the bootstrapping better describe the molecular gas population compared to just using the emission in the cataloged objects alone.

\section{COHRS survey in molecular clouds}
\label{A:full_survey_decomp}
Here we collect the maps of the sub-cubes from the full COHRS data overlaid with the identified cloud mask.

\begin{figure*}
\centering
\includegraphics[width=0.75\textwidth]{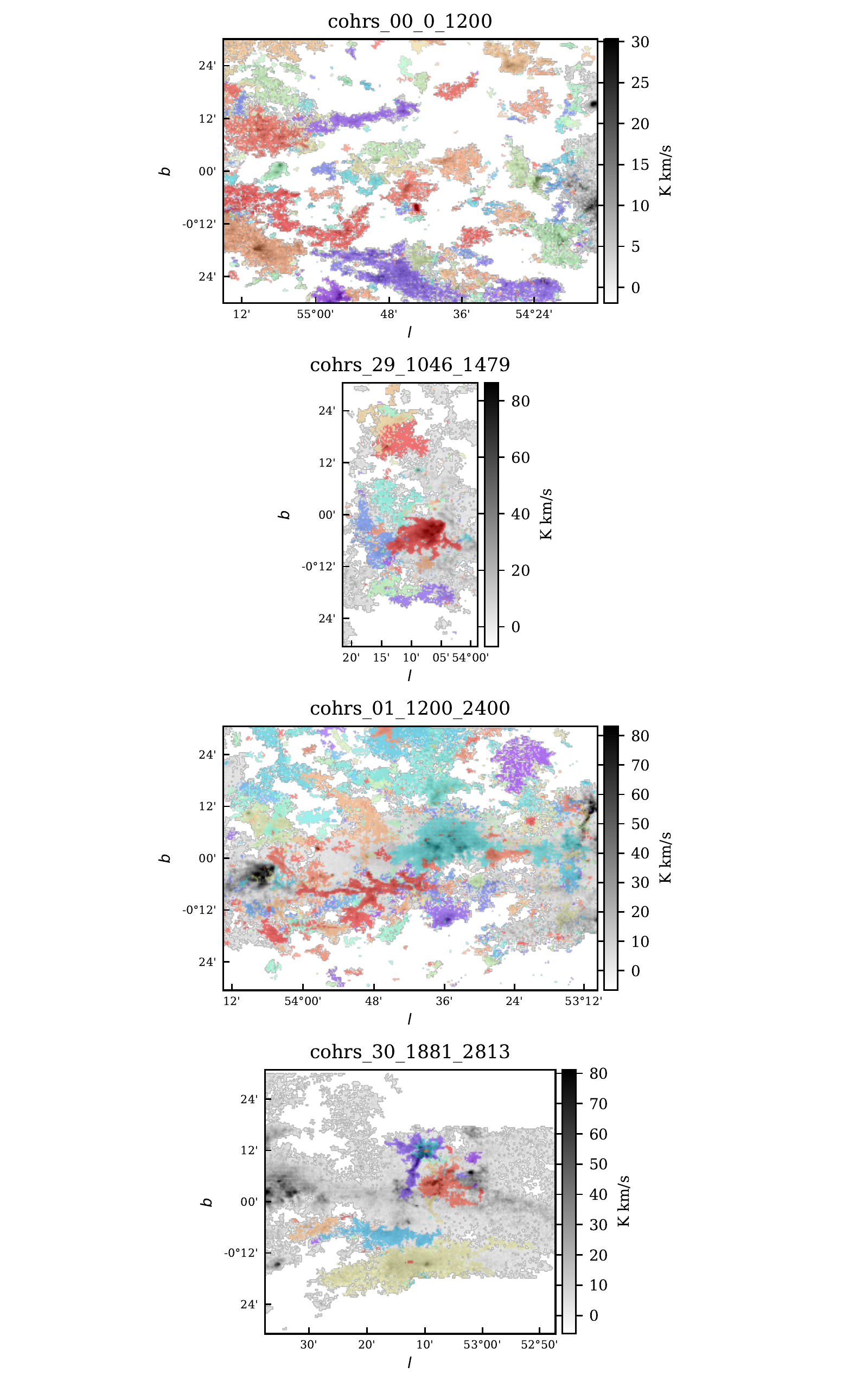}
\caption{Longitude-latitude integrated intensity map of COHRS given sub-cubes masked as explained in Section~\ref{S:scimes_cohrs}. In color the identified clouds are indicated.}
\label{F:cohrs_asgnpp_0}
\end{figure*}

\newpage
\clearpage

\begin{figure*}
\centering
\includegraphics[width=0.75\textwidth]{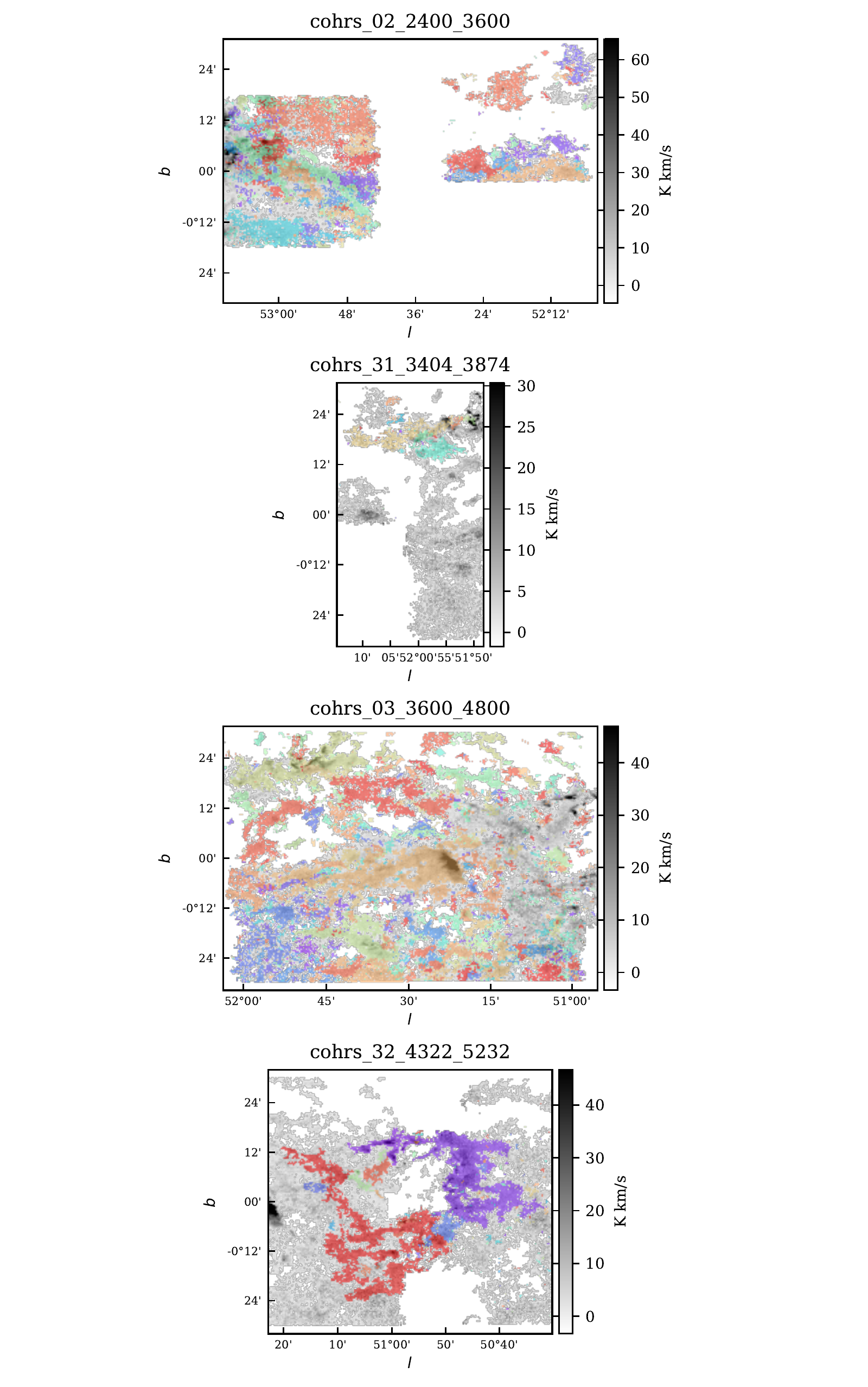}
\caption{Longitude-latitude integrated intensity map of COHRS given sub-cubes masked as explained in Section~\ref{S:scimes_cohrs}. In color the identified clouds are indicated.}
\label{F:cohrs_asgnpp_1}
\end{figure*}

\newpage
\clearpage

\begin{figure*}
\centering
\includegraphics[width=0.75\textwidth]{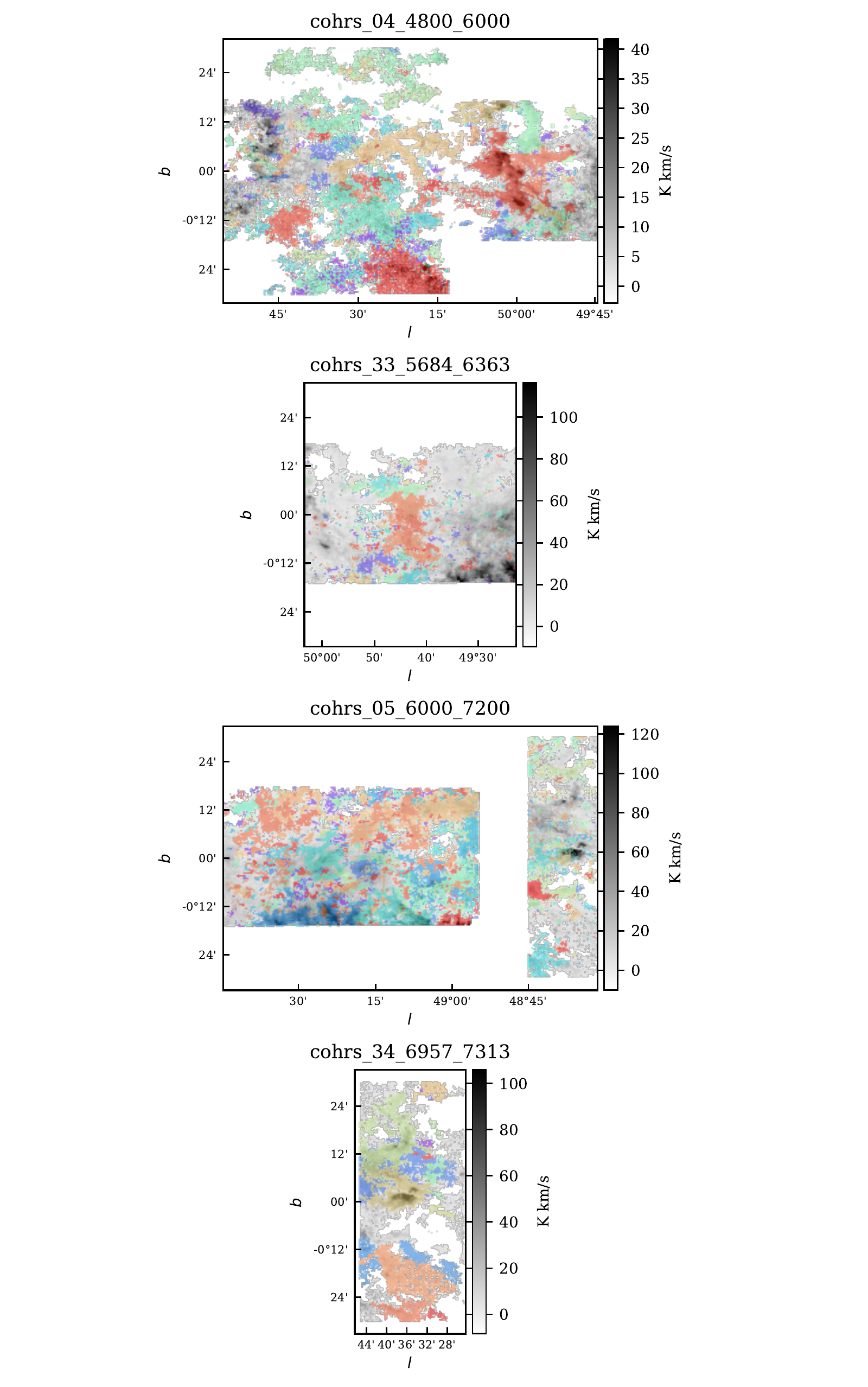}
\caption{Longitude-latitude integrated intensity map of COHRS given sub-cubes masked as explained in Section~\ref{S:scimes_cohrs}. In color the identified clouds are indicated.}
\label{F:cohrs_asgnpp_2}
\end{figure*}

\newpage
\clearpage

\begin{figure*}
\centering
\includegraphics[width=0.75\textwidth]{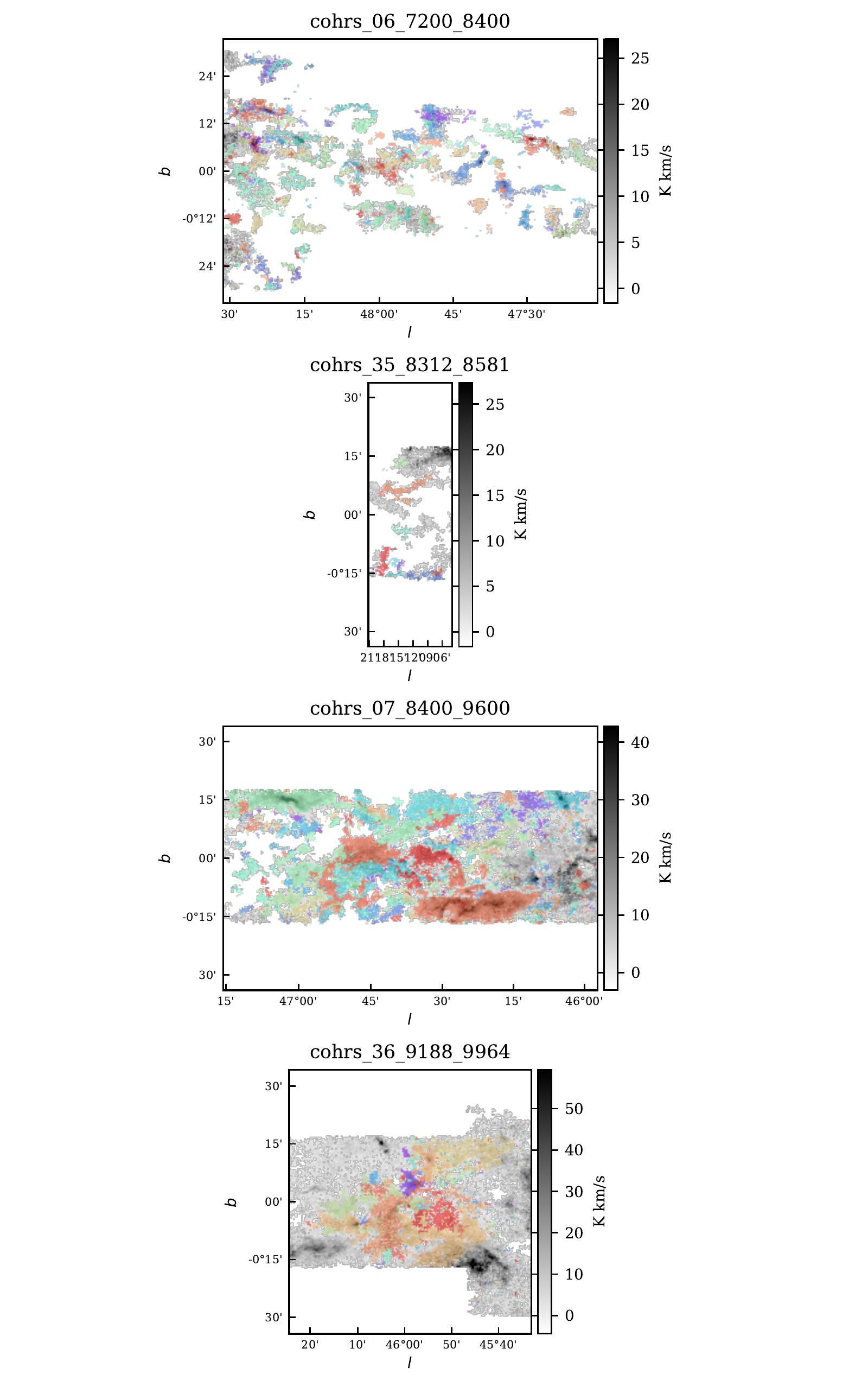}
\caption{Longitude-latitude integrated intensity map of COHRS given sub-cubes masked as explained in Section~\ref{S:scimes_cohrs}. In color the identified clouds are indicated.}
\label{F:cohrs_asgnpp_3}
\end{figure*}

\newpage
\clearpage

\begin{figure*}
\centering
\includegraphics[width=0.75\textwidth]{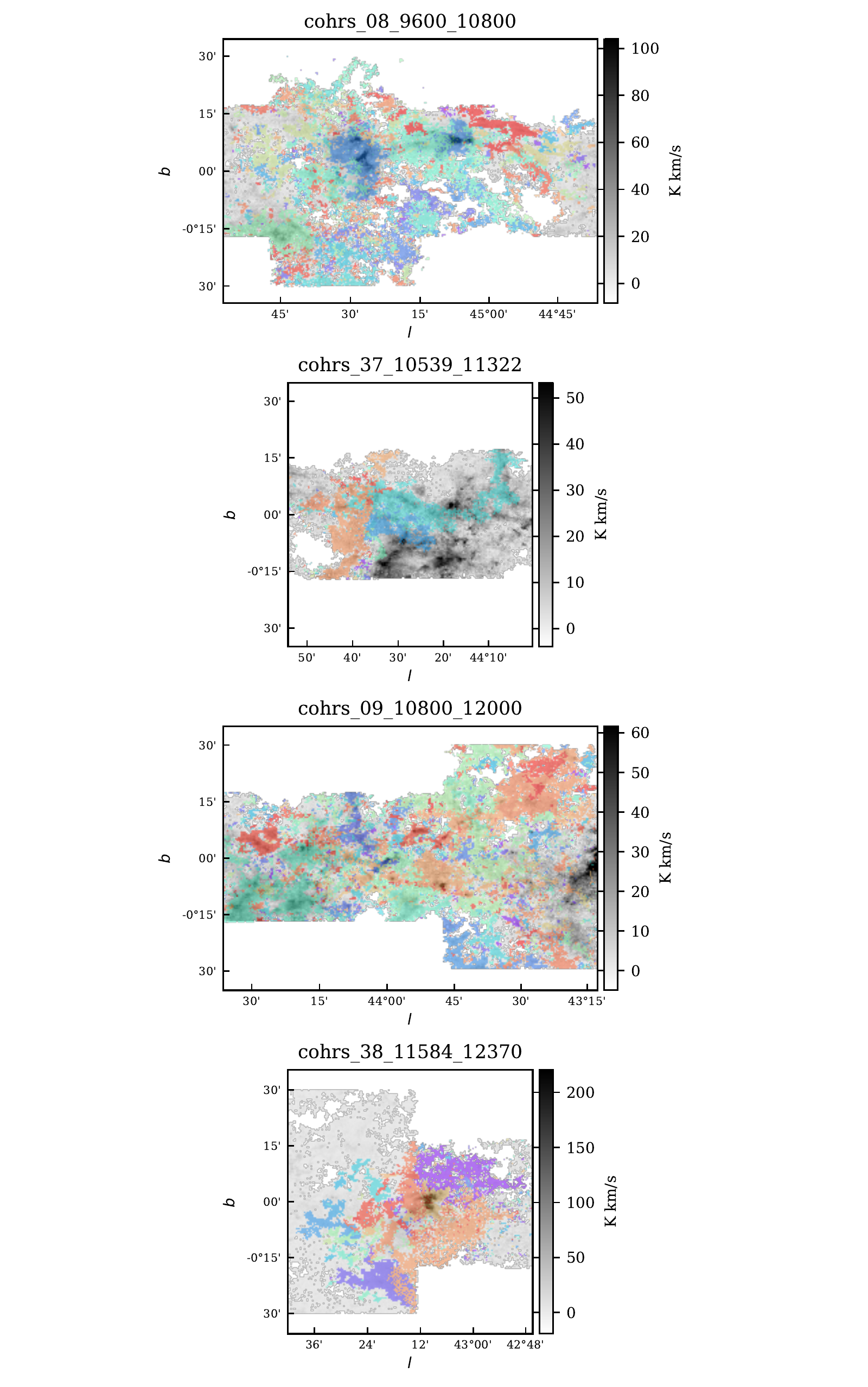}
\caption{Longitude-latitude integrated intensity map of COHRS given sub-cubes masked as explained in Section~\ref{S:scimes_cohrs}. In color the identified clouds are indicated.}
\label{F:cohrs_asgnpp_4}
\end{figure*}

\newpage
\clearpage

\begin{figure*}
\centering
\includegraphics[width=0.75\textwidth]{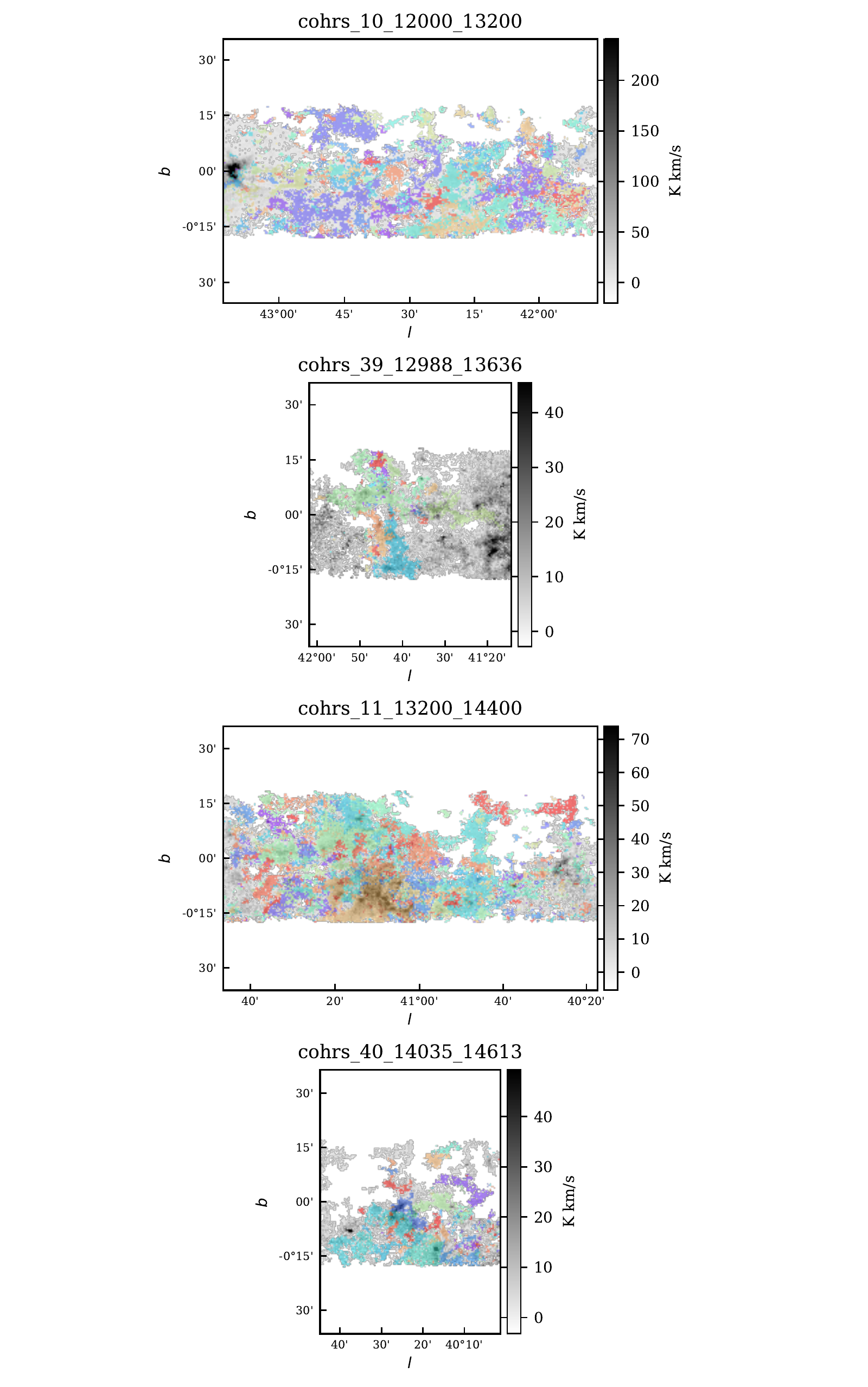}
\caption{Longitude-latitude integrated intensity map of COHRS given sub-cubes masked as explained in Section~\ref{S:scimes_cohrs}. In color the identified clouds are indicated.}
\label{F:cohrs_asgnpp_5}
\end{figure*}

\newpage
\clearpage

\begin{figure*}
\centering
\includegraphics[width=0.75\textwidth]{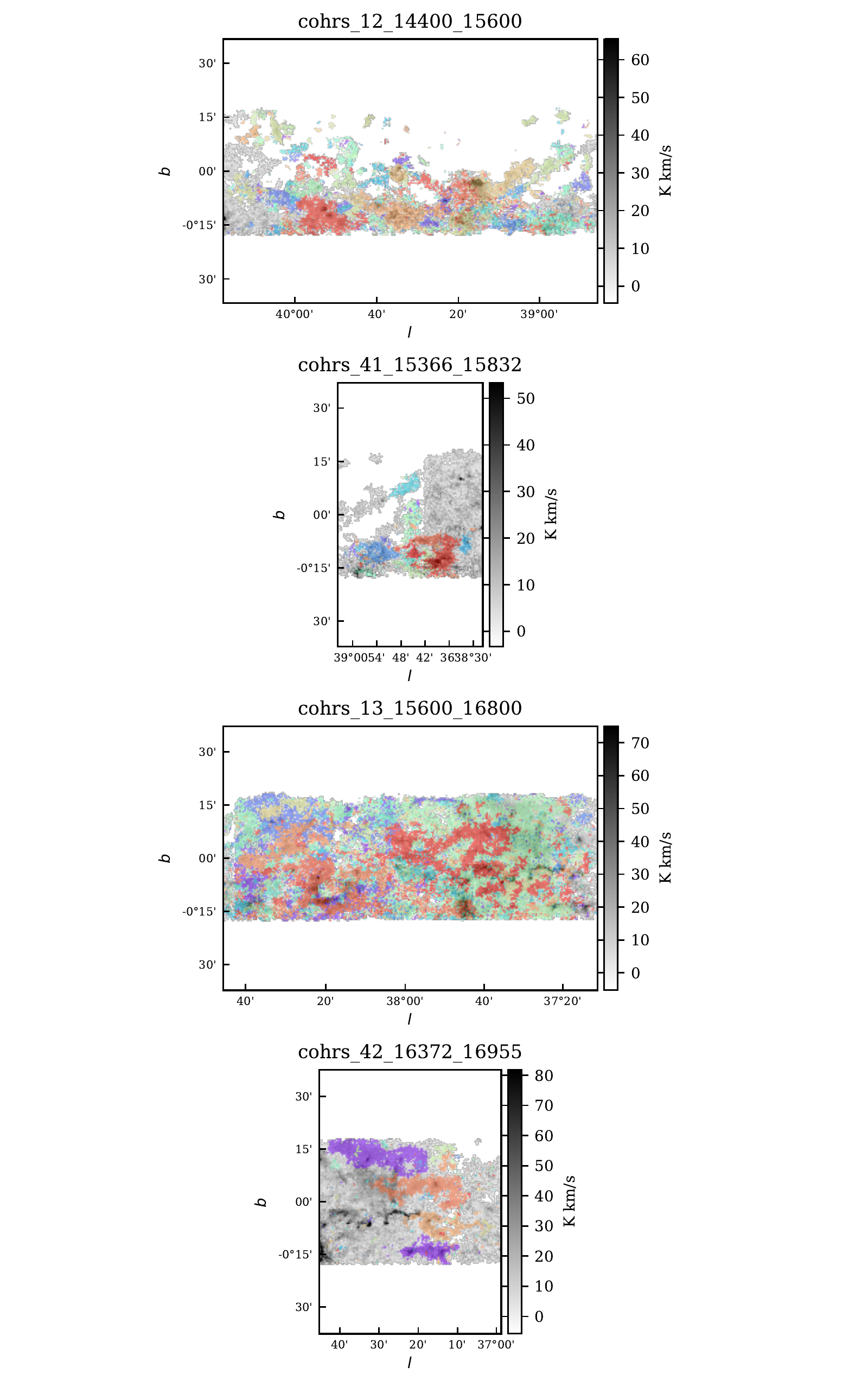}
\caption{Longitude-latitude integrated intensity map of COHRS given sub-cubes masked as explained in Section~\ref{S:scimes_cohrs}. In color the identified clouds are indicated.}
\label{F:cohrs_asgnpp_6}
\end{figure*}

\newpage
\clearpage

\begin{figure*}
\centering
\includegraphics[width=0.75\textwidth]{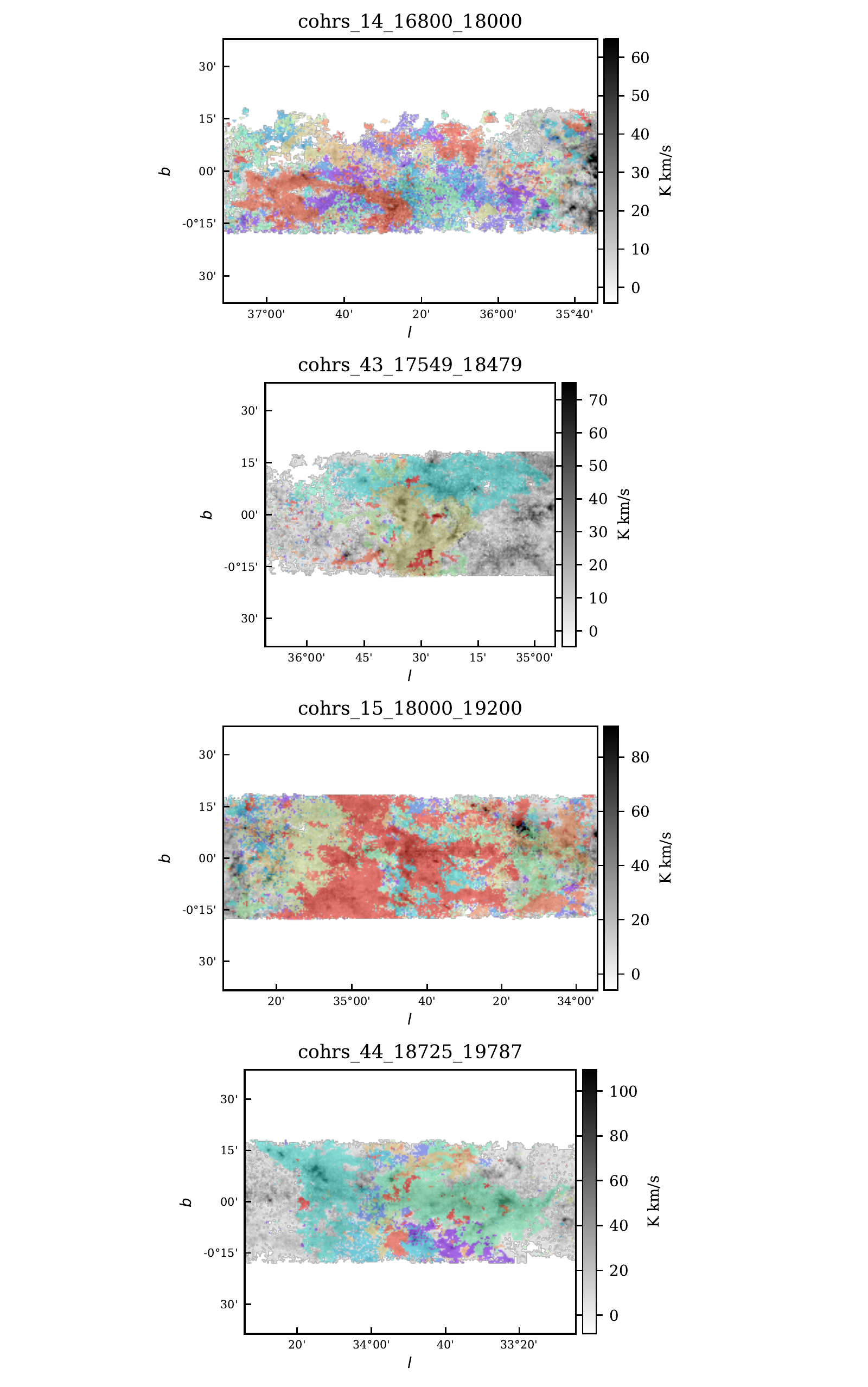}
\caption{Longitude-latitude integrated intensity map of COHRS given sub-cubes masked as explained in Section~\ref{S:scimes_cohrs}. In color the identified clouds are indicated.}
\label{F:cohrs_asgnpp_7}
\end{figure*}

\newpage
\clearpage

\begin{figure*}
\centering
\includegraphics[width=0.75\textwidth]{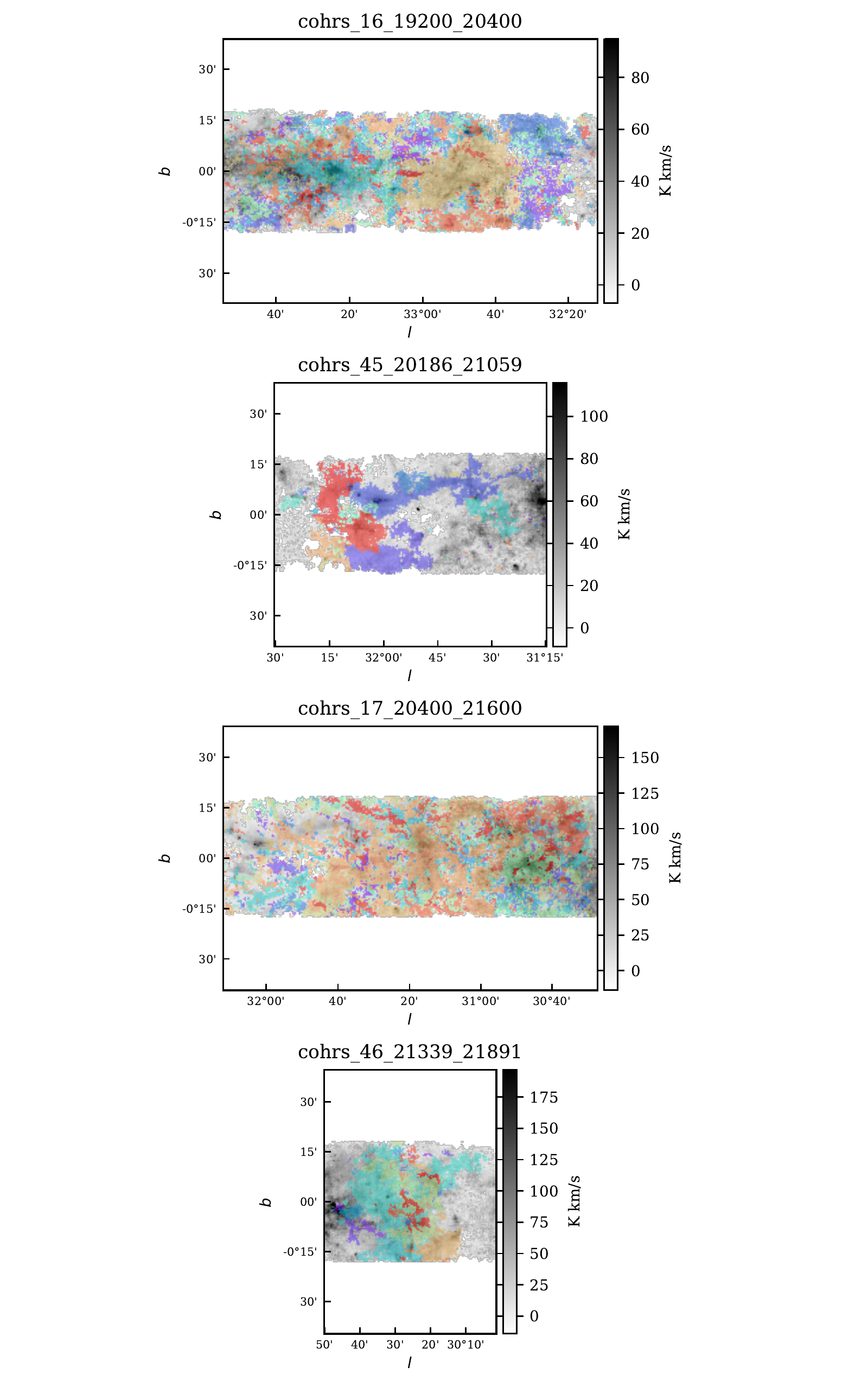}
\caption{Longitude-latitude integrated intensity map of COHRS given sub-cubes masked as explained in Section~\ref{S:scimes_cohrs}. In color the identified clouds are indicated.}
\label{F:cohrs_asgnpp_8}
\end{figure*}

\newpage
\clearpage

\begin{figure*}
\centering
\includegraphics[width=0.75\textwidth]{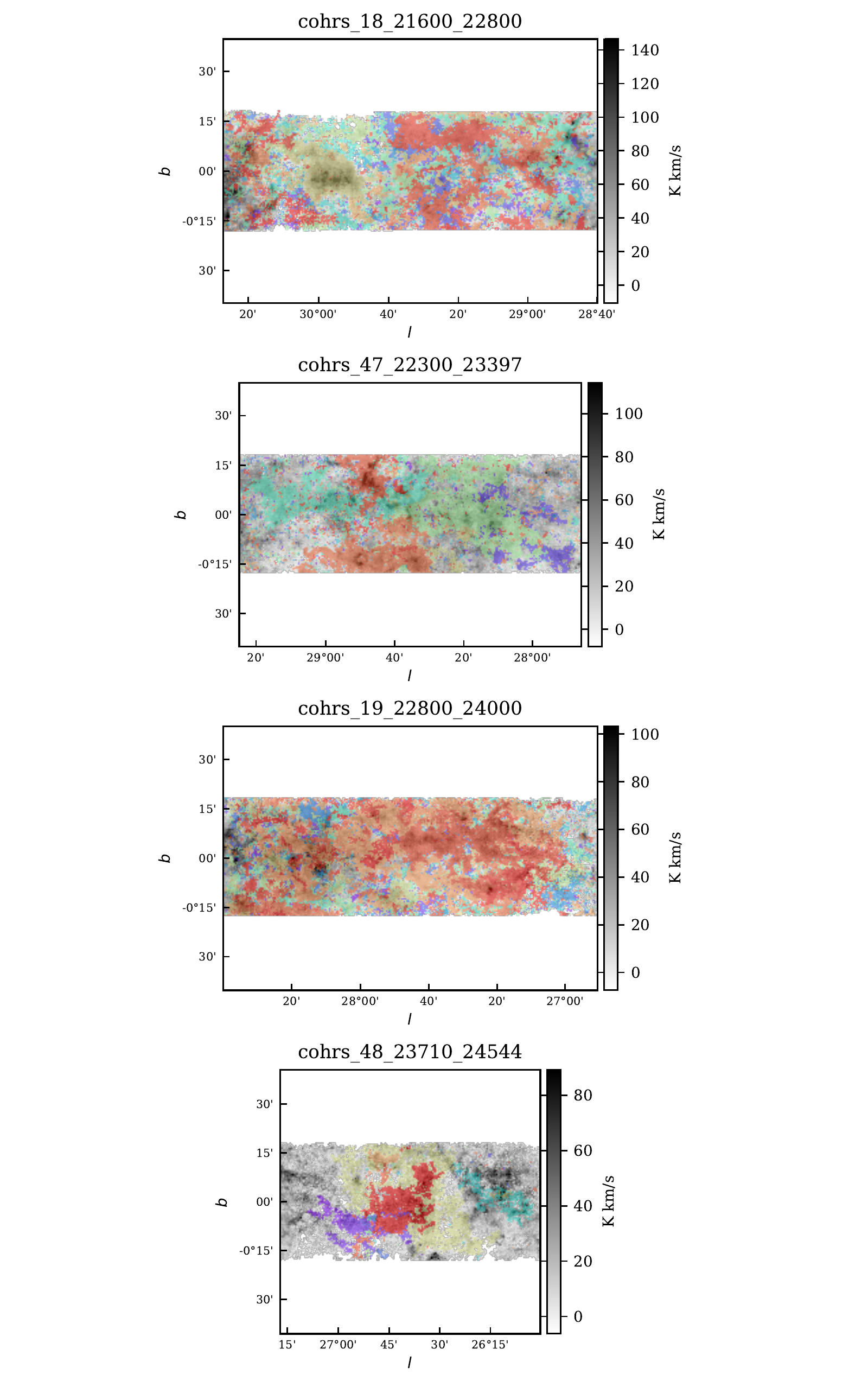}
\caption{Longitude-latitude integrated intensity map of COHRS given sub-cubes masked as explained in Section~\ref{S:scimes_cohrs}. In color the identified clouds are indicated.}
\label{F:cohrs_asgnpp_9}
\end{figure*}

\newpage
\clearpage

\begin{figure*}
\centering
\includegraphics[width=0.75\textwidth]{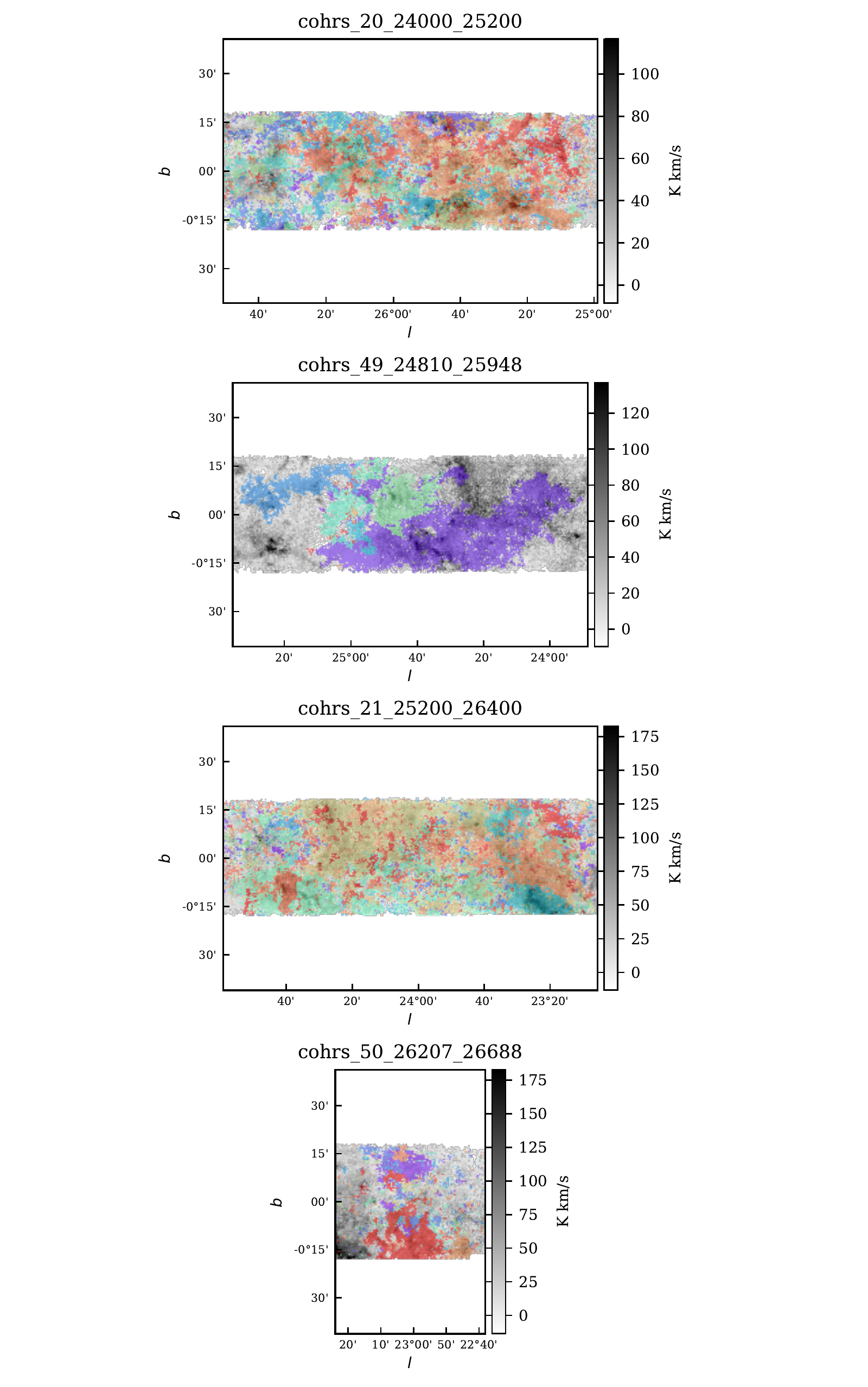}
\caption{Longitude-latitude integrated intensity map of COHRS given sub-cubes masked as explained in Section~\ref{S:scimes_cohrs}. In color the identified clouds are indicated.}
\label{F:cohrs_asgnpp_10}
\end{figure*}

\newpage
\clearpage

\begin{figure*}
\centering
\includegraphics[width=0.75\textwidth]{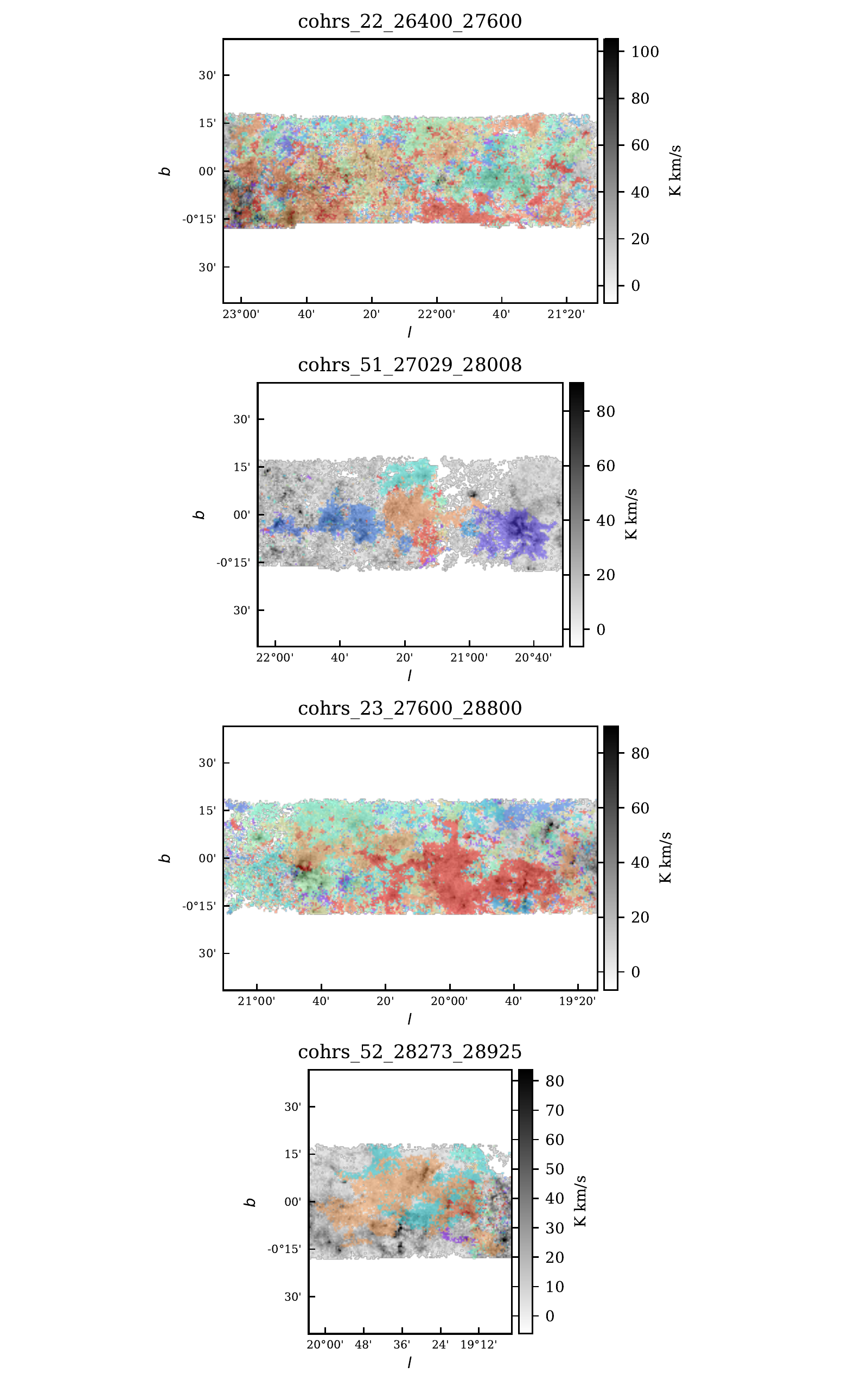}
\caption{Longitude-latitude integrated intensity map of COHRS given sub-cubes masked as explained in Section~\ref{S:scimes_cohrs}. In color the identified clouds are indicated.}
\label{F:cohrs_asgnpp_11}
\end{figure*}

\newpage
\clearpage

\begin{figure*}
\centering
\includegraphics[width=0.75\textwidth]{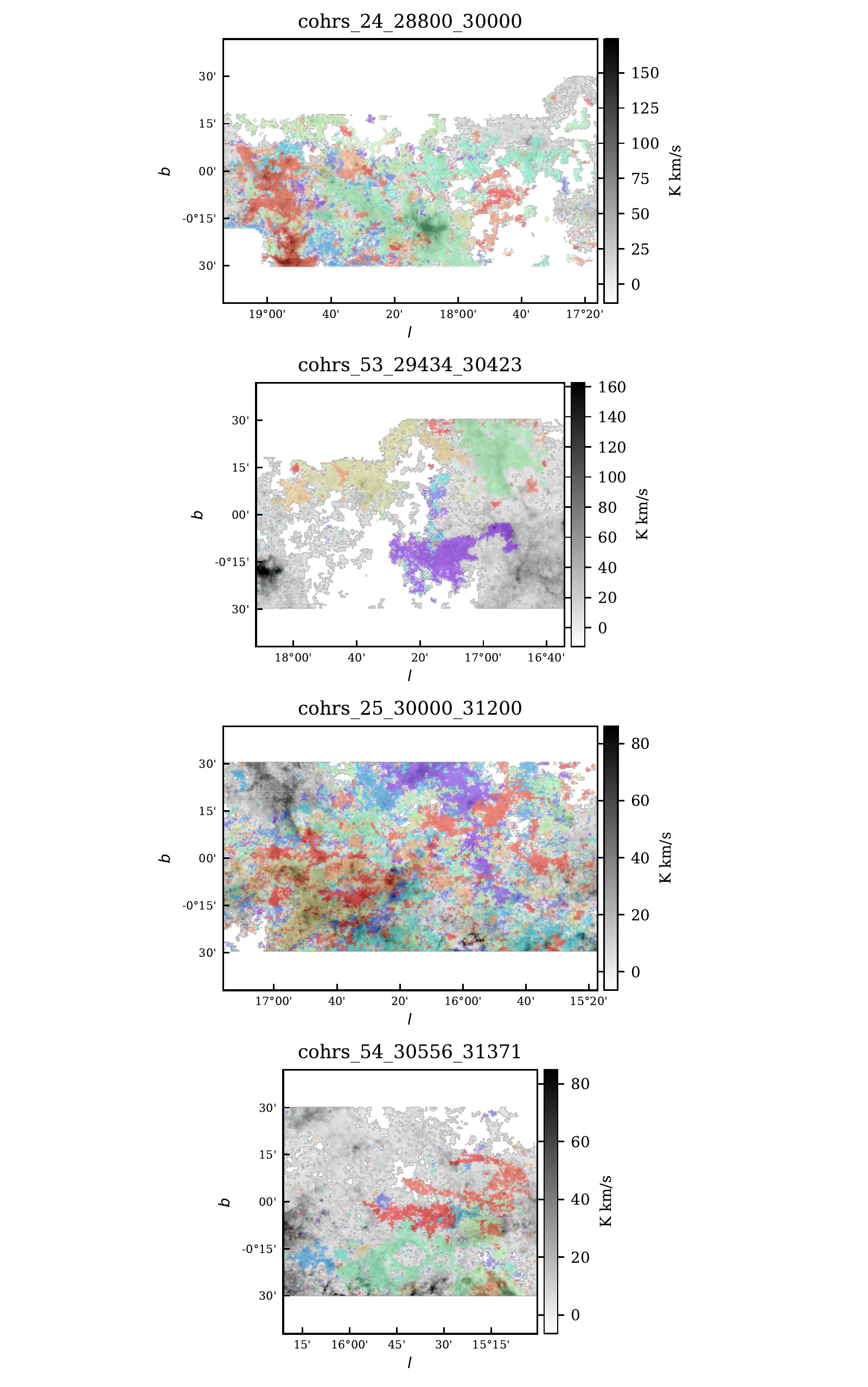}
\caption{Longitude-latitude integrated intensity map of COHRS given sub-cubes masked as explained in Section~\ref{S:scimes_cohrs}. In color the identified clouds are indicated.}
\label{F:cohrs_asgnpp_12}
\end{figure*}

\newpage
\clearpage

\begin{figure*}
\centering
\includegraphics[width=0.75\textwidth]{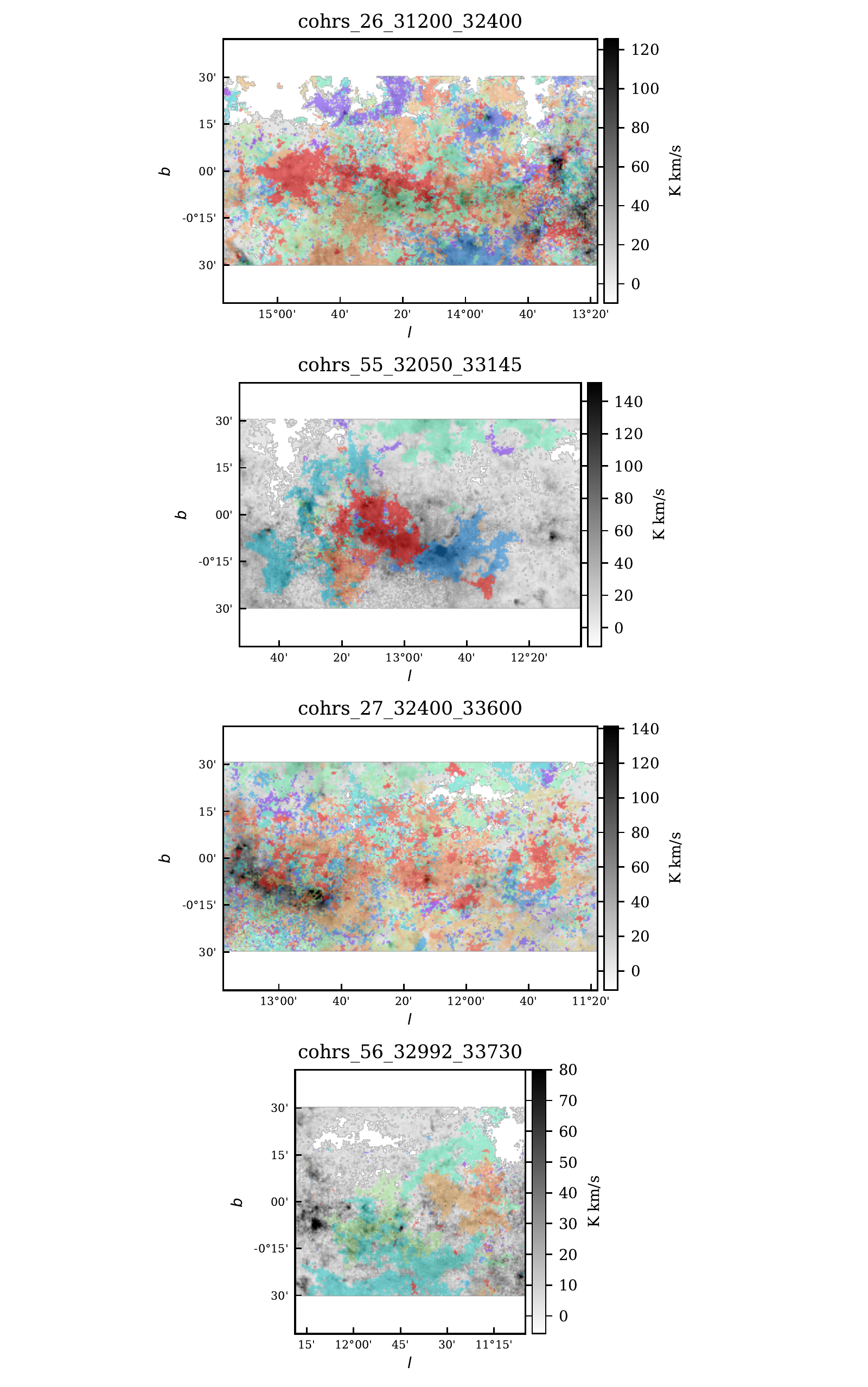}
\caption{Longitude-latitude integrated intensity map of COHRS given sub-cubes masked as explained in Section~\ref{S:scimes_cohrs}. In color the identified clouds are indicated.}
\label{F:cohrs_asgnpp_13}
\end{figure*}

\newpage
\clearpage

\begin{figure*}
\centering
\includegraphics[width=0.75\textwidth]{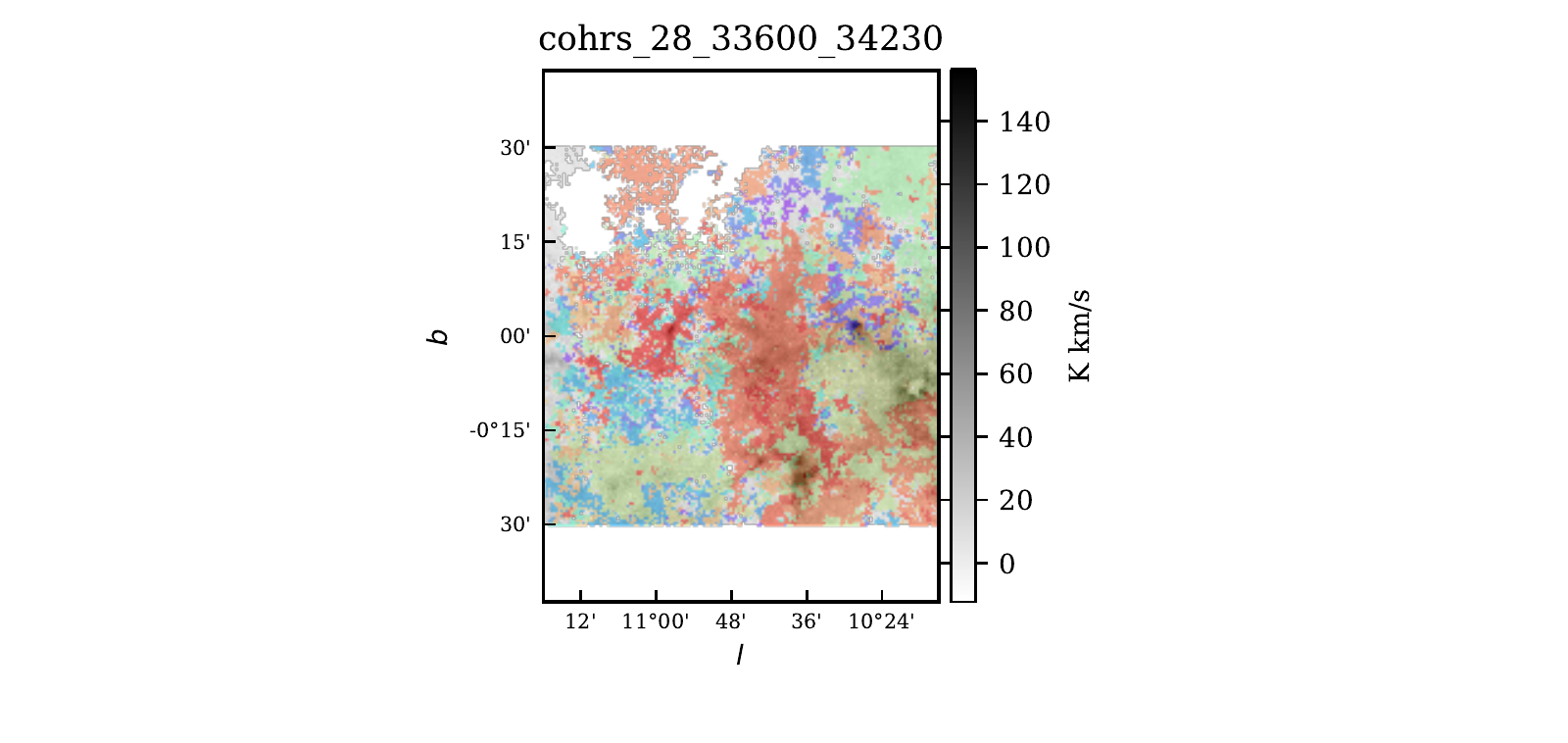}
\caption{Longitude-latitude integrated intensity map of COHRS given sub-cubes masked as explained in Section~\ref{S:scimes_cohrs}. In color the identified clouds are indicated.}
\label{F:cohrs_asgnpp_14}
\end{figure*}

\newpage
\clearpage

\begin{figure*}
\centering
\includegraphics[width=0.85\textwidth]{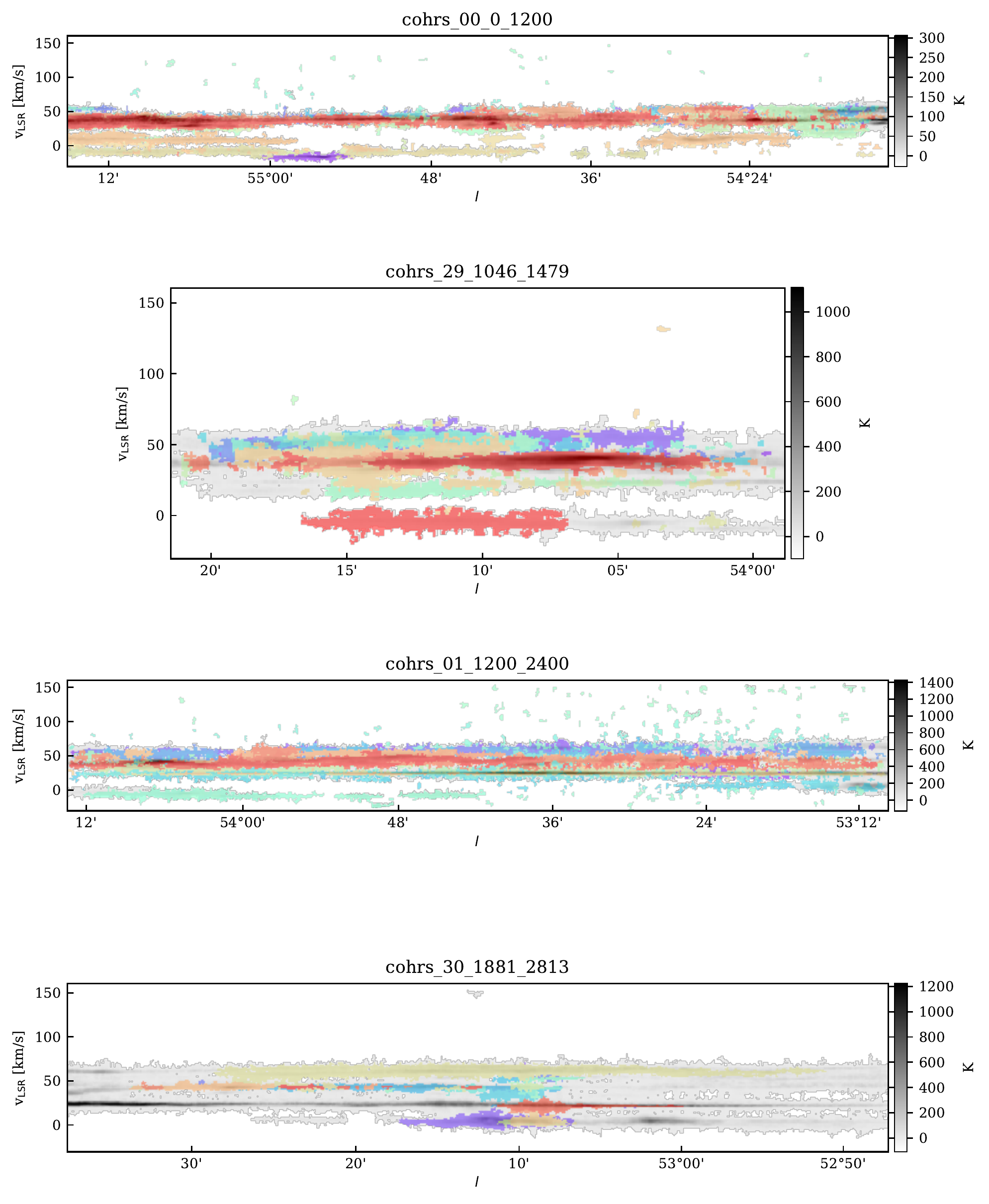}
\caption{Longitude-velocity integrated map of COHRS given sub-cubes masked as explained in Section~\ref{S:scimes_cohrs}. In color the identified clouds are indicated.}
\label{F:cohrs_asgnpv_0}
\end{figure*}

\newpage
\clearpage

\begin{figure*}
\centering
\includegraphics[width=0.85\textwidth]{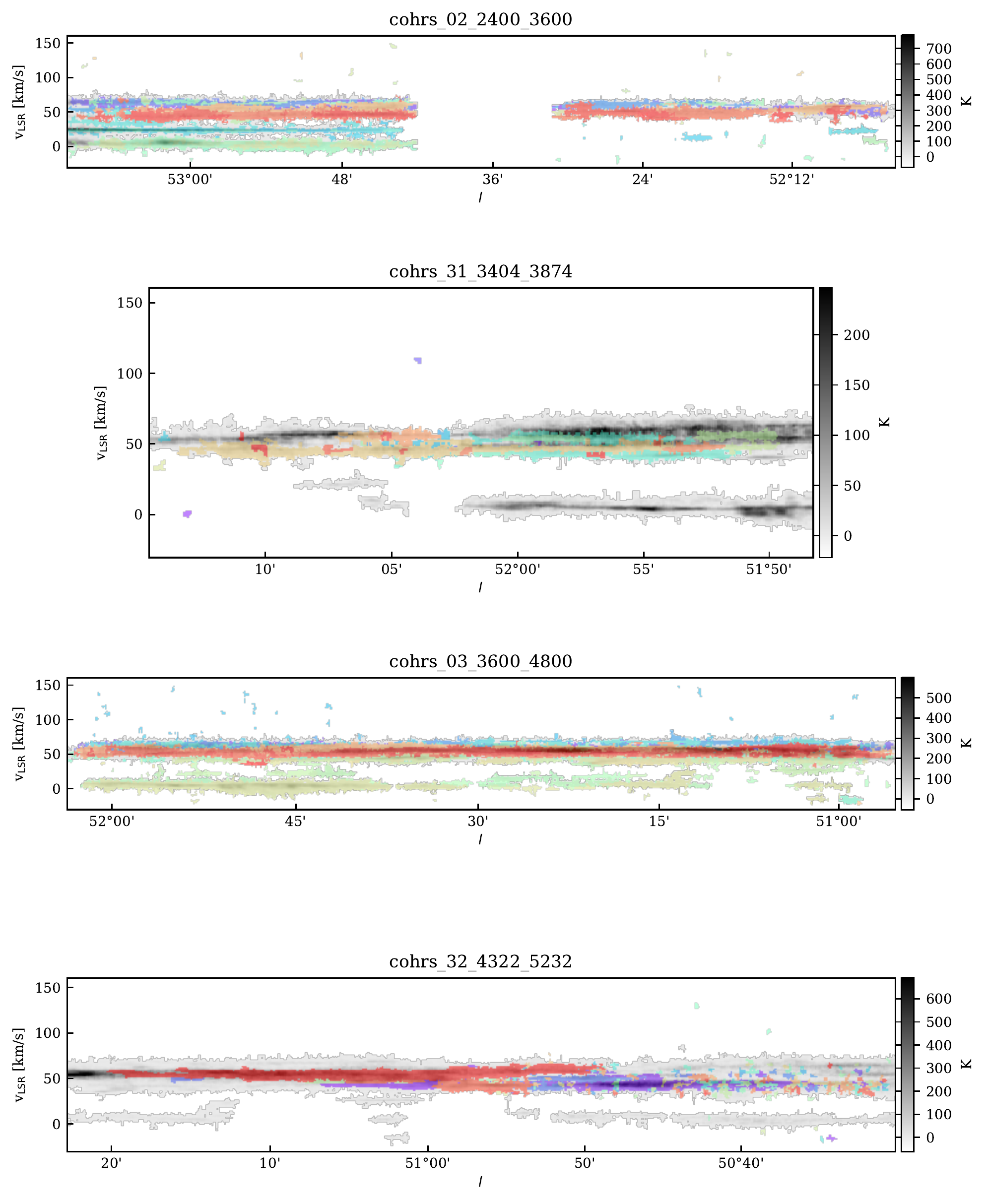}
\caption{Longitude-velocity integrated map of COHRS given sub-cubes masked as explained in Section~\ref{S:scimes_cohrs}. In color the identified clouds are indicated.}
\label{F:cohrs_asgnpv_1}
\end{figure*}

\newpage
\clearpage

\begin{figure*}
\centering
\includegraphics[width=0.85\textwidth]{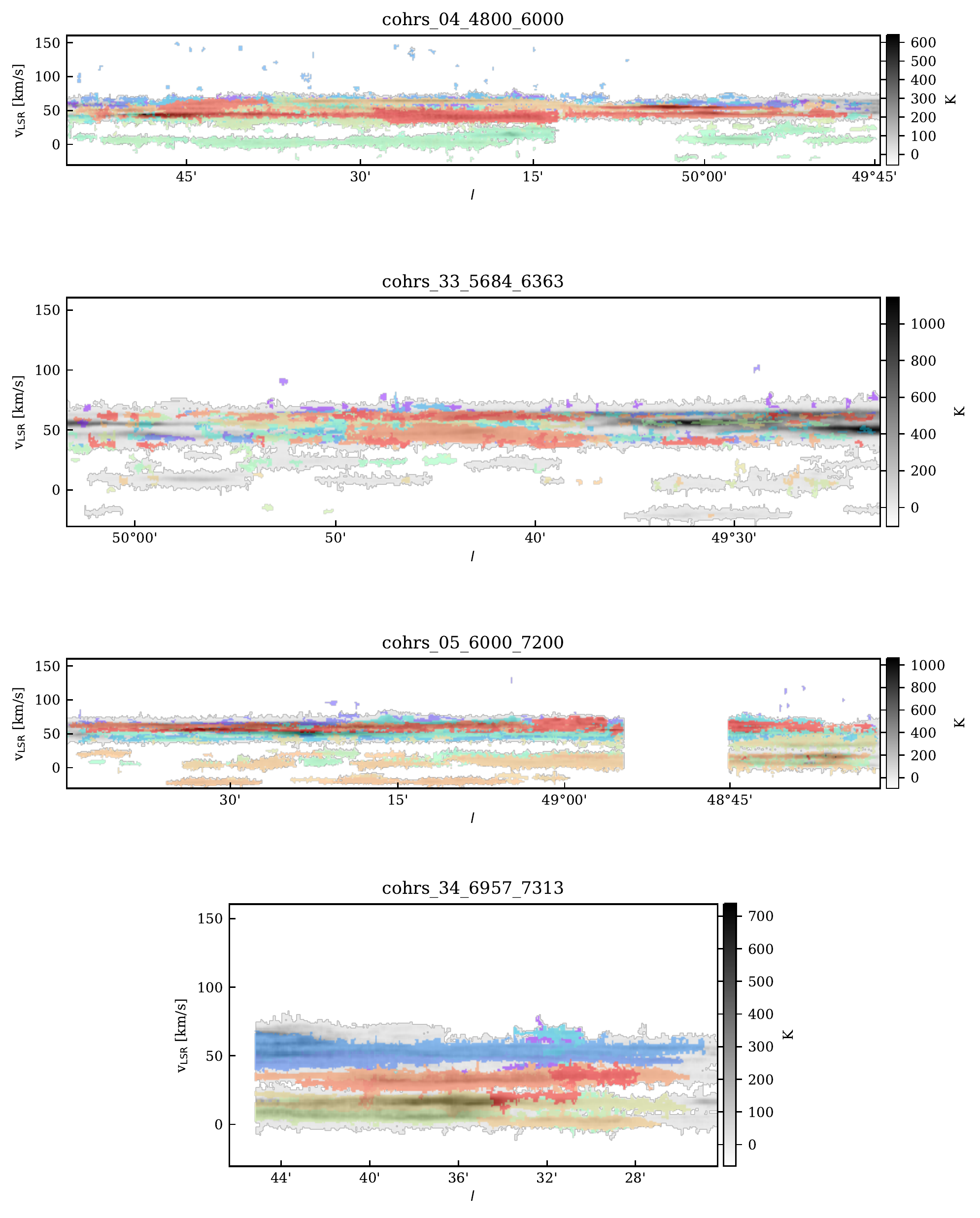}
\caption{Longitude-velocity integrated map of COHRS given sub-cubes masked as explained in Section~\ref{S:scimes_cohrs}. In color the identified clouds are indicated.}
\label{F:cohrs_asgnpv_2}
\end{figure*}

\newpage
\clearpage

\begin{figure*}
\centering
\includegraphics[width=0.85\textwidth]{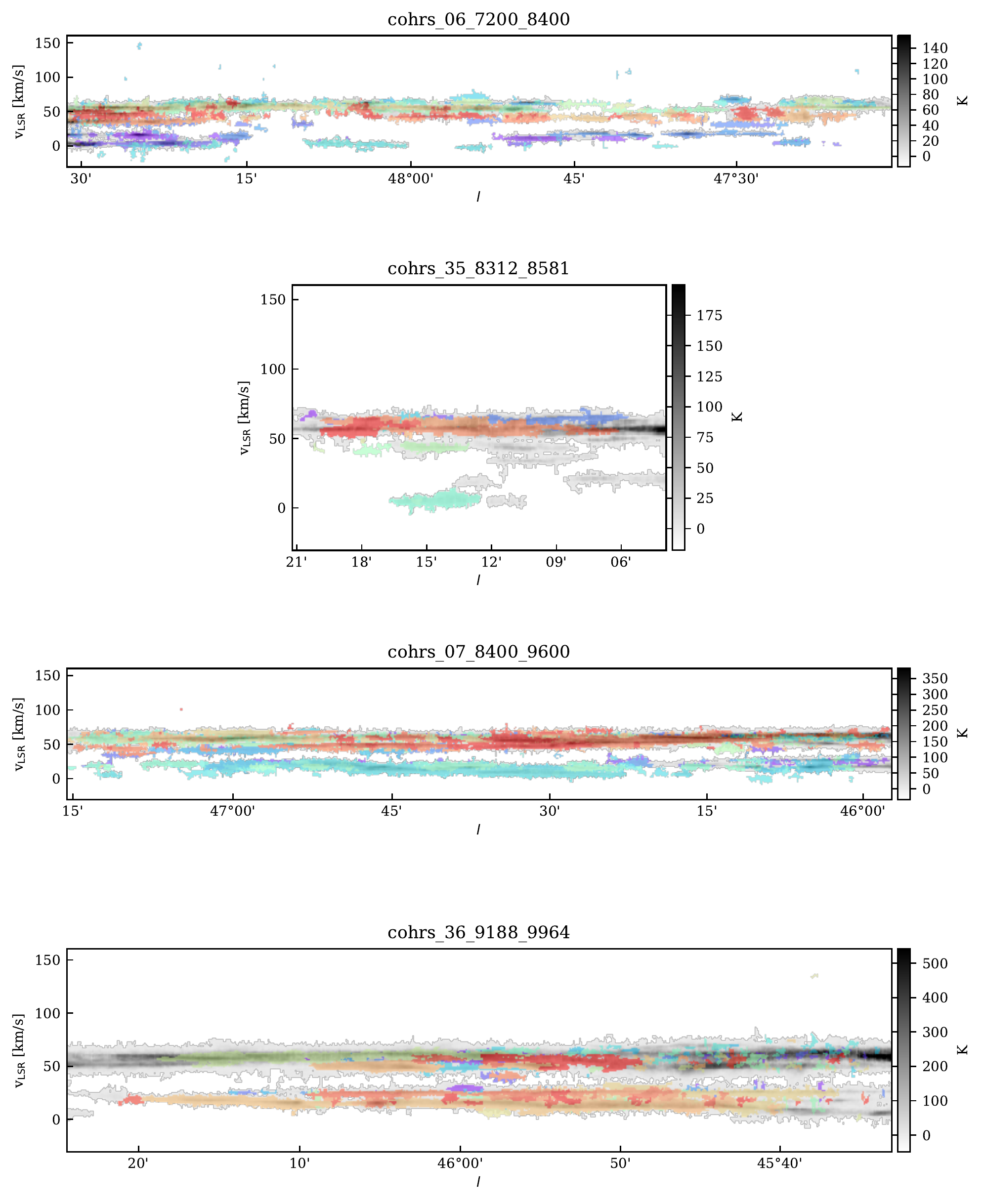}
\caption{Longitude-velocity integrated map of COHRS given sub-cubes masked as explained in Section~\ref{S:scimes_cohrs}. In color the identified clouds are indicated.}
\label{F:cohrs_asgnpv_3}
\end{figure*}

\newpage
\clearpage

\begin{figure*}
\centering
\includegraphics[width=0.85\textwidth]{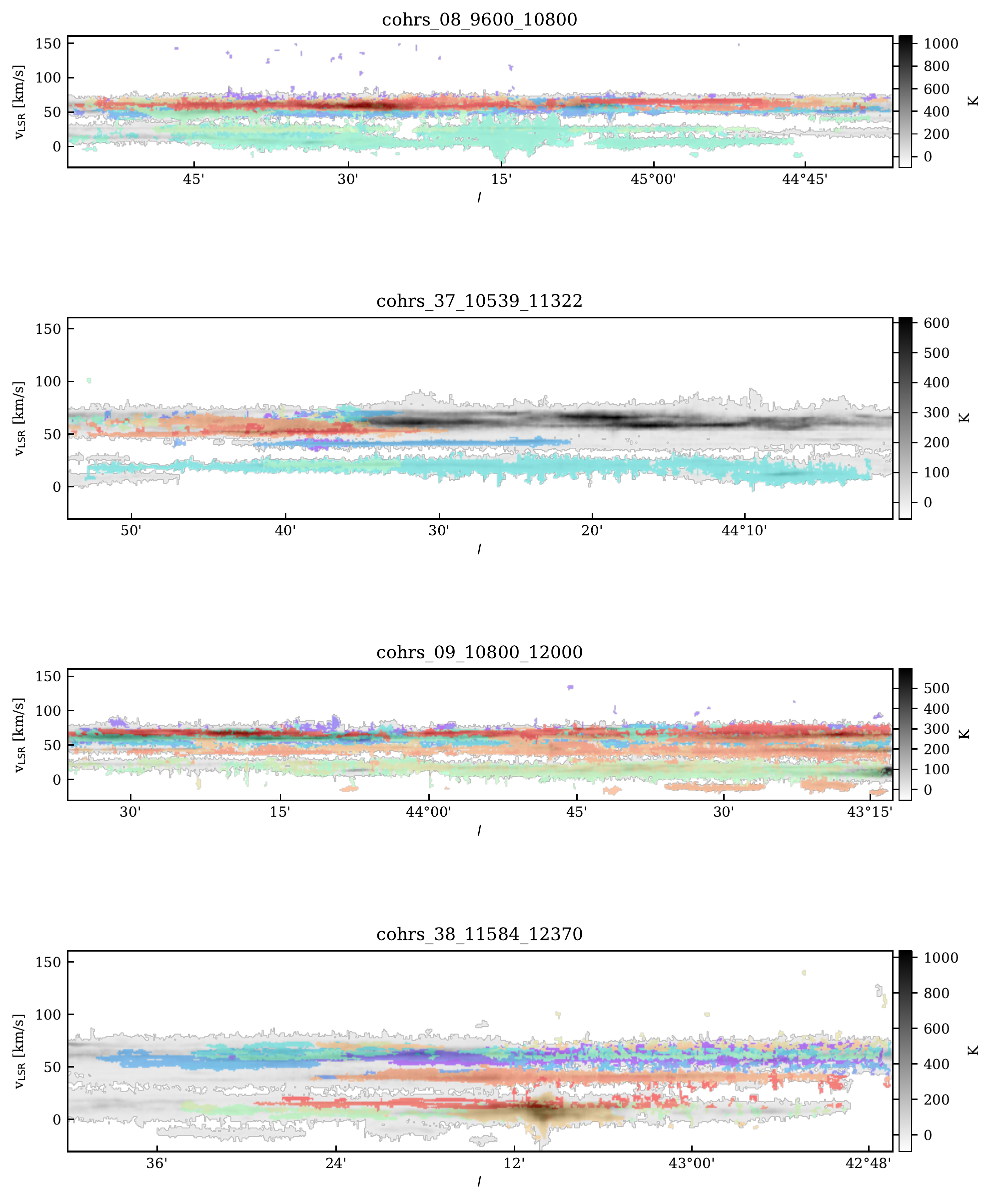}
\caption{Longitude-velocity integrated map of COHRS given sub-cubes masked as explained in Section~\ref{S:scimes_cohrs}. In color the identified clouds are indicated.}
\label{F:cohrs_asgnpv_4}
\end{figure*}

\newpage
\clearpage

\begin{figure*}
\centering
\includegraphics[width=0.85\textwidth]{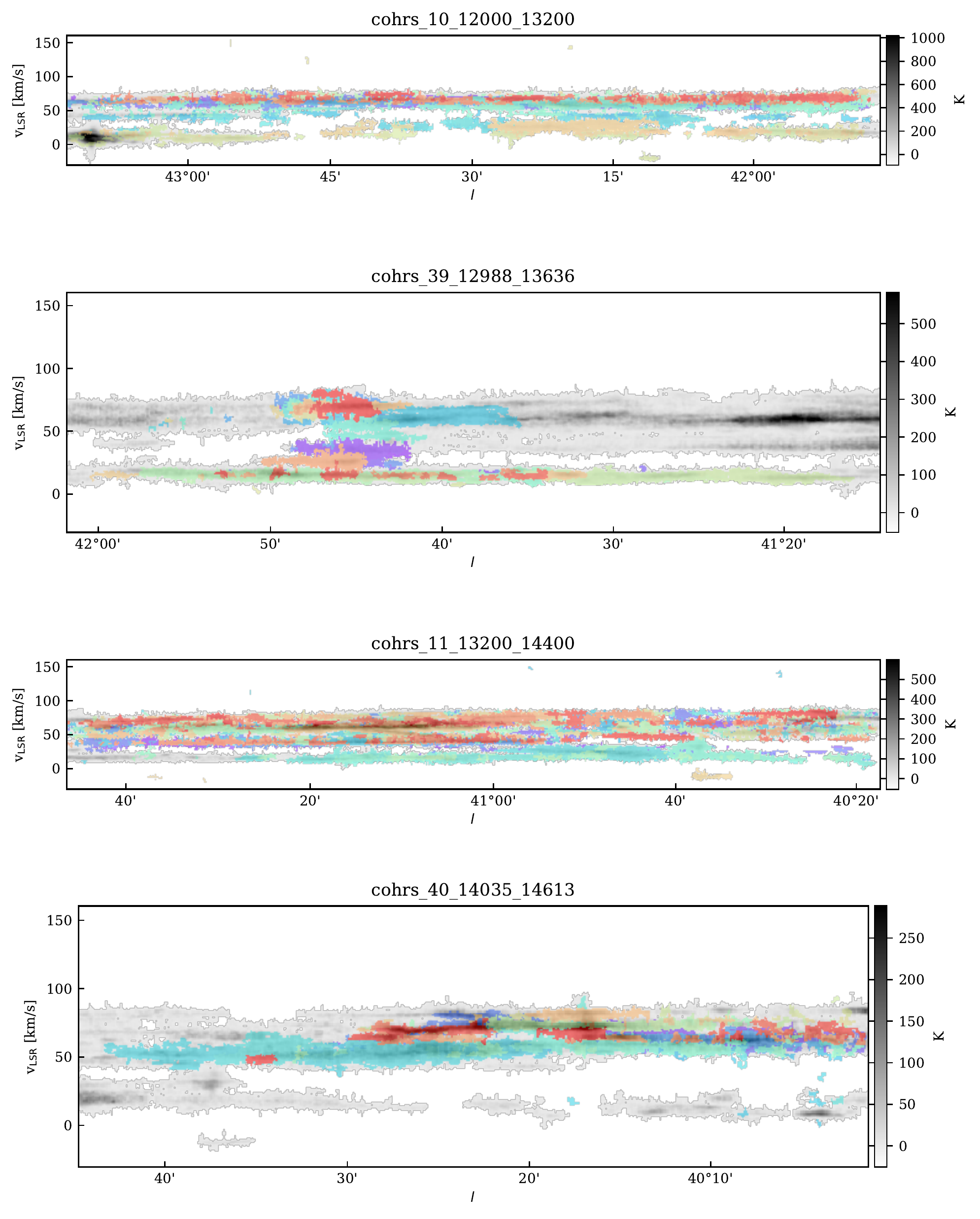}
\caption{Longitude-velocity integrated map of COHRS given sub-cubes masked as explained in Section~\ref{S:scimes_cohrs}. In color the identified clouds are indicated.}
\label{F:cohrs_asgnpv_5}
\end{figure*}

\newpage
\clearpage

\begin{figure*}
\centering
\includegraphics[width=0.85\textwidth]{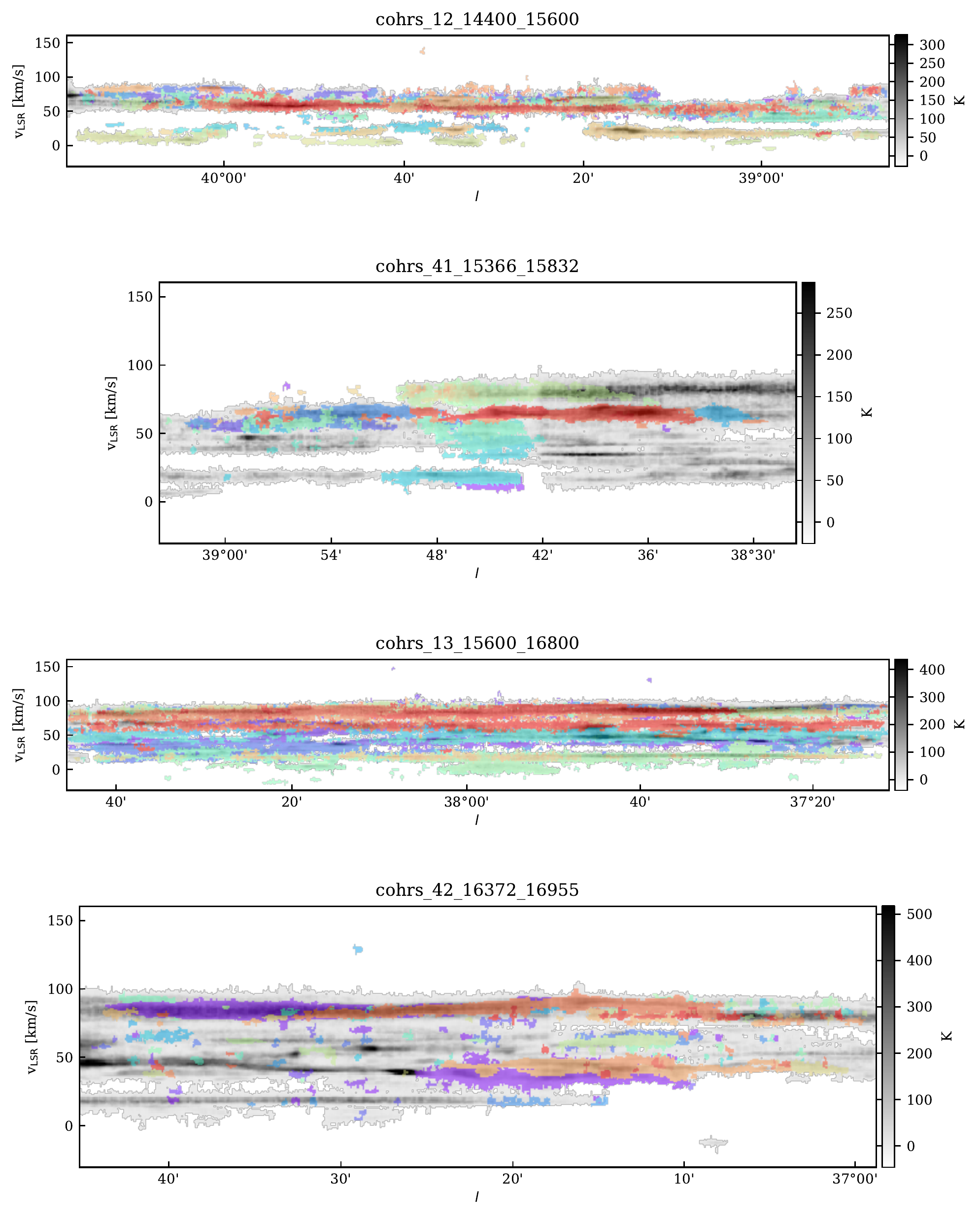}
\caption{Longitude-velocity integrated map of COHRS given sub-cubes masked as explained in Section~\ref{S:scimes_cohrs}. In color the identified clouds are indicated.}
\label{F:cohrs_asgnpv_6}
\end{figure*}

\newpage
\clearpage

\begin{figure*}
\centering
\includegraphics[width=0.85\textwidth]{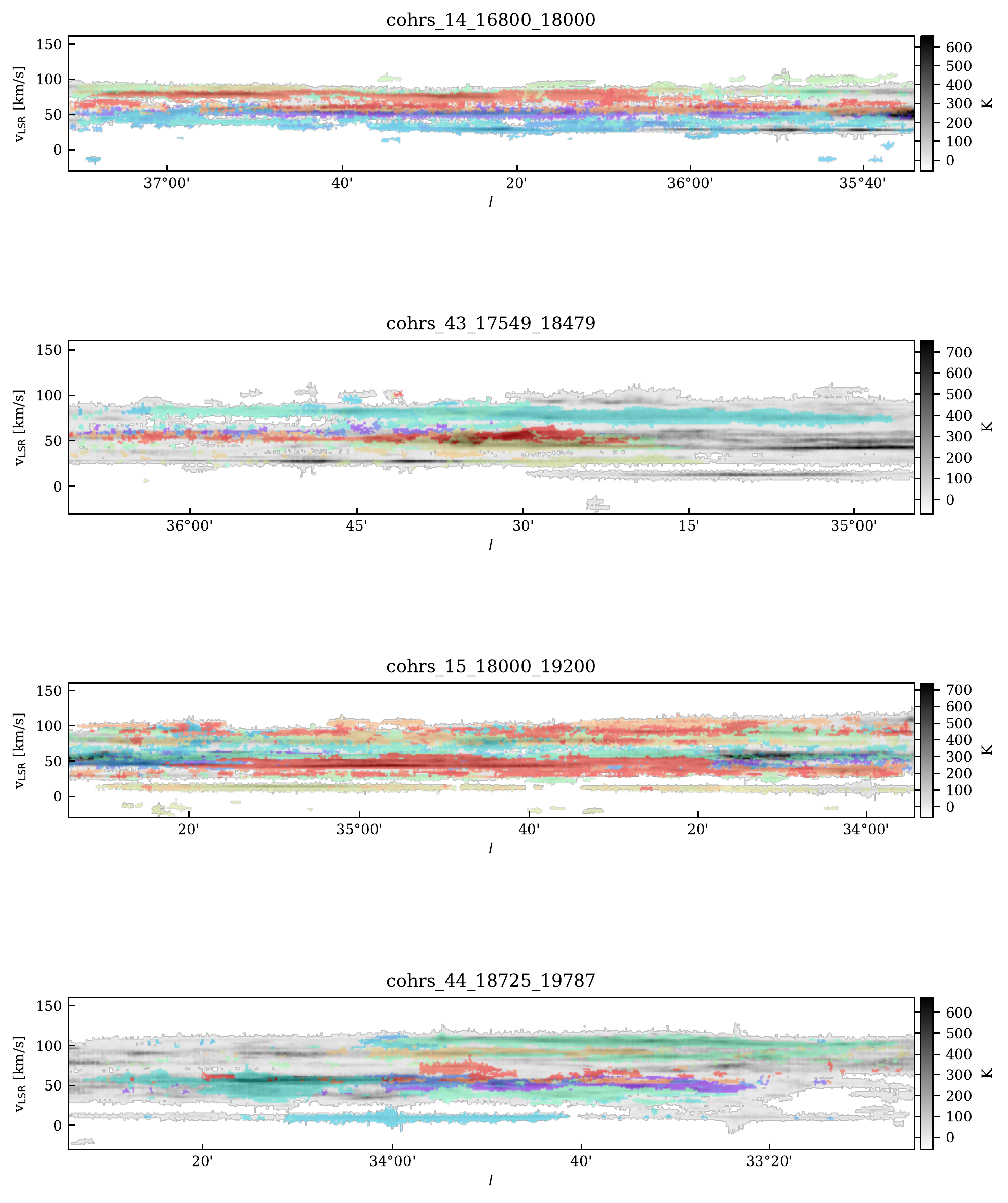}
\caption{Longitude-velocity integrated map of COHRS given sub-cubes masked as explained in Section~\ref{S:scimes_cohrs}. In color the identified clouds are indicated.}
\label{F:cohrs_asgnpv_7}
\end{figure*}

\newpage
\clearpage

\begin{figure*}
\centering
\includegraphics[width=0.85\textwidth]{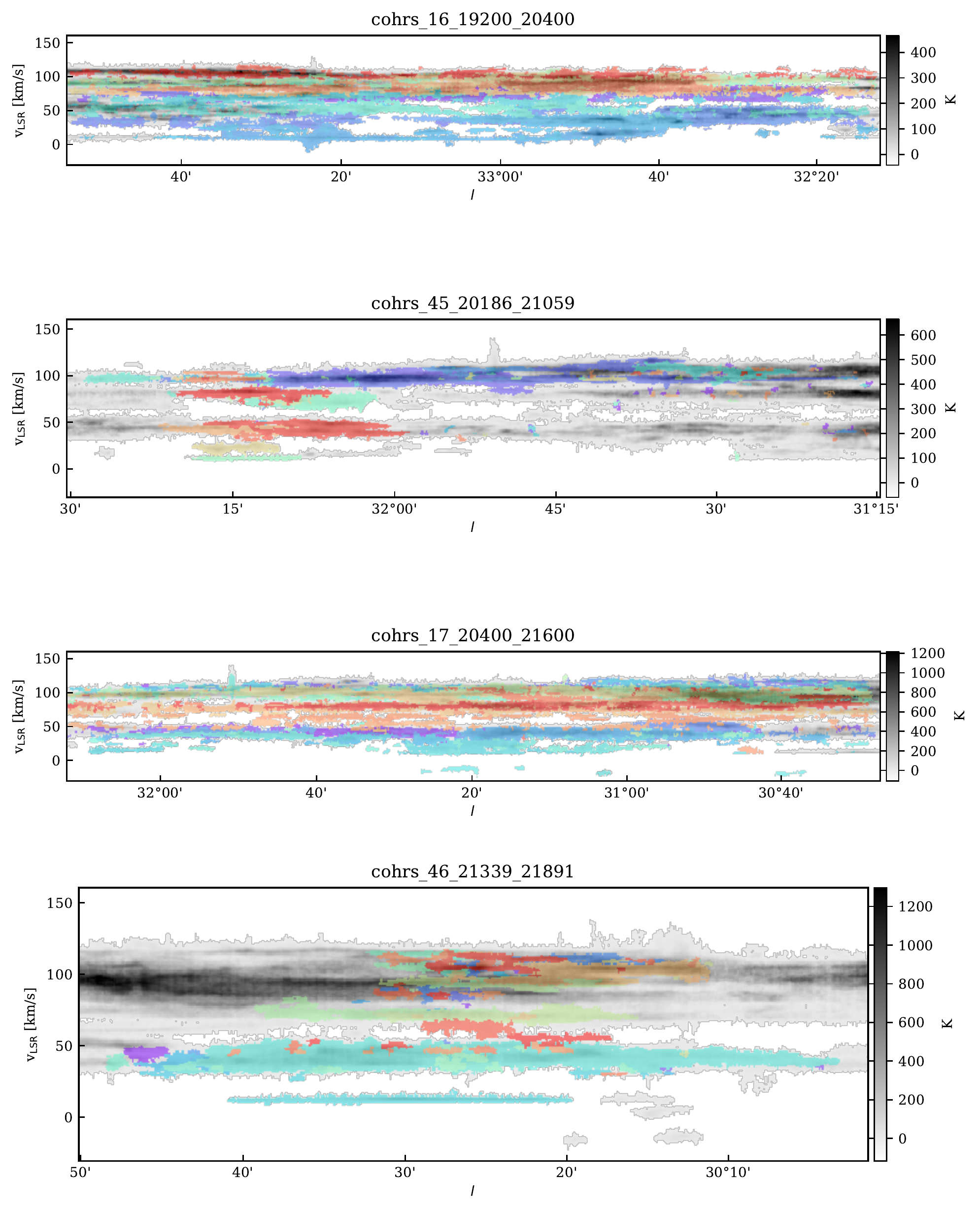}
\caption{Longitude-velocity integrated map of COHRS given sub-cubes masked as explained in Section~\ref{S:scimes_cohrs}. In color the identified clouds are indicated.}
\label{F:cohrs_asgnpv_8}
\end{figure*}

\newpage
\clearpage

\begin{figure*}
\centering
\includegraphics[width=0.85\textwidth]{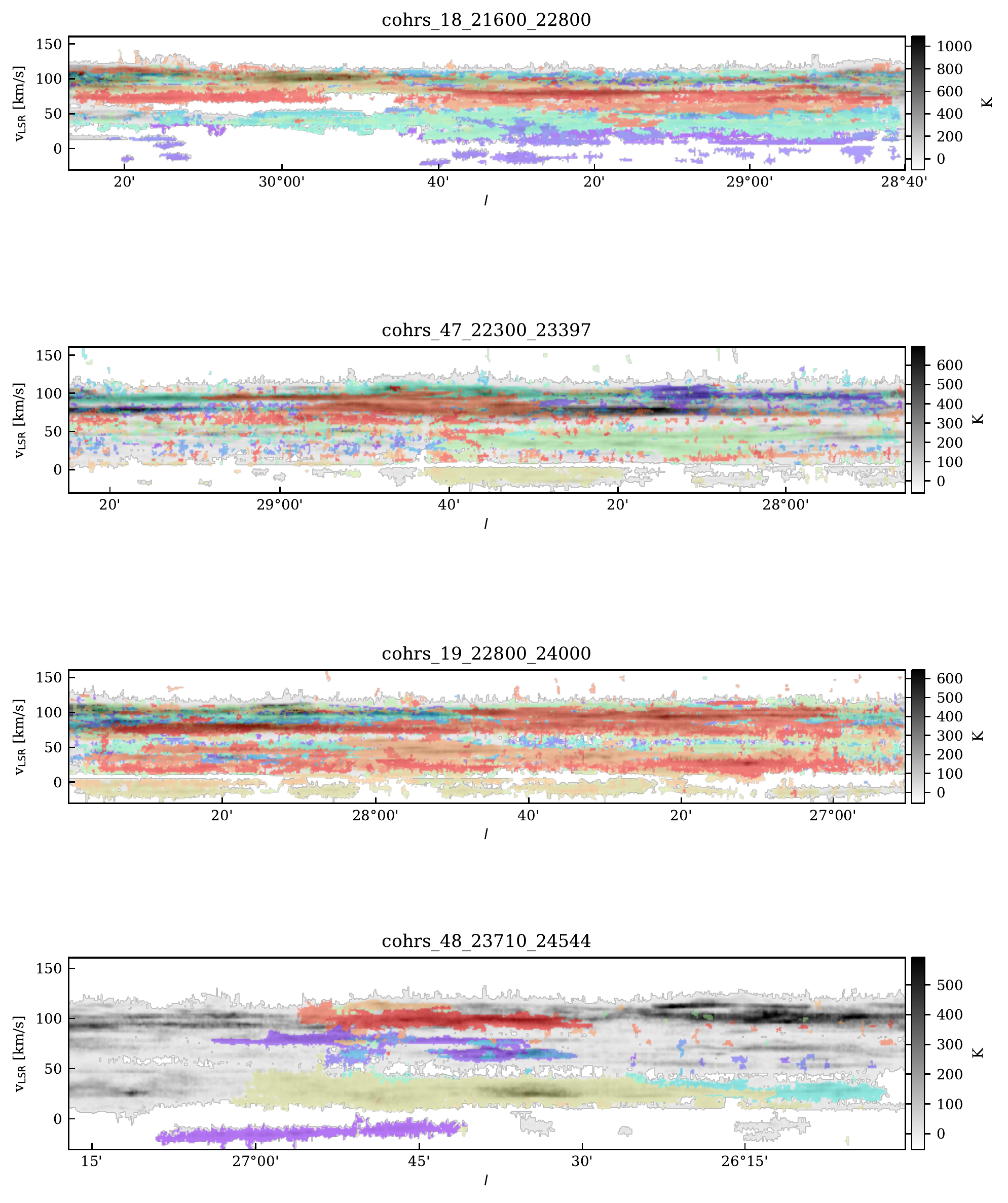}
\caption{Longitude-velocity integrated map of COHRS given sub-cubes masked as explained in Section~\ref{S:scimes_cohrs}. In color the identified clouds are indicated.}
\label{F:cohrs_asgnpv_9}
\end{figure*}

\newpage
\clearpage

\begin{figure*}
\centering
\includegraphics[width=0.85\textwidth]{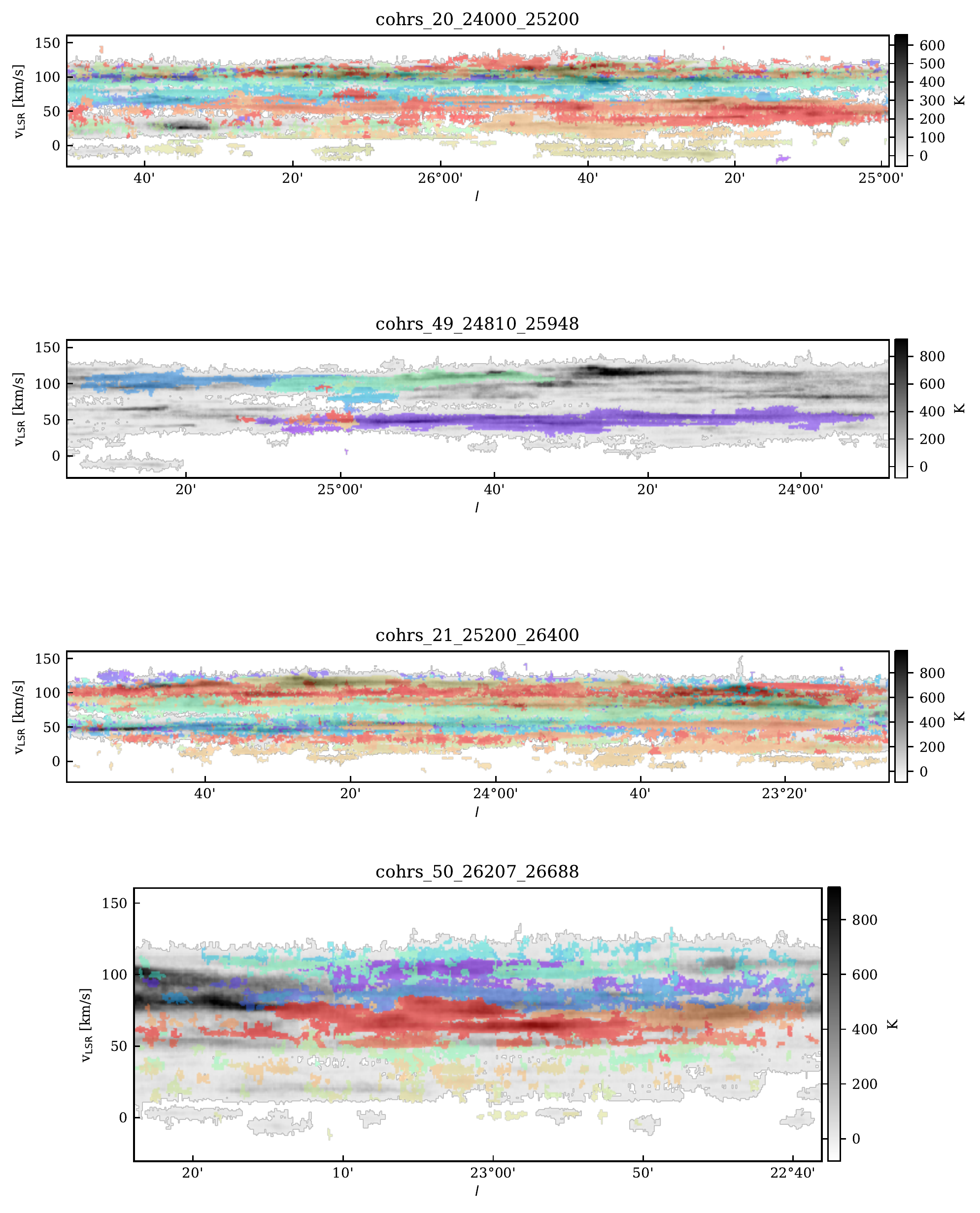}
\caption{Longitude-velocity integrated map of COHRS given sub-cubes masked as explained in Section~\ref{S:scimes_cohrs}. In color the identified clouds are indicated.}
\label{F:cohrs_asgnpv_10}
\end{figure*}

\newpage
\clearpage

\begin{figure*}
\centering
\includegraphics[width=0.85\textwidth]{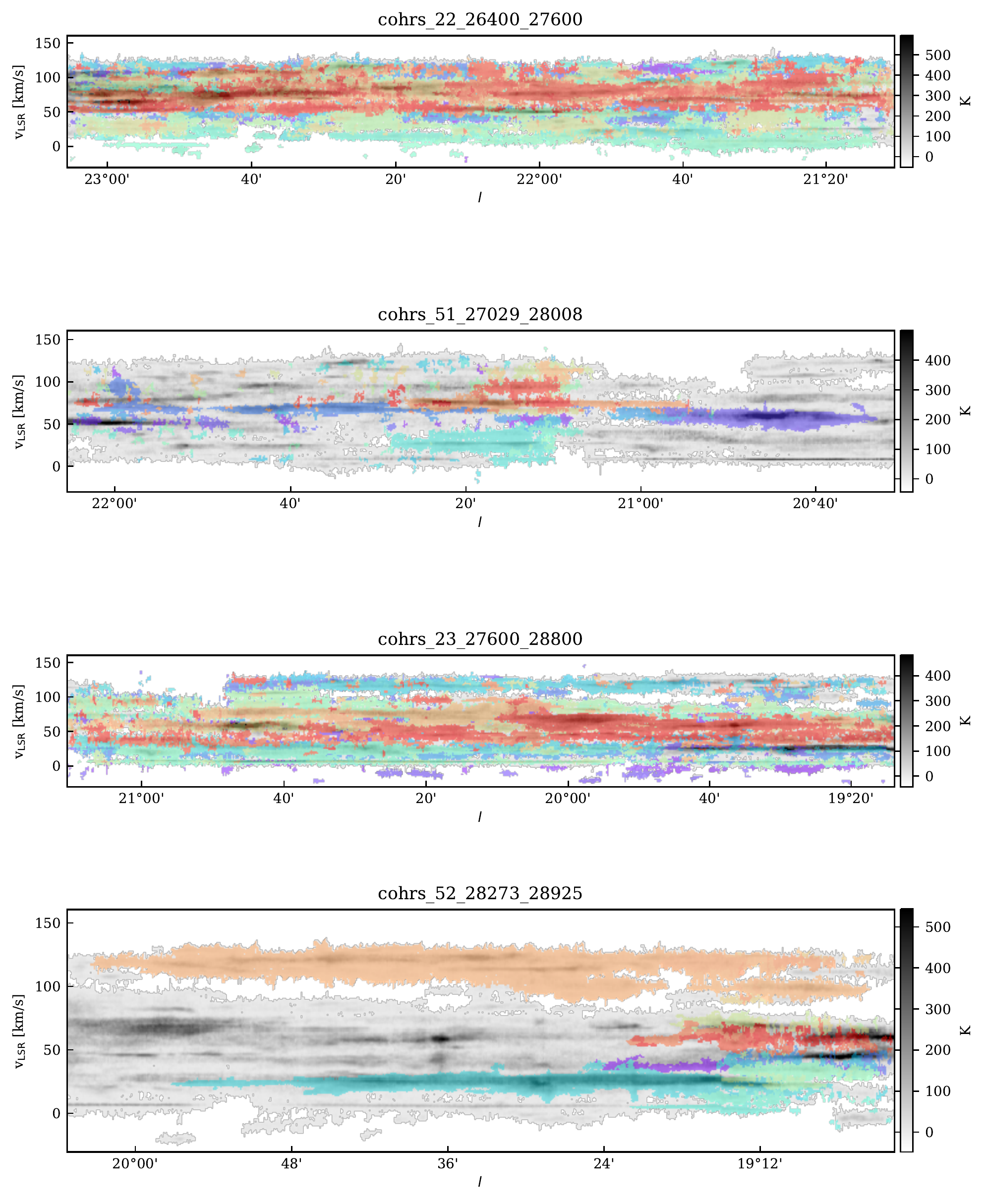}
\caption{Longitude-velocity integrated map of COHRS given sub-cubes masked as explained in Section~\ref{S:scimes_cohrs}. In color the identified clouds are indicated.}
\label{F:cohrs_asgnpv_11}
\end{figure*}

\newpage
\clearpage

\begin{figure*}
\centering
\includegraphics[width=0.85\textwidth]{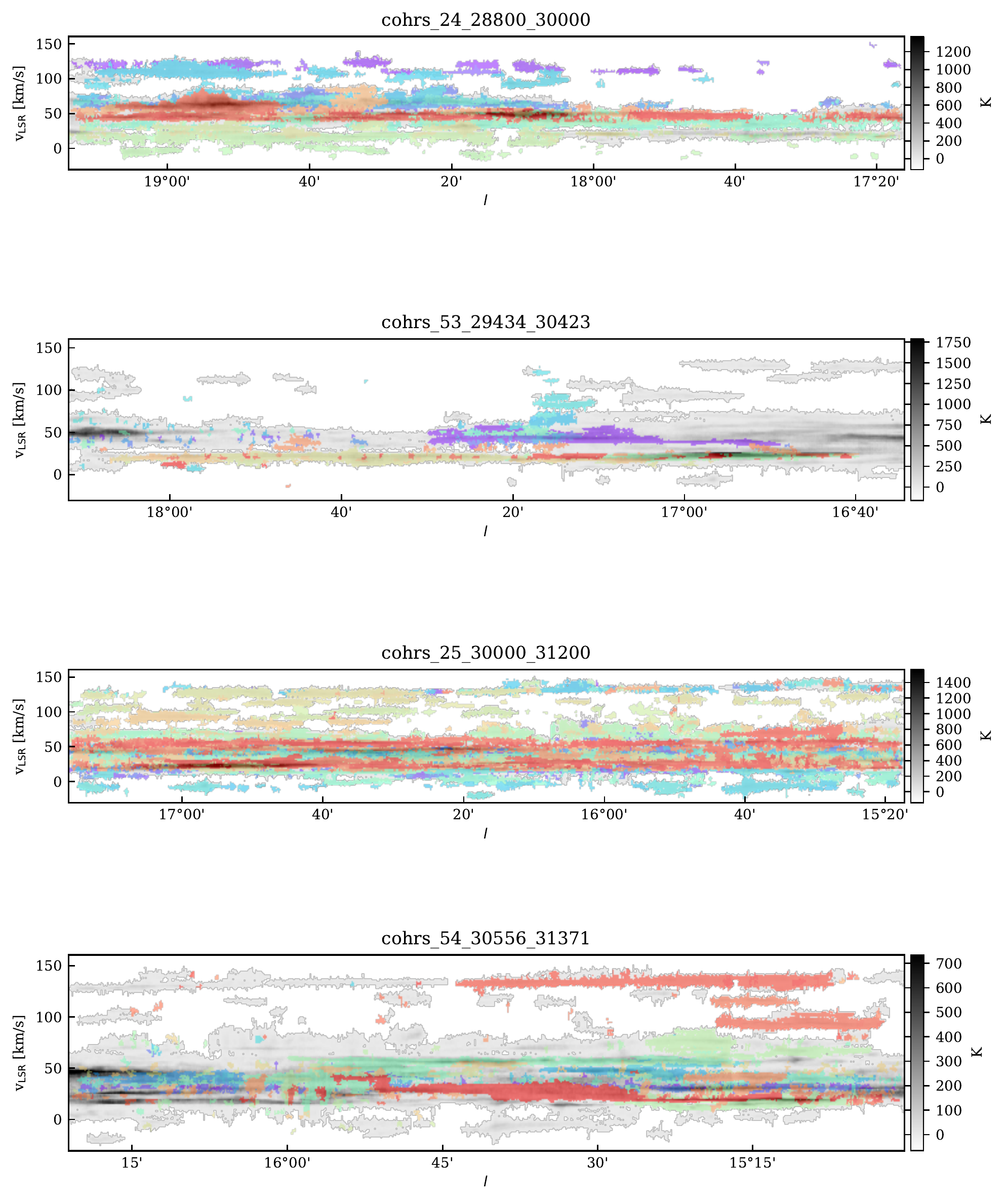}
\caption{Longitude-velocity integrated map of COHRS given sub-cubes masked as explained in Section~\ref{S:scimes_cohrs}. In color the identified clouds are indicated.}
\label{F:cohrs_asgnpv_12}
\end{figure*}

\newpage
\clearpage

\begin{figure*}
\centering
\includegraphics[width=0.85\textwidth]{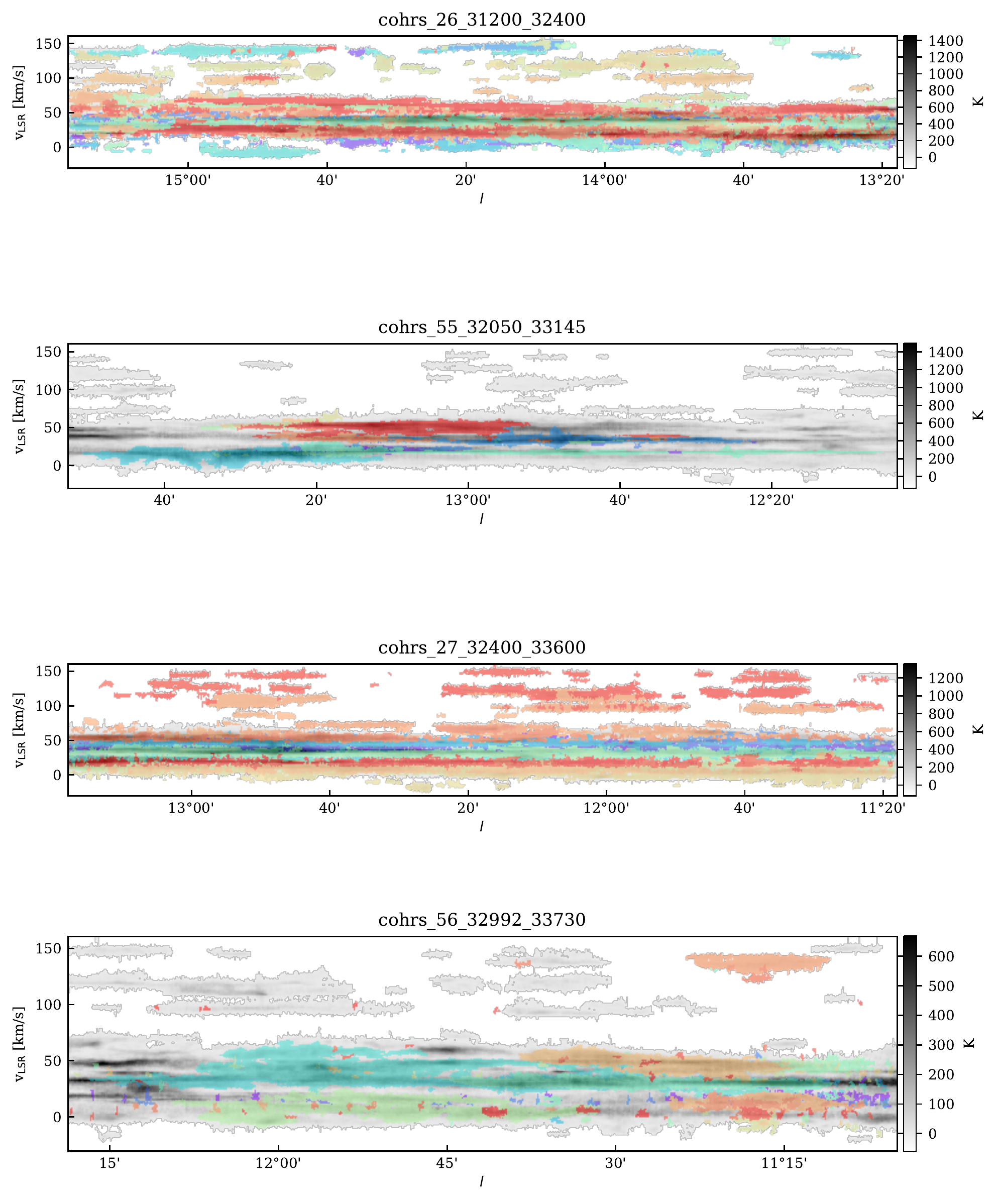}
\caption{Longitude-velocity integrated map of COHRS given sub-cubes masked as explained in Section~\ref{S:scimes_cohrs}. In color the identified clouds are indicated.}
\label{F:cohrs_asgnpv_13}
\end{figure*}

\newpage
\clearpage

\begin{figure*}
\centering
\includegraphics[width=0.85\textwidth]{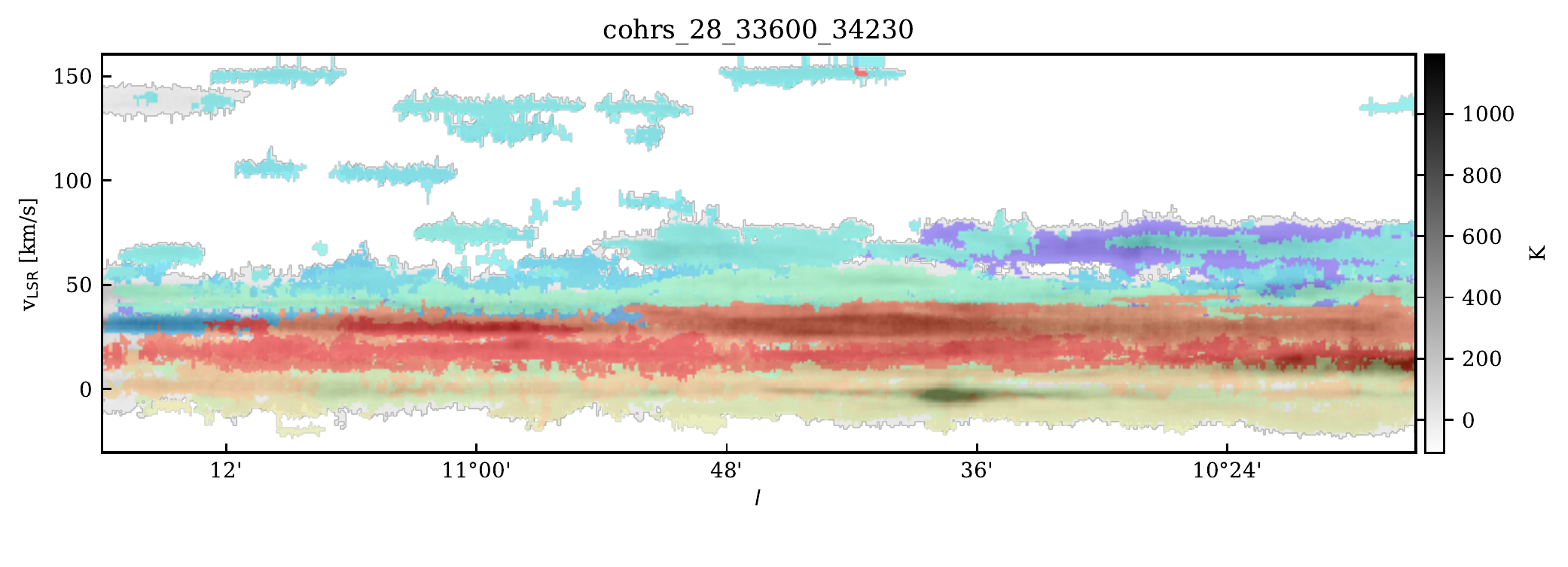}
\caption{Longitude-velocity integrated map of COHRS given sub-cubes masked as explained in Section~\ref{S:scimes_cohrs}. In color the identified clouds are indicated.}
\label{F:cohrs_asgnpv_14}
\end{figure*}

\newpage
\clearpage

\end{document}